\documentclass[11pt,english,preprint,aps,prd,showpacs,superscriptaddress,nofootinbib,tightenlines]{revtex4}
\usepackage[latin9]{inputenc}
\usepackage{babel}
\usepackage{array}
\usepackage{multirow}
\usepackage{amsmath}
\usepackage{amssymb}
\usepackage{graphicx}

\makeatletter

\providecommand{\tabularnewline}{\\}

\@ifundefined{textcolor}{}
{%
 \definecolor{BLACK}{gray}{0}
 \definecolor{WHITE}{gray}{1}
 \definecolor{RED}{rgb}{1,0,0}
 \definecolor{GREEN}{rgb}{0,1,0}
 \definecolor{BLUE}{rgb}{0,0,1}
 \definecolor{CYAN}{cmyk}{1,0,0,0}
 \definecolor{MAGENTA}{cmyk}{0,1,0,0}
 \definecolor{YELLOW}{cmyk}{0,0,1,0}
}

\usepackage{babel}
\usepackage{babel}

\usepackage{babel}

\providecommand{\tabularnewline}{\\}

\@ifundefined{textcolor}{}{%
 \definecolor{BLACK}{gray}{0}
 \definecolor{WHITE}{gray}{1}
 \definecolor{RED}{rgb}{1,0,0}
 \definecolor{GREEN}{rgb}{0,1,0}
 \definecolor{BLUE}{rgb}{0,0,1}
 \definecolor{CYAN}{cmyk}{1,0,0,0}
 \definecolor{MAGENTA}{cmyk}{0,1,0,0}
 \definecolor{YELLOW}{cmyk}{0,0,1,0}
}

\usepackage{amsfonts}
\usepackage{multirow}
\usepackage{mathrsfs}
\usepackage{epstopdf}
\usepackage{bm}
\usepackage{bbm}
\usepackage{color}
\usepackage{array}
\usepackage{diagbox}
\usepackage{float}
\usepackage{braket}
\usepackage{upgreek}
\usepackage{subcaption}
\usepackage{diagbox}
\usepackage{slashed}
\usepackage[figureright]{rotating}

\def\OMIT#1{}

\def\hlinew#1{%
  \noalign{\ifnum0=`}\fi\hrule \@height #1 \futurelet
   \reserved@a\@xhline}
\usepackage{array}
\newcommand{\PreserveBackslash}[1]{\let\temp=\\#1\let\\=\temp}
\newcolumntype{C}[1]{>{\PreserveBackslash\centering}p{#1}}
\newcolumntype{R}[1]{>{\PreserveBackslash\raggedleft}p{#1}}
\newcolumntype{L}[1]{>{\PreserveBackslash\raggedright}p{#1}}

\newcommand{\beq}{\begin{equation}}
\newcommand{\eeq}{\end{equation}}
\newcommand{\bqa}{\begin{eqnarray}}
\newcommand{\eqa}{\end{eqnarray}}
\newcommand{\bseq}{\begin{subequations}}
\newcommand{\eseq}{\end{subequations}}

\newcommand{\fverb}{\setbox\fverbbox=\hbox\bgroup\verb}
\newcommand{\fverbdo}{\egroup\medskip\noindent%
\fbox{\unhbox\fverbbox}\ }
\newcommand{\fverbit}{\egroup\item[\fbox{\unhbox\fverbbox}]}
\newbox\fverbbox

\newcommand{\Rmnum}[1]{\expandafter\@slowromancap\romannumeral #1@}

\usepackage{tikz}
\usetikzlibrary {shapes.misc}
\usetikzlibrary{arrows.meta}
\usetikzlibrary{calc}
\usetikzlibrary{
  decorations.pathmorphing,
  decorations.pathreplacing,
  decorations.markings
}
\usetikzlibrary{patterns}

\tikzset{
  every picture/.style={semithick, line cap=round},
  scalar/.style={dashed},
  fermion/.default=0.5,
  fermion/.style={postaction={decorate, decoration={
    markings,
    mark=at position #1 with {\arrow{Stealth[angle=30:7pt,inset=1.5pt]}},
    transform={xshift={3.5pt*cos(15)}}
  }}},
  antifermion/.default=0.5,
  antifermion/.style={postaction={decorate, decoration={
    markings,
    mark=at position #1 with {\arrowreversed{Stealth[angle=30:7pt,inset=1.5pt]}},
    transform={xshift={-3.5pt*cos(15)}}
  }}},
  gluon/.default=3pt,
  gluon/.style={decorate, decoration={
    coil,
    amplitude=0.5*#1,
    aspect=1,
    segment length=#1
  }},
  rgluon/.default=3pt,
  rgluon/.style={decorate, decoration={
    coil,
    amplitude=-0.5*#1,
    aspect=-1,
    segment length=#1
  }},
  gluonpre/.default=0pt,
  gluonpre/.style={decorate, decoration={
    coil,
    amplitude=1.5pt,
    aspect=1,
    segment length=3pt,
    pre length=#1
  }},
  rgluonpre/.default=0pt,
  rgluonpre/.style={decorate, decoration={
    coil,
    amplitude=-1.5pt,
    aspect=-1,
    segment length=3pt,
    pre length=#1
  }},
  crossmark/.style={cross out, draw=red, inner sep=2pt},
  counter/.style={path picture={
    \draw (path picture bounding box.south east) --
      (path picture bounding box.north west)
      (path picture bounding box.south west) --
      (path picture bounding box.north east);
  }},
  cnode/.default=8pt,
  cnode/.style={inner sep=0pt, minimum size=#1, circle}
}

\usepackage{babel}

\makeatother

\begin{document}

\title{%
Final-state rescattering mechanism of double-charm
baryon decays: $\mathcal{B}_{cc}\to\mathcal{B}_{c}P$}

\author{Xiao-Hui Hu}

\thanks{huxiaohui@cumt.edu.cn, corresponding author}

\affiliation{The College of Materials and physics, China University of mining
and technology, Xuzhou 221116, China \vspace{0.2cm}
 }

\affiliation{Lanzhou Center for Theoretical Physics and Key Laboratory of Theoretical
Physics of Gansu Province, Lanzhou University, Lanzhou 730000, China
\vspace{0.2cm}
 }

\author{Cai-Ping Jia}

\thanks{jiacp@pku.edu.cn, corresponding author}

\affiliation{Center for High Energy Physics, Peking University, Beijing 100871,
China\vspace{0.2cm}
 }

\author{Ye Xing}

\thanks{xingye\_guang@cumt.edu.cn, corresponding author}

\affiliation{The College of Materials and physics, China University of mining
and technology, Xuzhou 221116, China \vspace{0.2cm}
 }

\author{Fu-Sheng Yu}
\thanks{yufsh@lzu.edu.cn, corresponding author}

\affiliation{Lanzhou Center for Theoretical Physics and Key Laboratory of Theoretical
Physics of Gansu Province, Lanzhou University, Lanzhou 730000, China
\vspace{0.2cm}
 }

\affiliation{Frontiers Science Center for Rare Isotopes, and School of Nuclear
Science and Technology, Lanzhou University, Lanzhou 730000, China\vspace{0.2cm}
 }

\date{\today}

\begin{abstract}
The doubly charmed baryon, $\Xi_{cc}^{++}$, was first observed by
LHCb through the non-leptonic decay modes of $\Xi_{cc}^{++}\to\Lambda_{c}^{+}K^{-}\pi^{+}\pi^{+}$
and $\Xi_{c}^{+}\pi^{+}$. Following this discovery, researchers shifted
their focus to identifying other doubly charmed baryons, specifically
$\Xi_{cc}^{+}$ and $\Omega_{cc}^{+}$. In this study, we examine
the non-leptonic weak decays of doubly charmed baryons, denoted as
${\cal B}_{cc}\to{\cal B}_{c}P$, where ${\cal B}_{cc}$ represents
the doubly charmed baryons, specifically $(\Xi_{cc}^{++},\Xi_{cc}^{+},\Omega_{cc}^{+})$.
The notation ${\cal B}_{c}$ denotes the singly charmed baryons, specifically
$({\cal B}_{\bar{3}},{\cal B}_{6})$, while $P$ signifies the light
pseudoscalar. These terms are pertinent to the non-leptonic decay
modes under discussion. While the short-distance contributions can
be precisely estimated through theoretical calculations, addressing
the long-distance contributions for final-state-interaction effects
presents a significant challenge. In order to address this issue, we utilize the rescattering mechanism of final state interaction effects to compute the long-distance contributions. We initially derive the entire hadronic loop contributions for these two-body nonleptonic decays of doubly charmed baryons. In subsequent analyses, we are able to calculate relative strong phases. As a result, we can provide predictions for their decay asymmetry parameters and CP violations.
Furthermore, we employ experimental data from the LHCb collaboration, specifically the ratio \(Br(\Xi_{cc}^{++}\to\Xi_{c}^{\prime+}\pi^{+})/Br(\Xi_{cc}^{++}\to\Xi_{c}^{+}\pi^{+})=(1.41\pm0.17\pm0.10)\), to ascertain the model parameters \(\eta=0.9\pm0.2\). Consequently, we present the predictions of branching ratios and decay asymmetry parameters for 67 distinct decay processes and $CP$ violations for the singly Cabibbo suppressed channels.
This not only strengthens the validity of our theoretical predictions, but also provides a more comprehensive theoretical framework for the future identification of other doubly charmed baryons.
\end{abstract}
\maketitle

\section{Introduction}
~\label{sec:Introduction} 
Over the past two decades, a significant
number of new hadrons with the heavy quark have been identified experimentally~\cite{ParticleDataGroup:2024cfk}.
These encompass both exotic states and conventional heavy baryons,
offering an optimal opportunity to explore and establish traditional
baryon spectroscopy. Until now, most candidates of conventional baryons
are accommodated into the singly heavy baryons, while the experimental
observations for doubly heavy baryons are still scarce. In 2002, the
SELEX collaboration initially reported the detection of $\Xi_{cc}^{+}$.
However, subsequent experiments have yet to corroborate this discovery~\cite{Ratti:2003ez,Aubert:2006qw,Chistov:2006zj,Aaij:2013voa}.
In 2017, the LHCb Collaboration made a remarkable discovery, revealing
a significant structure, $\Xi_{cc}^{++}(3621)$, within the invariant
mass spectrum of the $\Lambda_{c}^{+}K^{-}\pi^{+}\pi^{+}$~\cite{Aaij:2017ueg}.
Also, the existence of $\Xi_{cc}^{++}$ was confirmed by them through
the decay channel $\Xi_{cc}^{++}\to\Xi_{c}^{+}\pi^{+}$~\cite{Aaij:2018gfl}.
The LHCb Collaboration has conducted precise measurements of its mass
and lifetime~\cite{LHCb:2018zpl,LHCb:2019qed}, and also measured
the branching fraction of $\Xi_{cc}^{++}\to\Xi_{c}^{\prime+}\pi^{+}$
decay relative to $\Xi_{cc}^{++}\to\Xi_{c}^{+}\pi^{+}$ decay as $Br(\Xi_{cc}^{++}\to\Xi_{c}^{\prime+}\pi^{+})/Br(\Xi_{cc}^{++}\to\Xi_{c}^{+}\pi^{+})={\cal B}^{\prime}/{\cal B}=(1.41\pm0.17\pm0.10)$~\cite{LHCb:2022rpd}.
The other doubly heavy baryons, such as $\Xi_{cc}^{+}$~\cite{LHCb:2021eaf,LHCb:2019gqy},
$\Omega_{cc}^{+}$~\cite{LHCb:2021rkb}, $\Xi_{bc}^{+}$~\cite{LHCb:2022fbu},
$\Xi_{bc}^{0}$~\cite{LHCb:2020iko,LHCb:2021xba} and $\Omega_{bc}^{0}$~\cite{LHCb:2021xba},
have also been investigated by the LHCb Collaboration; however, no
signals have been detected to date. The doubly charmed tetraquark
$T_{cc}^{+}(3875)$ was also identified by the LHCb Collaboration,
which may provide insights into the nature of doubly heavy baryons
~\cite{LHCb:2021vvq,LHCb:2021auc}. Prior to the experimental discovery
of $\Xi_{cc}^{++}$, theoretical studies had already suggested that
the most likely discovery of these doubly charmed baryons would occur
via two decay channels: $\Xi_{cc}^{++}\to\Lambda_{c}^{+}K^{-}\pi^{+}\pi^{+}$
and $\Xi_{cc}^{++}\to\Xi_{c}^{+}\pi^{+}$~\cite{Yu:2017zst}. Therefore,
pre-theoretical studies of their decays are crucial for the experimental
research of doubly heavy baryons.

Numerous theoretical investigations have been conducted on the production,
mass spectra, and both strong and weak or radiative decays of doubly
heavy baryons, utilizing a variety of methodologies~\cite{Kiselev:2001fw,Ebert:1996ec,Tong:1999qs,Ebert:2002ig,Gershtein:2000nx,Roberts:2007ni,Ortiz-Pacheco:2023kjn,Valcarce:2008dr,Lu:2017meb,Yu:2022lel,Li:2022oth,Eakins:2012jk,Shah:2017liu,Soto:2020pfa,Savage:1990di,Song:2022csw,Cohen:2006jg,Wei:2015gsa,Aliev:2012nn,Liu:2009jc,Brown:2014ena,Padmanath:2015jea,Mathur:2018rwu,Mathur:2018epb,Albertus:2009ww,White:1991hz,Li:2017ndo,Yu:2017zst,Ebert:2004ck,Roberts:2008wq,Branz:2010pq,Albertus:2010hi,Qin:2021zqx,Bahtiyar:2018vub,Chen:2016spr,Cheng:2021qpd,Silvestre-Brac:1996myf,Yoshida:2015tia,Eakins:2012fq,Xiao:2017udy,Mehen:2017nrh,Ma:2017nik,Xiao:2017dly,Yan:2018zdt,He:2021iwx,Chen:2022fye,Song:2023cyk,Hu:2017dzi,Zhao:2023yuk}.
Theoretical studies on the weak decays of doubly heavy baryons have
been mostly focused on semileptonic weak decays, while nonleptonic
weak decays have received relatively less attention. This is mainly
due to the large number of nonleptonic decay channels. In addition,
unlike semi-leptonic decays, non-leptonic decays include not only
factorizable contributions but also non-factorizable ones, which are
difficult to systematically study using QCD methods. At present, there
appears to be a notable dearth of research on the nonleptonic weak
decays of doubly heavy baryons, which also lacks a systematic approach~\cite{Wang:2017azm,Shi:2017dto,Li:2017ndo,Sharma:2017txj,Gerasimov:2019jwp,Cheng:2020wmk,Gutsche:2018msz,Dhir:2018twm}.
Aside from an SU(3) symmetry analysis \cite{Wang:2017azm,Shi:2017dto}
and a phenomenological study \cite{Li:2017ndo}, there is no research
on the non-factorizable contribution in doubly charmed baryon decays
that is based on QCD or is model-independent. The predominant technique
for evaluating such a contribution is the pole-model~\cite{Sharma:2017txj,Gerasimov:2019jwp,Cheng:2020wmk,Gutsche:2018msz,Dhir:2018twm}.
For example, the factorizable and non-factorizable contributions from
the pole-model give a ratio ${\cal B}^{\prime}/{\cal B}$ in the range
$0.81-0.83$~\cite{Sharma:2017txj,Gerasimov:2019jwp}. This implies
that the branching fraction of the decay into $\Xi_{c}^{+\prime}$
is suppressed, in direct contradiction with experiment data~\cite{LHCb:2022rpd}.
In addition, when the interference between the W-emission and W-exchange
contributions is considered, the ratio ${\cal B}^{\prime}/{\cal B}$
increases to about $6.74$~\cite{Cheng:2020wmk}. By integrating
the factorizable amplitudes derived from heavy quark effective theory
with the W-exchange contribution ascertained through light-cone sum
rules (LCSR), the ratio ${\cal B}^{\prime}/{\cal B}$ can be determined
to be $1.42\pm0.78$~\cite{Shi:2022kfa}, which is align with the
experimental results, within the acceptable margin of error. So the
investigation of non-leptonic weak decays is profoundly impacted by
non-factorizable contributions, such as the W-exchange contribution.
In the prior research~\cite{Yu:2017zst}, the authors calculated
the short-distance contributions of decay amplitudes using a factorization
approach. Conversely, the long-distance contributions were initially
examined in double-charm-baryon decays, taking into account the rescattering
mechanism. It is found that these long-distance contributions are
significantly enhanced and play a crucial role in the non-leptonic
weak decays. Subsequently, they utilized this framework to examine
the two-body non-leptonic weak decays of doubly heavy baryons ${\cal B}_{cc}\to{\cal B}_{c}P(V)$~\cite{Han:2021azw,Jiang:2018oak},
${\cal B}_{cc}\to{\cal B}D^{(*)}$~\cite{Li:2020qrh}, and ${\cal B}_{bc}\to{\cal B}_{b}P$~\cite{Han:2021gkl}.
The long-distance contributions were calculated using the cutting
rule for the imaginary component of the hadronic triangle diagram.
{It is worth noting that final-state interaction effects have been reasonably addressed for singly charmed baryon decays in recent work~\cite{Jia:2024pyb}, which explores long-distance contributions within the framework of hadronic loop integrals. This method computes both the real and imaginary parts of the amplitudes comprehensively, providing strong phases that enhance our understanding of decay asymmetries and $CP$ asymmetries.}
In this paper, we will {adopt this method to improve the understanding of double-charmed baryon} 
by calculating the cumulative triangle-diagram contribution including
real and imaginary parts.

\begin{figure}[t]
\centering\includegraphics[width=0.5\textwidth]{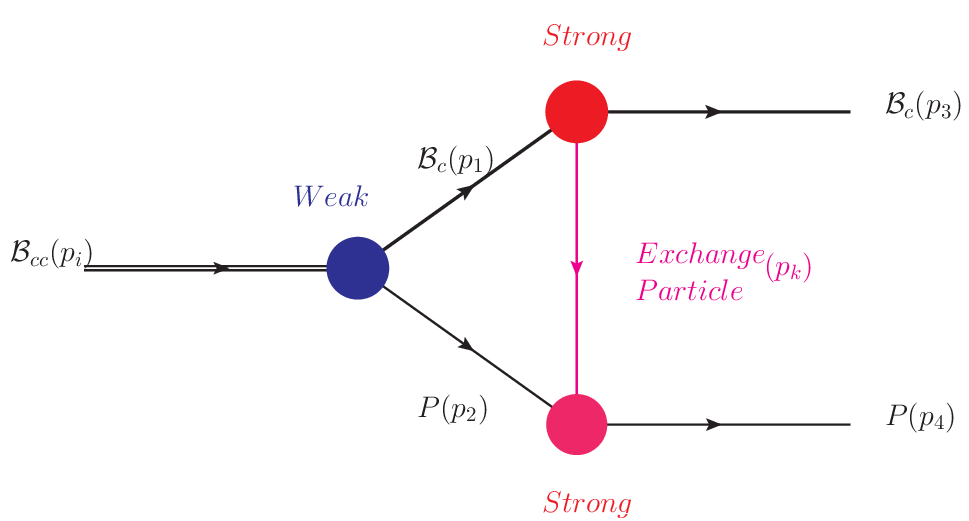} \caption{The diagram description of the rescattering mechanism at the hadron
level. $p_{i}$, $p_{1}$ and $p_{2}$ are the momentum of the initial
and final states, $p_{1}$, $p_{2}$ and $p_{k}$ are the ones of
intermediate states. The blue ball represents the weak vertex, and
the red one denotes the strong vertex.}
\label{fig:triangle} 
\end{figure}


In the following, we will explain the theoretically improved framework
of two-body nonleptonic decay of double-charmed baryons ${\cal B}_{cc}\to{\cal B}_{c}P$. 
\begin{itemize}
\item[(i)] Firstly, the factorizable contributions are computed using the factorization
method, which comprises two components: the non-perturbative transition
matrix element $\langle{\cal B}_{cc}|(V-A)_{\mu}|{\cal B}_{c}\rangle$,
characterized by form factors, and the matrix element $\langle P|(V-A)_{\mu}|0\rangle$,
represented by the decay constant of meson $P$. The non-perturbative
form factors have been assessed in a variety of theoretical studies,
utilizing LCSR \cite{Shi:2019fph,Hu:2019bqj,Hu:2022xzu,Aliev:2022tvs,Aliev:2022maw},
light-front quark model (LFQM)~ \cite{Wang:2017mqp,Zhao:2018mrg,Cheng:2020wmk,Ke:2019lcf,Ke:2022gxm,Hu:2020mxk},
QCD sum rules~\cite{Shi:2019hbf}, QCD factorization~\cite{Sharma:2017txj,Gerasimov:2019jwp},
diquark effective theory~\cite{Shi:2020qde}, constituent quark model~\cite{Gutsche:2018msz,Gutsche:2019iac},
and heavy quark effective theory~\cite{Sharma:2017txj,Dhir:2018twm}.
In this work, we adopt the theoretical predictions of form factors
within LFQM \cite{Wang:2017mqp} as inputs. The decay constants of
mesons have been accurately calculated and measured~\cite{Workman:2022ynf,Choi:2015ywa,Feldmann:1998vh}. 
\item[(ii)] Secondly, we will show how the computation of non-factorizable long-distance
contributions is achieved via the rescattering mechanism briefly.
The final states of the nonleptonic weak decays of the doubly charmed
baryons, ${\cal B}_{c}$ and $P$, involve their mutual scattering
via the exchange of one particle, resulting in a triangle diagram
at the hadron level, as depicted in Fig.~\ref{fig:triangle}. {The triangle diagram mechanism has been widely studied in recent works
about exotic hadrons \cite{Mikhasenko:2015oxp,Liu:2024uxn,Wu:2019rog,Hsiao:2019ait,Ling:2021qzl,Burns:2022uiv,Pan:2023hrk,Guo:2019twa,Guo:2013sya,Jiang:2017tdc,Bayar:2016ftu,Guo:2017jvc,Chen:2019mgp,Liu:2023cwk,Bayar:2023azy,Xie:2022lyw,Lin:2017mtz}.} To
circumvent the issue of double-counting, contributions from both short
and long distances are segregated into the tree emission process ${\cal B}_{cc}(p_{i})\to{\cal B}_{c}(p_{1})P(p_{2})$
and final state interaction effects. In this study, we employ the
effective Lagrangian at the hadronic level of the strong vertex in
the rescattering mechanism to compute the final state interaction
effects. Previous literature~\cite{Yu:2017zst,Han:2021azw,Jiang:2018oak,Li:2020qrh,Han:2021gkl}
have utilized the optical theorem and Cutkosky cutting rules to compute
triangle diagrams. However, these calculations only yield the imaginary
component of the amplitude, making it challenging to accurately represent
the strong phase. In this paper, we corrected this issue by employing
the comprehensive analytical expression of loop integrals, which allows
us to ascertain both the magnitude and strong phase of triangle diagrams.
While in the calculations of the final states interaction effects,
the non-perturbative parameters such as the cut-off $\Lambda$ in
the loop calculation will bring large theoretical uncertainties on
the branching ratios. Unlike the case of B meson~\cite{Ablikim:2002ep,Cheng:2004ru},
there are no enough data to determine the non-perturbative parameters
of doubly charmed baryons. Therefore, the main problem in the calculations
of long-distance contribution is how to control the theoretical uncertainties.
Due to the ratios of the branching fractions are not sensitive to
the non-perturbative parameters, we will calculate the ratios to control
the uncertainties of the theoretical prediction. 
\end{itemize}
The remainder of this paper is organized as follows. In the section
\ref{sec:framework}, we will introduce the theoretical framework
of the final states rescattering mechanism and factorization approach.
In the section \ref{sec:results}, the input parameters, numerical
results of observations including branching ratios, decay asymmetry parameters and CP violation, and the corresponding theoretical analysis are presented.
In the section \ref{sec:summary}, we give a brief summary. In the
Appendix \ref{app:amp}, the amplitudes of each modes are listed.
In the Appendix \ref{app:strong}, the strong coupling constants of $g_{VVP}$, $g_{VPP}$, $g_{PBB}$ and $g_{VBB}$ are collected.

\section{Theoretical framework}
\label{sec:framework} 
In this section, we will present a topological
analysis of each nonleptonic weak decay channel of doubly charmed
baryons, as detailed in Sec.~\ref{subsec:topo}. Subsequently,
we present the general framework for calculating the short-distance
and long-distance dynamics of all topological diagrams in Sec.~\ref{subsec:short}
and Sec.~\ref{subsec:long}, respectively.

\subsection{Topological analysis}

\label{subsec:topo} 
\begin{figure}[htp]
\includegraphics[width=1\textwidth]{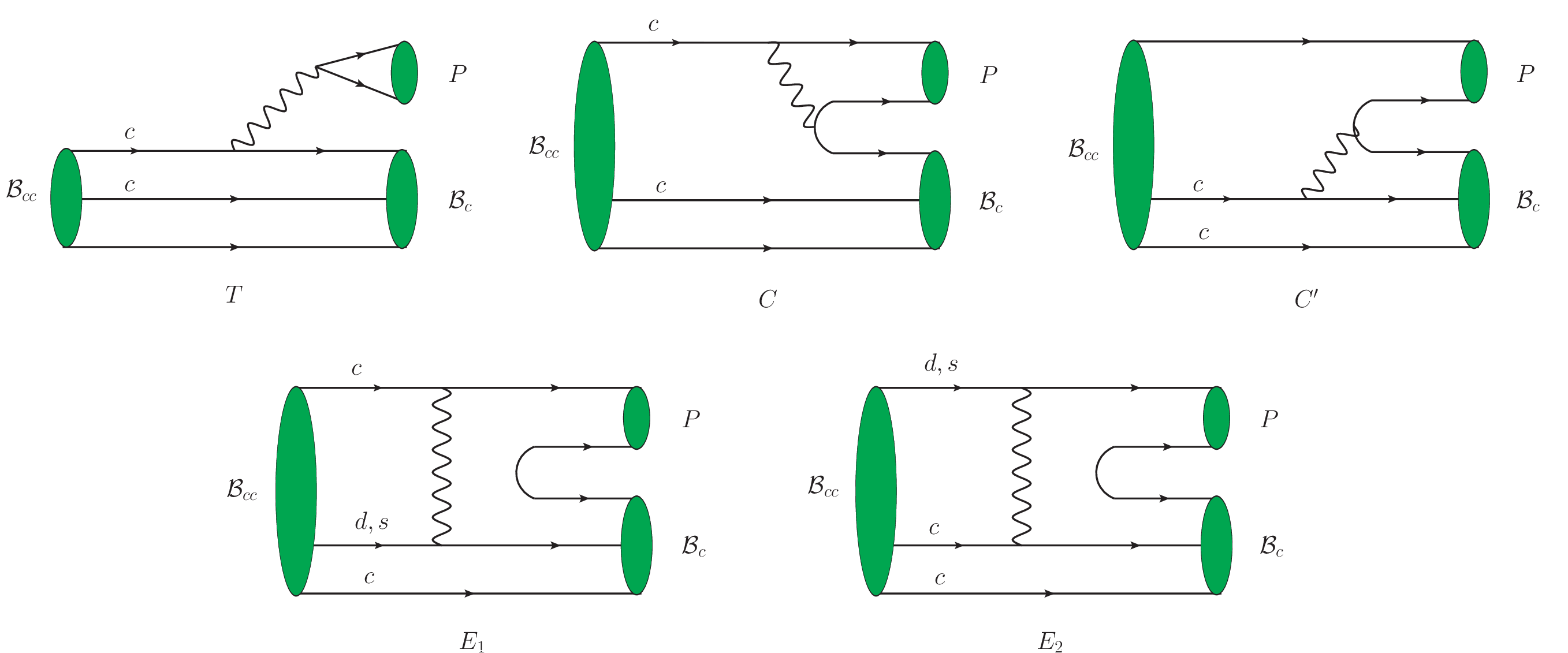} \caption{The five tree level topological diagrams for two body non leptonic
decays ${\cal B}_{cc}\to{\cal B}_{c}P$ of the doubly charmed baryons
${\cal B}_{cc}=(\Xi_{cc}^{++},\Xi_{cc}^{+},\Omega_{cc}^{+})$. }
\label{fig:treetopo} 
\end{figure}

In this work, we will investigate the exclusive channels of nonleptonic
weak decay of doubly charmed baryons $\mathcal{B}_{cc}\to\mathcal{B}_{c}P$.
The initial state is characterized by a doubly charmed baryon triplet
$\mathcal{B}_{cc}=(\Xi_{cc}^{++},\Xi_{cc}^{+},\Omega_{cc}^{+})$,
while the final state encompasses the singly charmed baryon $\mathcal{B}_{c}=(\mathcal{B}_{\bar{3}},\mathcal{B}_{6})$,
and the light pseudoscalar mesons $P=(\pi,K,\eta_{1,8})$, as shown
in Appendix~\ref{app:strong}. Based on the symmetry analysis of the
initial and final states of particles, we will study forty-nine decay
processes.

The tree-level contributions to the nonleptonic weak decay of doubly
charmed baryons $\mathcal{B}_{cc}\to\mathcal{B}_{c}P$ are shown in
Fig.~\ref{fig:treetopo}, where each diagram contains both short-distance
and long-distance contributions. Due to their distinct topologies,
the five diagrams can be categorized into three types: $T$, $C(C^{\prime})$
and $E_{1}(E_{2})$. $T$ represents the color-favored external $W$
emission diagram. Both $C$ and $C^{\prime}$ are color-suppressed
internal $W$-emission diagrams. The difference between them is whether
both quarks of the final light meson come from the weak vertex. In
the $C$ diagram, all constituent quarks of the final light meson
come from the weak vertex. In the $C^{\prime}$ diagram, one constituent
quark of the final light meson comes from the initial state baryon,
and only its antiquark comes from the weak vertex. The contributions
of $W$-exchange can be divided into two cases. In the $E_{1}$ diagram,
the light quark which comes from the decay of charmed quark is taken
by the final light meson. In the $E_{2}$ diagram, it is taken by
the final charmed baryon.

The topology analysis for each nonleptonic weak decay channel can
be concluded in Tab.~\ref{tab:topologies}. The topological amplitudes
for the channels with the sextet single charmed baryons are distinguished
from anti-triplet baryons by adding a $tilde$, e.g. $\tilde{T}$.
The short-distance dominated modes are listed in the first part of
Tab.~\ref{tab:topologies}. The following three parts are the long-distance
dynamics dominated modes, which can be divided into three groups based
on the CKM matrix elements: the Cabibbo-favored (CF) decays induced
by $c\to su\bar{d}$, (ii) the singly Cabibbo-suppressed (SCS) ones
induced by $c\to du\bar{d}$ or $c\to su\bar{s}$, the doubly Cabibdbo-suppressed
(DCS) ones induced by $c\to du\bar{s}$. The topological diagram $T$,
dominated by the factorizable short-distance contributions~\cite{Lu:2009cm},
can be determined using the factorization hypothesis. While the contribution
of the $C(C^{\prime})$ diagram is markedly diminished at the charm
scale, primarily due to the color factor $a_{2}(m_{c})=C_{1}(m_{c})+C_{2}(m_{c})/N_{c}$~\cite{Yu:2017zst}.
At the charm quark mass scale, the long-distance contribution in the
$E_{1}(E_{2})$ diagram exceeds that of the short-distance contribution,
which is suppressed by at least one order~\cite{Lu:2009cm}. Based
on the topology analysis, we can determine that the ratio $Br(\Xi_{cc}^{++}\to\Xi_{c}^{\prime+}\pi^{+})/Br(\Xi_{cc}^{++}\to\Xi_{c}^{+}\pi^{+})$
is approximate to $0.5$. This finding contrasts with the experimental
data~\cite{LHCb:2022rpd}. Since the non-factorizable long-distance
contributions may predominantly influence the calculation of the $C(C^{\prime})$
diagram, potentially exerting a significant impact on the final result
of the decay branching ratio. So we need more investigation on the
nonfactorizable long distance contributions. The inclusion of non-factorizable
contributions, such as final state interaction effects mediated by
loop mechanisms, into our model allows for the improvement of theoretical
predictions through parameter and assumption refinement. This ensures
a more precise congruence with experimental data. The calculation
of the factorizable short-distance and non-factorizable long-distance
contributions will be introduced in detail in the following subsections.

\begin{table}
\caption{The topological analysis of each nonleptonic weak decay channel.}
\label{tab:topologies} %
\tiny
\begin{tabular}{|llllll|}
\hline \hline 
Decay  & Topology  & Decay  & Topology  & Decay  & Topology \tabularnewline
\hline 
$\Xi_{cc}^{++}\to\Xi_{c}^{+}\pi^{+}$  & $\lambda_{sd}(T+C^{\prime})$  & $\Xi_{cc}^{+}\to\Xi_{c}^{0}\pi^{+}$  & $\lambda_{sd}(T-E_{2})$  & $\Omega_{cc}^{+}\to\Omega_{c}^{0}\pi^{+}$  & $\lambda_{sd}\tilde{T}$ \tabularnewline
$\Xi_{cc}^{++}\to\Xi_{c}^{\prime+}\pi^{+}$  & $\frac{1}{\sqrt{2}}\lambda_{sd}\big(\tilde{T}+\tilde{C}^{\prime}\big)$  & $\Xi_{cc}^{+}\to\Xi_{c}^{\prime0}\pi^{+}$  & $\frac{1}{\sqrt{2}}\lambda_{sd}\big(\tilde{T}+\tilde{E}_{2}\big)$  & $\Omega_{cc}^{+}\to\Xi_{c}^{0}\pi^{+}$  & $-\lambda_{d}T-\lambda_{s}E_{2}$ \tabularnewline
$\Xi_{cc}^{++}\to\Sigma_{c}^{+}\pi^{+}$  & $\frac{1}{\sqrt{2}}\lambda_{d}\big(\tilde{T}+\tilde{C}^{\prime}\big)$  & $\Xi_{cc}^{+}\to\Sigma_{c}^{0}\pi^{+}$  & $\lambda_{d}(\tilde{T}+\tilde{E}_{2})$  & $\Omega_{cc}^{+}\to\Xi_{c}^{\prime0}\pi^{+}$  & $\frac{1}{\sqrt{2}}\big(\lambda_{d}\tilde{T}+\lambda_{s}\tilde{E}_{2}\big)$ \tabularnewline
$\Xi_{cc}^{++}\to\Lambda_{c}^{+}\pi^{+}$  & $\lambda_{d}(T+C^{\prime})$  & $\Xi_{cc}^{+}\to\Xi_{c}^{\prime0}K^{+}$  & $\frac{1}{\sqrt{2}}\big(\lambda_{s}\tilde{T}+\lambda_{d}\tilde{E}_{2}\big)$  & $\Omega_{cc}^{+}\to\Omega_{c}^{0}K^{+}$  & $\lambda_{s}\big(\tilde{T}+\tilde{E}_{2}\big)$ \tabularnewline
$\Xi_{cc}^{++}\to\Xi_{c}^{\prime+}K^{+}$  & $\frac{1}{\sqrt{2}}\lambda_{s}\big(\tilde{T}+\tilde{C}^{\prime}\big)$  & $\Xi_{cc}^{+}\to\Xi_{c}^{0}K^{+}$  & $\lambda_{s}T+\lambda_{d}E_{2}$  & $\Omega_{cc}^{+}\to\Xi_{c}^{0}K^{+}$  & $\lambda_{ds}(-T+E_{2})$ \tabularnewline
$\Xi_{cc}^{++}\to\Xi_{c}^{+}K^{+}$  & $\lambda_{s}(T+C^{\prime})$  & $\Xi_{cc}^{+}\to\Sigma_{c}^{0}K^{+}$  & $\lambda_{ds}\tilde{T}$  & $\Omega_{cc}^{+}\to\Xi_{c}^{\prime0}K^{+}$  & $\frac{1}{\sqrt{2}}\lambda_{ds}\big(\tilde{T}+\tilde{E}_{2}\big)$ \tabularnewline
$\Xi_{cc}^{++}\to\Lambda_{c}^{+}K^{+}$  & $\lambda_{ds}(T+C^{\prime})$  & $\Xi_{cc}^{++}\to\Sigma_{c}^{+}K^{+}$  & $\frac{1}{\sqrt{2}}\lambda_{ds}\big(\tilde{T}+\tilde{C}^{\prime}\big)$  &  & \tabularnewline
\hline 
$\Xi_{cc}^{++}\to\Sigma_{c}^{++}\bar{K}^{0}$  & $\tilde{C}$  & $\Xi_{cc}^{+}\to\Xi_{c}^{\prime+}\pi^{0}$  & $\frac{1}{2}(-\tilde{C}^{\prime}+\tilde{E}_{2})$  & $\Xi_{cc}^{+}\to\Xi_{c}^{+}\eta_{8}$  & $\frac{1}{\sqrt{6}}(C^{\prime}-2E_{1}-E_{2})$ \tabularnewline
$\Xi_{cc}^{+}\to\Omega_{c}^{0}K^{+}$  & $\tilde{E}_{2}$  & $\Xi_{cc}^{+}\to\Xi_{c}^{\prime+}\eta_{1}$  & $\frac{1}{\sqrt{6}}(\tilde{C}^{\prime}+\tilde{E}_{1}+\tilde{E}_{2})$  & $\Omega_{cc}^{+}\to\Xi_{c}^{\prime+}\bar{K}^{0}$  & $\frac{1}{\sqrt{2}}(\tilde{C}+\tilde{C}^{\prime})$ \tabularnewline
$\Xi_{cc}^{+}\to\Sigma_{c}^{+}\bar{K}^{0}$  & $\frac{1}{\sqrt{2}}\big(\tilde{C}+\tilde{E}_{1}\big)$  & $\Xi_{cc}^{+}\to\Xi_{c}^{\prime+}\eta_{8}$  & $\frac{1}{2\sqrt{3}}(\tilde{C}^{\prime}-2\tilde{E}_{1}+\tilde{E}_{2})$  & $\Omega_{cc}^{+}\to\Xi_{c}^{+}\bar{K}^{0}$  & $-C+C^{\prime}$ \tabularnewline
$\Xi_{cc}^{+}\to\Lambda_{c}^{+}\bar{K}^{0}$  & $-C+E_{1}$  & $\Xi_{cc}^{+}\to\Xi_{c}^{+}\pi^{0}$  & $-\frac{1}{\sqrt{2}}(C^{\prime}+E_{2})$  &  & \tabularnewline
$\Xi_{cc}^{+}\to\Sigma_{c}^{++}K^{-}$  & $\tilde{E}_{1}$  & $\Xi_{cc}^{+}\to\Xi_{c}^{+}\eta_{1}$  & $\frac{1}{\sqrt{3}}(C^{\prime}+E_{1}-E_{2})$  &  & \tabularnewline
\hline 
{$\Xi_{cc}^{++}\to\Sigma_{c}^{++}\pi^{0}$ }  & {$-\frac{1}{\sqrt{2}}\tilde{C}$}  & {$\Xi_{cc}^{+}\to\Xi_{c}^{\prime+}K^{0}$ }  & {$\frac{1}{\sqrt{2}}(\lambda_{s}\tilde{C}^{\prime}\!\!+\!\!\lambda_{d}\tilde{E}_{1})$
}  & {$\Omega_{cc}^{+}\to\Sigma_{c}^{++}K^{-}$ }  & {$\lambda_{s}\tilde{E}_{1}$ }\tabularnewline
{$\Xi_{cc}^{++}\to\Sigma_{c}^{++}\eta_{1}$ }  & {$\frac{1}{\sqrt{3}}(\lambda_{d}\!\!+\!\!\lambda_{s})\tilde{C}$}  & {$\Xi_{cc}^{+}\to\Xi_{c}^{+}K^{0}$ }  & {$\lambda_{s}C^{\prime}\!\!+\!\!\lambda_{d}E_{1}$ }  & {$\Omega_{cc}^{+}\to\Xi_{c}^{\prime+}\pi^{0}$ }  & {$\frac{1}{2}(-\lambda_{d}\tilde{C}\!\!+\!\!\lambda_{s}\tilde{E}_{2})$ }\tabularnewline
{$\Xi_{cc}^{++}\to\Sigma_{c}^{++}\eta_{8}$ }  & {$\frac{1}{\sqrt{6}}(\lambda_{d}-2\lambda_{s})\tilde{C}$ }  & {$\Xi_{cc}^{+}\to\Lambda_{c}^{+}\pi^{0}$ }  & {$\frac{\lambda_{d}}{\sqrt{2}}(C\!\!-\!\!C^{\prime}\!\!-\!\!E_{1}\!\!-\!\!E_{2})$
}  & {$\Omega_{cc}^{+}\to\Xi_{c}^{+}\pi^{0}$ }  & {$\frac{1}{\sqrt{2}}(\lambda_{d}C-\lambda_{s}E_{2})$ }\tabularnewline
\multirow{2}{*}{{$\Xi_{cc}^{+}\to\Sigma_{c}^{+}\pi^{0}$ } } & \multirow{2}{*}{$\frac{\lambda_{d}}{2}(\!\!-\!\!\tilde{C}\!\!-\!\!\tilde{C}^{\prime}\!\!+\!\!\tilde{E}_{1}\!\!+\!\!\tilde{E}_{2})$}  & \multirow{2}{*}{{$\Xi_{cc}^{+}\to\Lambda_{c}^{+}\eta_{1}$ } } & {$\frac{1}{\sqrt{3}}[\lambda_{d}(C^{\prime}\!\!-\!\!C\!\!+\!\!E_{1}\!\!-\!\!E_{2})$}  & \multirow{2}{*}{{$\Omega_{cc}^{+}\to\Xi_{c}^{\prime+}\eta_{1}$ } } & {$\frac{1}{\sqrt{6}}[\lambda_{s}(\tilde{C}\!\!+\!\!\tilde{C}^{\prime}\!\!+\!\!\tilde{E}_{1}\!\!+\!\!\tilde{E}_{2})$ }\tabularnewline
 &  &  & {$\qquad-{\lambda_{s}}C${]}}  &  & {$\qquad+{\lambda_{d}}\tilde{C}${]}} \tabularnewline
\multirow{2}{*}{{$\Xi_{cc}^{+}\to\Sigma_{c}^{+}\eta_{1}$ } } & {$\frac{1}{\sqrt{6}}[\lambda_{d}(\tilde{C}\!\!+\!\!\tilde{C}^{\prime}\!\!+\!\!\tilde{E}_{1}\!\!+\!\!\tilde{E}_{2})$}  & \multirow{2}{*}{{$\Xi_{cc}^{+}\to\Lambda_{c}^{+}\eta_{8}$ } } & {$\frac{1}{\sqrt{6}}[\lambda_{d}(C^{\prime}\!\!-\!\!C\!\!+\!\!E_{1}\!\!-\!\!E_{2})$
}  & \multirow{2}{*}{{$\Omega_{cc}^{+}\to\Xi_{c}^{\prime+}\eta_{8}$ } } & {$\frac{1}{2\sqrt{3}}[\lambda_{s}({\tilde{E}_{2}}\!\!-\!\!2\tilde{C}\!\!-\!\!2\tilde{C}^{\prime}\!\!-\!\!2\tilde{E}_{1})$ }\tabularnewline
 & {$\qquad+{\lambda_{s}}\tilde{C}${]}}  &  & {$\qquad+{2\lambda_{s}}C${]}}  &  & {$\qquad+\lambda_{d}\tilde{C}${]}}\tabularnewline
\multirow{2}{*}{{$\Xi_{cc}^{+}\to\Sigma_{c}^{+}\eta_{8}$ } } & {$\frac{1}{2\sqrt{3}}[\lambda_{d}(\tilde{C}\!\!+\!\!\tilde{C}^{\prime}\!\!+\!\!\tilde{E}_{1}\!\!+\!\!\tilde{E}_{2})$
}  & \multirow{2}{*}{{$\Omega_{cc}^{+}\to\Xi_{c}^{+}\eta_{8}$ } } & {$\frac{1}{\sqrt{6}}[\lambda_{s}(2C\!\!-\!\!2C^{\prime}\!\!+\!\!2E_{1}\!\!-\!\!E_{2})$ } & \multirow{2}{*}{{$\Omega_{cc}^{+}\to\Xi_{c}^{+}\eta_{1}$ } } & {$\frac{1}{\sqrt{3}}[\lambda_{s}(C^{\prime}\!\!-\!\!C\!\!+\!\!E_{1}\!\!-\!\!E_{2})$}\tabularnewline
 & {$\qquad-2{\lambda_{s}}\tilde{C}${]}}  &  &{$ \qquad-{\lambda_{d}}C${]}} &  & {$\qquad-{\lambda_{d}}C${]}}\tabularnewline
{$\Xi_{cc}^{+}\to\Sigma_{c}^{++}\pi^{-}$ }  & {$\lambda_{d}\tilde{E}_{1}$ }  & {$\Omega_{cc}^{+}\to\Sigma_{c}^{+}\bar{K}^{0}$ }  & {$\frac{1}{\sqrt{2}}(\lambda_{d}\tilde{C}^{\prime}\!\!+\!\!\lambda_{s}\tilde{E}_{1})$
}  & {$\Omega_{cc}^{+}\to\Lambda_{c}^{+}\bar{K}^{0}$ }  & {$\lambda_{d}C^{\prime}\!\!+\!\!\lambda_{s}E_{1}$ } \tabularnewline\hline
{$\Xi_{cc}^{++}\to\Sigma_{c}^{++}K^{0}$ }  & {$\tilde{C}$ }  & {$\Omega_{cc}^{+}\to\Sigma_{c}^{+}\eta_{8}$ }  & {$\frac{1}{2\sqrt{3}}\big(-2\tilde{C}^{\prime}+\tilde{E}_{1}+\tilde{E}_{2}\big)$
}  & {$\Omega_{cc}^{+}\to\Sigma_{c}^{++}\pi^{-}$ }  & {$\tilde{E}_{1}$ }\tabularnewline
{$\Xi_{cc}^{+}\to\Sigma_{c}^{+}K^{0}$ }  & {$\frac{1}{\sqrt{2}}\big(\tilde{C}+\tilde{C}^{\prime}\big)$
}  & {$\Omega_{cc}^{+}\to\Lambda_{c}^{+}\pi^{0}$ }  & {$-\frac{1}{\sqrt{2}}(E_{1}+E_{2})$ }  & {$\Omega_{cc}^{+}\to\Xi_{c}^{\prime+}K^{0}$ }  & {$\frac{1}{\sqrt{2}}\big(\tilde{C}+\tilde{E}_{1}\big)$ }\tabularnewline
{$\Xi_{cc}^{+}\to\Lambda_{c}^{+}K^{0}$ }  & {$-C+C^{\prime}$ }  & {$\Omega_{cc}^{+}\to\Lambda_{c}^{+}\eta_{1}$ }  & {$\frac{1}{\sqrt{3}}(C^{\prime}+E_{1}-E_{2})$ }  & {$\Omega_{cc}^{+}\to\Xi_{c}^{+}K^{0}$ }  & {$-C+E_{1}$ }\tabularnewline
{$\Omega_{cc}^{+}\to\Sigma_{c}^{+}\pi^{0}$ }  & {$\frac{1}{2}(-\tilde{E}_{1}+\tilde{E}_{2})$ }  & {$\Omega_{cc}^{+}\to\Lambda_{c}^{+}\eta_{8}$ }  & {$-\frac{1}{\sqrt{6}}(2C^{\prime}-E_{1}+E_{2})$ }  &  & \tabularnewline
{$\Omega_{cc}^{+}\to\Sigma_{c}^{+}\eta_{1}$ }  & {$\frac{1}{\sqrt{6}}\big(\tilde{C}^{\prime}+\tilde{E}_{1}+\tilde{E}_{2}\big)$
}  & {$\Omega_{cc}^{+}\to\Sigma_{c}^{0}\pi^{+}$ }  & {$\tilde{E}_{2}$ }  &  & \tabularnewline
\hline 
\hline
\end{tabular}
\end{table}

\subsection{Calculation of short-distance amplitudes under the factorization
hypothesis}

\label{subsec:short}

At the tree level, the non-leptonic weak decay of doubly charmed baryons
are induced by the decay of the charm quark. The effective Hamiltonian
can be represented as follows, 
\begin{align}
\mathcal{H}_{eff}=\frac{G_{F}}{\sqrt{2}}\sum_{q^{\prime}=d,s}^ {}V_{cq^{\prime}}^{\ast}V_{uq}[C_{1}(\mu)O_{1}(\mu)+C_{2}(\mu)O_{2}(\mu)]+h.c..\label{eq:hamilton}
\end{align}
Here $V_{cq^{\prime}}$ and $V_{uq}$ represent the Cabibbo-Kobayashi-Maskawa
(CKM) matrix elements. And the four-fermion operators $O_{1}$ and
$O_{2}$ can be expressed as, 
\begin{align}
O_{1}=(\bar{u}_{\alpha}q_{\beta})_{V-A}(\bar{q}^{\prime}_{\beta}c_{\alpha})_{V-A},\quad O_{2}=(\bar{u}_{\alpha}q_{\alpha})_{V-A}(\bar{q}^{\prime}_{\beta}c_{\beta})_{V-A},
\end{align}
with the color indices $\alpha$ and $\beta$. And $C_{1,2}(\mu)$
are the relevant Wilson coefficients. After inserting the effective
Hamiltonian as Eq.~(\ref{eq:hamilton}), the amplitude of the non-leptonic
weak decay of doubly charmed baryons $\mathcal{B}_{cc}\to\mathcal{B}_{c}P$
can be evaluated with the hadronic matrix element, 
\begin{align}
\langle\mathcal{B}_{c}P|\mathcal{H}_{eff}|\mathcal{B}_{cc}\rangle=\frac{G_{F}}{\sqrt{2}}V_{cq^{\prime}}^{\ast}V_{uq}\sum_{i=1,2}C_{i}\langle\mathcal{B}_{c}P|O_{i}|\mathcal{B}_{cc}\rangle.\label{eq:hme}
\end{align}
In this work, we neglect the penguin operators for the strong suppression
of CKM matrix elements in charmed hadron decays. \label{sec:short}
According to the factorization hypothesis, the matrix elements $\langle\mathcal{B}_{c}M|O_{i}|\mathcal{B}_{cc}\rangle$
in Eq.~(\ref{eq:hme}) can be factorized into two parts. The first
part is parameterized with the decay constant of the emitted meson,
while the second part can be evaluated using the heavy-light transition
form factors. The short-distance contribution of the $T$ diagram
can be expressed as: 
\begin{align}
\langle\mathcal{B}_{c}M|\mathcal{H}_{eff}|\mathcal{B}_{cc}\rangle_{SD}^{T}=\frac{G_{F}}{\sqrt{2}}V_{cq^{\prime}}^{\ast}V_{uq}a_{1}(\mu)\langle M|\bar{u}\gamma^{\mu}(1-\gamma_{5})q|0\rangle\langle\mathcal{B}_{c}|\bar{q}^{\prime}\gamma_{\mu}(1-\gamma_{5})c|\mathcal{B}_{cc}\rangle,\label{eq:hadronic1}
\end{align}
here $a_{1}(\mu)=C_{1}(\mu)+C_{2}(\mu)/3$ is the effective Wilson
coefficients. For color suppressed $C$ diagram, its short-distance
contribution can be given via its relation to the $T$ diagram after
Fierz transformation. And under the charm scale $\mu=m_{c}$, the
Wilson coefficients are taken as $C_{1}(\mu)=1.21$ and $C_{2}(\mu)=-0.42$~\cite{Li:2012cfa}.
$M$ represents both pseudoscalar meson($P$) and vector meson($V$).
The vector meson also contributes to long-distance dynamics as an
intermediate state in the next subsections.

By utilizing the heavy-light transition form factors $f_{1,2,3}$
and $g_{1,2,3}$, the transition matrix elements of ${\cal B}_{cc}\to{\cal B}_{c}$
can be effectively parameterized as, 
\begin{eqnarray}
 &  & \langle{\cal B}_{c}(p^{\prime},s_{z}^{\prime})|\bar{q}^{\prime}\gamma_{\mu}(1-\gamma_{5})c|{\cal B}_{cc}(p,s_{z})\rangle\nonumber \\
 &  & =\bar{u}(p^{\prime},s_{z}^{\prime})\Big[\gamma_{\mu}f_{1}(q^{2})+i\sigma_{\mu\nu}\frac{q^{\nu}}{M_{\mathcal{B}_{cc}}}f_{2}(q^{2})+\frac{q^{\mu}}{M_{\mathcal{B}_{cc}}}f_{3}(q^{2})\Big]u(p,s_{z})\nonumber \\
 &  & -\bar{u}(p^{\prime},s_{z}^{\prime})\Big[\gamma_{\mu}g_{1}(q^{2})+i\sigma_{\mu\nu}\frac{q^{\nu}}{M_{\mathcal{B}_{cc}}}g_{2}(q^{2})+\frac{q^{\mu}}{M_{\mathcal{B}_{cc}}}g_{3}(q^{2})\Big]\gamma_{5}u(p,s_{z}).\label{eq:ff}
\end{eqnarray}
Here $q=p-p^{\prime}$, $M_{\mathcal{B}_{cc}}$ is the mass of doubly
charmed baryons.

The first matrix element in Eq.~(\ref{eq:hadronic1}) are defined
using the decay constants of the emitted mesons $P$ and $V$, denoted
by $f_{P}$ and $f_{V}$, respectively. 
\begin{align}
\langle P(p)|\bar{u}\gamma^{\mu}(1-\gamma_{5})q|0\rangle & =-if_{P}p^{\mu},\label{eq:Pdecay}\\
\langle V(p)|\bar{u}\gamma^{\mu}(1-\gamma_{5})q|0\rangle & =m_{V}f_{V}\epsilon^{\ast\mu},\label{eq:Vdecay}
\end{align}
where $\epsilon^{\mu}$ represents the polarization vector of the
vector meson.

Substituting Eqs.~(\ref{eq:ff}-\ref{eq:Vdecay}) into Eq.~(\ref{eq:hadronic1}),
the short-distance amplitudes of the weak decays ${\cal B}_{cc}\to{\cal B}_{c}P$
and $\mathcal{B}_{c}V$ can be written as, 
\begin{align}
\mathcal{A}({\cal B}_{cc}\to{\cal B}_{c}P) & =i\bar{u}_{\mathcal{B}_{c}}(A_{SD}+B_{SD}\gamma_{5})u_{\mathcal{B}_{cc}},\label{eq:B2BP}\\
\mathcal{A}({\cal B}_{cc}\to{\cal B}_{c}V) & =\epsilon^{\ast\mu}\bar{u}_{\mathcal{B}_{c}}\left[A_{SD}^{1}\gamma_{\mu}\gamma_{5}+A_{SD}^{2}\frac{p_{\mu}(\mathcal{B}_{c})}{M_{\mathcal{B}_{cc}}}\gamma_{5}+B_{SD}^{1}\gamma_{\mu}+B_{SD}^{2}\frac{p_{\mu}(\mathcal{B}_{c})}{M_{\mathcal{B}_{cc}}}\right]u_{\mathcal{B}_{cc}}.\label{eq:B2BV}
\end{align}
Here the short distance amplitudes $A_{SD},~B_{SD}$ and $A_{SD}^{1,2},B_{SD}^{1,2}$ are specifically designated to encapsulate the strong interaction information via factorization.
\begin{align}
A_{SD} & =\lambda f_{P}(M_{\mathcal{B}_{cc}}-M_{\mathcal{B}_{c}})f_{1}(m^{2}),\hspace{3cm}B_{SD}=\lambda f_{P}(M_{\mathcal{B}_{cc}}+M_{\mathcal{B}_{c}})g_{1}(m^{2}),\\
A_{SD}^{1} & =-\lambda f_{V}m\left(g_{1}(m^{2})+g_{2}(m^{2})\frac{M_{\mathcal{B}_{cc}}-M_{\mathcal{B}_{c}}}{M_{\mathcal{B}_{cc}}}\right),\ \ \ \ A_{SD}^{2}=-2\lambda f_{V}mg_{2}(m^{2}),\\
B_{SD}^{1} & =\lambda f_{V}m\left(f_{1}(m^{2})-f_{2}(m^{2})\frac{M_{\mathcal{B}_{cc}}+M_{\mathcal{B}_{c}}}{M_{\mathcal{B}_{cc}}}\right),\ \ \ \ \ \ \ \ B_{SD}^{2}=2\lambda f_{V}mf_{2}(m^{2}),
\end{align}
where the parameter $\lambda$ is $\frac{G_{F}}{\sqrt{2}}V_{CKM}a_{1,2}(\mu)$,
and $m$ is the mass of pseudoscalar or vector meson. We exclude $f_{3}$
and $g_{3}$ terms in Eq.~(\ref{eq:ff}) due to $m^{2}/M_{\mathcal{B}_{cc}}^{2}$
suppression. 

\subsection{Calculation of long-distance contributions using the final states
rescattering mechanism}

\label{subsec:long}


\begin{figure}[h]
    \centering
    \begin{minipage}{0.24\linewidth}
    \centering
    \subfloat[][]{
        \includegraphics[scale=0.35]{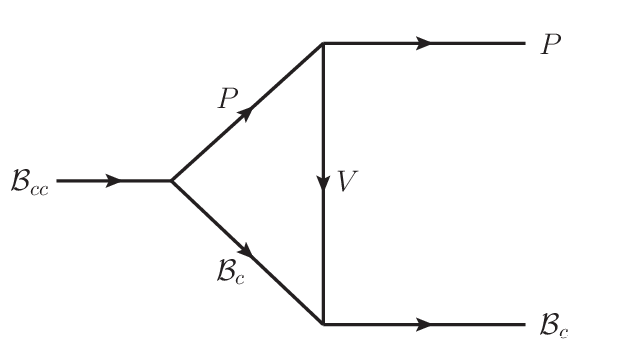}}
    \end{minipage}
    \begin{minipage}{0.24\linewidth}
    \centering
        \subfloat[][]{\includegraphics[scale=0.35]{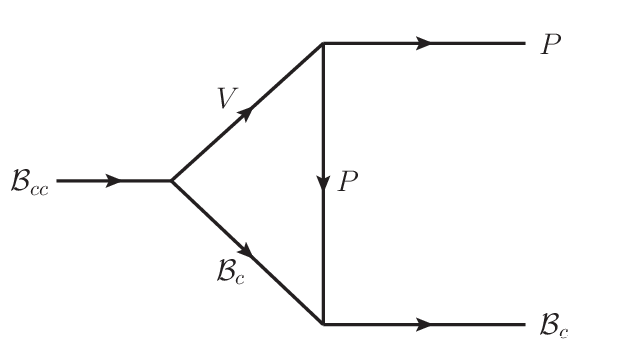}}
    \end{minipage}
    \begin{minipage}{0.24\linewidth}
    \centering
        \subfloat[][]{\includegraphics[scale=0.35]{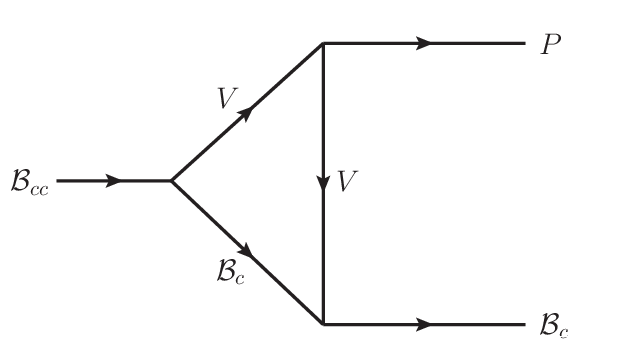}}
    \end{minipage}   

    \begin{minipage}{0.24\linewidth}
    \centering
        \subfloat[][]{\includegraphics[scale=0.35]{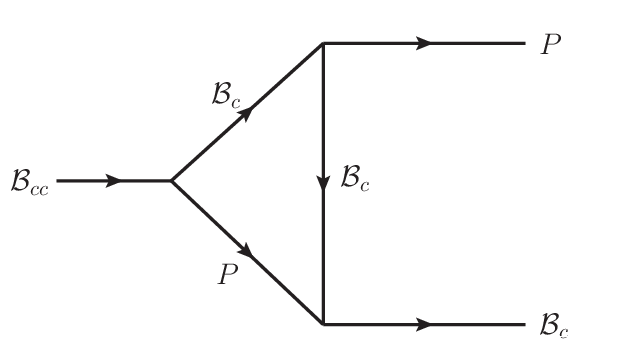}}
    \end{minipage}     
    \begin{minipage}{0.24\linewidth}
    \centering
        \subfloat[][]{\includegraphics[scale=0.35]{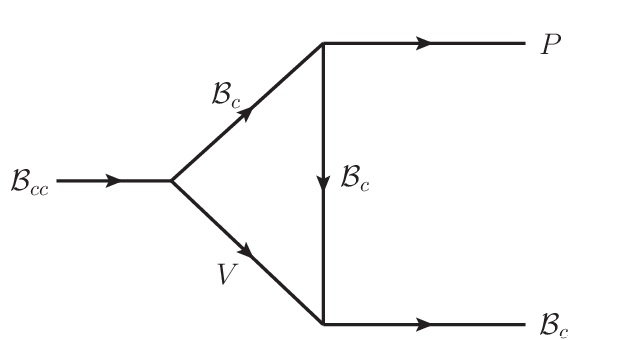}}
    \end{minipage}
    \begin{minipage}{0.24\linewidth}
    \centering
        \subfloat[][]{\includegraphics[scale=0.35]{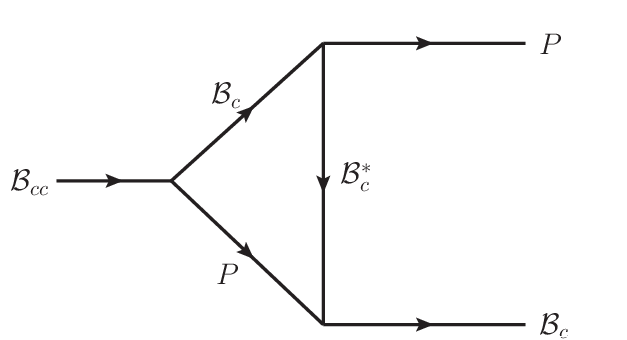}}
    \end{minipage}  
    \begin{minipage}{0.24\linewidth}
    \centering
        \subfloat[][]{\includegraphics[scale=0.35]{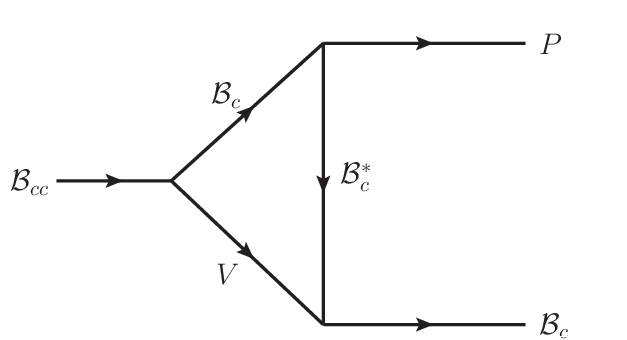}}
    \end{minipage}  
    \caption{The long-distance rescattering contributions to ${\cal {B}}_{cc}\to{\cal {B}}_{c}P$ manifested at hadron level via singly particle exchange, generating distinct triangle diagrams with intermediate states: (a) $\{P,{\cal {B}}_{c};V\}$, (b) $\{V,{\cal {B}}_{c};P\}$, (c) $\{V,{\cal {B}}_{c};V\}$, (d) $\{{\cal {B}}_{c},P;{\cal {B}}_{c}\}$, (e) $\{{\cal {B}}_{c},V;{\cal {B}}_{c}\}$, (f) $\{{\cal {B}}_{c},P;{\cal {B}}_{c}^{*}\}$, (g) $\{{\cal {B}}_{c},V;{\cal {B}}_{c}^{*}\}$. And the corresponding amplitudes of these triangle diagrams have been given by Eqs.~(\ref{eq:fig3a}-\ref{eq:fig3g}).}
    \label{fig:trianglesev}
\end{figure}


The long-distance contributions are large and very difficult to evaluate.
In this work, we employ the rescattering mechanism of final states
to perform the calculation of the long-distance contributions, as
done in Ref.~\cite{Yu:2017zst}. The rescattering mechanism of final
states can be constructed through the rescattering of two intermediate
particles, as depicted in Fig.~\ref{fig:triangle}. In the following,
we explain the detail of our calculation of the amplitudes of the
nonleptonic decays ${\cal B}_{cc}\to{\cal B}_{c}P$. These decays
can proceed as shown in Fig.~\ref{fig:trianglesev}. At the quark level,
the first weak vertex of the triangle diagram is induced by {the external W emission diagram $T$  to avoid double counting, which is dominated by the factorization contribution.}
The upcoming scattering process can occur
via either an $s$ channel or a $t/u$ channel. The diagram of $s$
channel would make a sizable contribution when the mass of the exchanged
particle is the sequel to the ones of the mother particle ${\cal B}_{cc}$.
While the heaviest discovered singly charmed baryon is approximately
$500$ MeV lighter than $m_{\Xi_{cc}^{+}}$. So the contribution of
the $s$-channel diagram is supposed to be highly suppressed by the
off-shell effect and can be safely neglected. As a result, we only
consider the contribution of the $t/u$-channel triangle diagram,
as depicted in Fig.~\ref{fig:trianglesev}. The remaining two strong interaction vertices, as depicted in Fig.~\ref{fig:trianglesev}, can be assessed utilizing the hadronic strong interaction Lagrangians~\cite{Aliev:2010yx,Yan:1992gz,Casalbuoni:1996pg,Meissner:1987ge,Li:2012bt,Aliev:2010nh} presented as following.

\begin{align}
\mathcal{L}_{VPP} & =ig_{VP_{1}P_{2}}[V^{\mu}P_{1}\partial_{\mu}P_{2}-V^{\mu}P_{2}\partial_{\mu}P_{1}],\label{eq:LVPP}\\
\mathcal{L}_{VVP} & =\frac{g_{V_{1}V_{2}P}}{\sqrt{m_{V_{1}}m_{V_{2}}}}\varepsilon_{\mu\nu\alpha\beta}Tr[\partial^{\mu}V_{1}^{\nu}\partial^{\alpha}PV_{2}^{\beta}],\label{eq:LVVP}\\
\mathcal{L}_{P{\mathcal{B}}_{c}{\mathcal{B}}_{c}} & =g_{P{\mathcal{B}}_{c}{\mathcal{B}}_{c}}Tr[\overline{{\mathcal{B}}}_{c}i\gamma_{5}P{\mathcal{B}}_{c}],\label{eq:PBB}\\
\mathcal{L}_{V{\mathcal{B}}_{c}{\mathcal{B}}_{c}} & =f_{1V{\mathcal{B}}_{c}{\mathcal{B}}_{c}}Tr[\overline{\mathcal{B}}_{c}\gamma_{\mu}V^{\mu}{\mathcal{B}}_{c}]+\frac{f_{2V{\mathcal{B}}_{c}{\mathcal{B}}_{c}}}{m_{{\mathcal{B}}_{c}}+m_{{\mathcal{B}}_{c}}}Tr[\overline{\mathcal{B}}_{c}\sigma_{\mu\nu}\partial^{\mu}V^{\nu}{\mathcal{B}}_{c}],\label{eq:VBB}\\
\mathcal{L}_{P{\mathcal{B}}_{c}{\mathcal{B}}_{c}^{*}} & ={g_{P{\mathcal{B}}_{c}{\mathcal{B}}_{c}^{*}}}\epsilon^{ijk}(\overline{\mathcal{B}}_{c})^{j}_{i}({\mathcal{B}}_{c}^{*})^{mkl}_{\mu}\partial^{\mu}P^{i}_{m},\label{eq:PBBS}\\
\mathcal{L}_{V{\mathcal{B}}_{c}{\mathcal{B}}_{c}^{*}} & =-i{g_{V{\mathcal{B}}_{c}\bar{\mathcal{B}}_{c}^{*}}}[\overline{\mathcal{B}}_{c}^{*\mu}\gamma^{5}\gamma^{\nu}{\mathcal{B}}_{c}+\overline{\mathcal{B}}_{c}\gamma^{5}\gamma^{\nu}{\mathcal{B}}_{c}^{*}](\partial_{\mu}V_{\nu}-\partial_{\nu}V_{\mu}).\label{eq:LVBBS}
\end{align}

There are various ways to compute
the triangle amplitude~\cite{Li:2002pj,Ablikim:2002ep,Li:1996cj,Dai:1999cs,Locher:1993cc,Cheng:2004ru,Lu:2005mx}.
The main point of distinction between them is their approach towards
the integration of hadronic loops. In work~\cite{Han:2021azw}, the
authors adopting the optical theorem and Cutkosky cutting rule as
in Ref.~\cite{Cheng:2004ru}, only calculate the absorptive (imaginary)
part of these diagrams. The real component of the triangle diagram can be theoretically derived from the dispersion relation.
However, due to significant ambiguity, it is challenging to reliably describe the amplitude across the entire region. Consequently, this effect is often neglected in numerous studies. To accurately determine the complete amplitude of the triangle diagram, the most direct method involves calculating the loop integrals. In this study, we employ the Passarino-Veltman method to simplify the tensor integrals and utilize the method of integral by parts to derive expressions with master integrals.

To clarify, we want to determine the amplitude
of the decay mode for ${\cal {B}}_{cc}(p_{i})\to{\cal {B}}_{c}(p_{4})P(p_{3})$. The Fig.~\ref{fig:trianglesev}(a)
shows a triangle diagram with intermediate states $\{P,{\cal {B}}_{c};V\}$,
involving a weak vertex ${\cal {B}}_{cc}(p_{i})\to{\cal {B}}_{c}(p_{2})P(p_{1})$, as well
as a rescattering amplitude of ${\cal {B}}_{c}(p_{2})P(p_{1})\to{\cal {B}}_{c}(p_{4})P(p_{3})$.
The factorization approach in Eq.~(\ref{eq:B2BP}) can calculate
the weak vertex, while the rescattering amplitude is computed using
the hadronic effective Lagrangian. We use the symbol $\mathcal{M}[P_{1},P_{2};P_{k}]$ to represent a
triangle amplitude in this work. The amplitudes of the diagrams shown in Fig.~\ref{fig:trianglesev}
can be expressed as follows, 
\begin{eqnarray}
{\cal {M}}[P,{\cal {B}}_{c};V] & = & -\int\frac{d^{4}p_{1}}{(2\pi)^{4}}\bar{u}(p_{4},s_{4})(f_{1{\cal {B}}_{c}{\cal {B}}_{c}V}\gamma_{\nu}-\frac{if_{2{\cal {B}}_{c}{\cal {B}}_{c}V}}{m_{2}+m_{4}}\sigma_{\mu\nu}p_{k}^{\mu})(\slashed p_{2}+m_{2})\nonumber \\
 &  & \times(A_{SD}+B_{SD}\gamma_{5})u(p_{i},s_{i})g_{VPP}(-g^{\alpha\nu}+\frac{p_{k}^{\alpha}p_{k}^{\nu}}{m_{k}^{2}})(p_{1}+p_{3})_{\alpha}{\cal {PF}},\label{eq:fig3a}\\
{\cal {M}}[V,{\cal {B}}_{c};P] & = & -\int\frac{d^{4}p_{1}}{(2\pi)^{4}}g_{P{\cal {B}}_{c}{\cal {B}}_{c}}\bar{u}(p_{4},s_{4})\gamma_{5}({\slashed p}_{2}+m_{2})\nonumber \\
 &  & \times(-g^{\rho\alpha}+\frac{p_{1}^{\rho}p_{1}^{\alpha}}{m_{1}^{2}})(A_{1}\gamma_{\rho}\gamma_{5}+A_{2}\frac{p_{2\rho}}{m_{i}}\gamma_{5}+B_{1}\gamma_{\rho}+B_{2}\frac{p_{2\rho}}{m_{i}})u(p_{i},s_{i})\nonumber \\
 &  & \times g_{VPP}(p_{3}-p_{k})_{\alpha}{\cal PF},\label{eq:fig3b}\\
{\cal {M}}[V,{\cal {B}}_{c};V] & = & -i\int\frac{d^{4}p_{1}}{(2\pi)^{4}}\bar{u}(p_{4},s_{4})(f_{1{\cal {B}}_{c}{\cal {B}}_{c}V}\gamma^{\nu}-\frac{if_{2{\cal {B}}_{c}{\cal {B}}_{c}V}}{m_{2}+m_{4}}\sigma^{\mu\nu}p_{k\mu})\nonumber \\
 &  & \times(\slashed p_{2}+m_{2})(A_{1}\gamma^{\rho}\gamma_{5}+A_{2}\frac{p_{2}^{\rho}}{m_{i}}\gamma_{5}+B_{1}\gamma^{\rho}+B_{2}\frac{p_{2}^{\rho}}{m_{i}})u(p_{i},s_{i})\nonumber \\
 &  & \times\frac{g_{VVP}}{\sqrt{m_{1}m_{k}}}\epsilon^{\alpha\beta\delta\sigma}p_{k\alpha}p_{3\delta}(-g_{\rho\sigma}+\frac{p_{1\rho}p_{1\sigma}}{m_{1}^{2}})(-g_{\nu\beta}+\frac{p_{k\nu}p_{k\beta}}{m_{k}^{2}}){\cal {PF}},\label{eq:fig3c}\\
 {\cal {M}}[V,{\cal {B}}_{c};V] & = & i\int\frac{d^{4}p_{1}}{(2\pi)^{4}}\bar{u}(p_{4},s_{4})(f_{1{\cal {B}}_{c}{\cal {B}}_{c}V}\gamma^{\nu}-\frac{if_{2{\cal {B}}_{c}{\cal {B}}_{c}V}}{m_{2}+m_{4}}\sigma^{\mu\nu}p_{k\mu})\nonumber \\
 &  & \times(\slashed p_{2}+m_{2})(A_{1}\gamma^{\rho}\gamma_{5}+A_{2}\frac{p_{2}^{\rho}}{m_{i}}\gamma_{5}+B_{1}\gamma^{\rho}+B_{2}\frac{p_{2}^{\rho}}{m_{i}})u(p_{i},s_{i})\nonumber \\
 &  & \times\frac{4g_{VVP}}{f_{P}}\epsilon^{\alpha\beta\delta\sigma}p_{k\alpha}p_{1\delta}(-g_{\rho\sigma}+\frac{p_{1\rho}p_{1\sigma}}{m_{1}^{2}})(-g_{\nu\beta}+\frac{p_{k\nu}p_{k\beta}}{m_{k}^{2}}){\cal {PF}},\label{eq:fig3cJ}\\
{\cal {M}}[{\cal {B}}_{c},P;{\cal {B}}_{c}] & = & \int\frac{d^{4}p_{1}}{(2\pi)^{4}}g_{P{\cal {B}}_{c}{\cal {B}}_{c}}\bar{u}(p_{4},s_{4})\gamma_{5}(\slashed p_{k}+m_{k})g_{P{\cal {B}}_{c}{\cal {B}}_{c}}\gamma_{5}(\slashed p_{1}+m_{1})\nonumber \\
 &  & \times(A_{SD}+B_{SD}\gamma_{5})u(p_{i},s_{i}){\cal PF},\label{eq:fig3d}\\
{\cal {M}}[{\cal {B}}_{c},V;{\cal {B}}_{c}] & = & -\int\frac{d^{4}p_{1}}{(2\pi)^{4}}\bar{u}(p_{4},s_{4})(f_{1V{\cal {B}}_{c}{\cal {B}}_{c}}\gamma_{\nu}-\frac{if_{2V{\cal {B}}_{c}{\cal {B}}_{c}}}{m_{k}+m_{4}}\sigma_{\mu\nu}p_{2}^{\mu})\nonumber \\
 &  & \times(\slashed p_{k}+m_{k})g_{P{\cal {B}}_{c}{\cal {B}}_{c}}\gamma_{5}(\slashed p_{1}+m_{1})(-g^{\nu\rho}+\frac{p_{2}^{\nu}p_{2}^{\rho}}{m_{2}^{2}})\nonumber \\
 &  & \times(A_{1}\gamma_{\rho}\gamma_{5}+A_{2}\frac{p_{1\rho}}{m_{i}}\gamma_{5}+B_{1}\gamma_{\rho}+B_{2}\frac{p_{1\rho}}{m_{i}})u(p_{i},s_{i}){\cal PF},\label{eq:fig3e}\\
{\cal {M}}[{\cal {B}}_{c},P;{\cal {B}}_{c}^{*}] & = & -\int\frac{d^{4}p_{1}}{(2\pi)^{4}}{g_{P{\cal {B}}_{c}{\cal {B}}_{c}^{*}}}{g_{P{\cal {B}}_{c}{\cal {B}}_{c}^{*}}}\bar{u}(p_{4},s_{4})p_{2}^{\mu}p_{3}^{\nu}(\slashed p_{k}+m_{k})\nonumber \\
 &  & \times\{-g_{\mu\nu}+\frac{\gamma_{\mu}\gamma_{\nu}}{3}+\frac{2p_{k\mu}p_{k\nu}}{3m_{k}^{2}}-\frac{p_{k\mu}\gamma_{\nu}-p_{k\nu}\gamma_{\mu}}{3m_{k}}\}(\slashed p_{1}+m_{1})\nonumber \\
 &  & \times(A_{SD}+B_{SD}\gamma_{5})u(p,s){\cal PF},\label{eq:fig3f}
\end{eqnarray}
\begin{eqnarray}
{\cal {M}}[{\cal {B}}_{c},V;{\cal {B}}_{c}^{*}] & = & \int\frac{d^{4}p_{1}}{(2\pi)^{4}}[{g_{V{\cal {B}}_{c}{\cal {B}}_{c}^{*}}}\bar{u}(p_{4},s_{4})\gamma^{5}\gamma_{\nu}(-p_{2}^{\rho}g^{\nu\sigma}+p_{2}^{\nu}g^{\rho\sigma})]\nonumber \\
 &  & {g_{P{\cal {B}}_{c}{\cal {B}}_{c}^{*}}}(\slashed p_{k}+m_{k})\{-g_{\rho\mu}+\frac{\gamma_{\rho}\gamma_{\mu}}{3}+\frac{2p_{k\rho}p_{k\mu}}{3m_{k}^{2}}-\frac{p_{k\rho}\gamma_{\mu}-p_{k\mu}\gamma_{\rho}}{3m_{k}}\}p_{3\mu}\nonumber \\
 &  & \times(\slashed p_{1}+m_{1})(A_{1}\gamma_{\sigma}\gamma_{5}+A_{2}\frac{p_{1\sigma}}{m_{i}}\gamma_{5}+B_{1}\gamma_{\sigma}+B_{2}\frac{p_{1\sigma}}{m_{i}})u(p,s){\cal PF}.\label{eq:fig3g}
\end{eqnarray}
where the note ${\cal {PF}}$ represents the multiply of the propagators
and form factor, 
\begin{eqnarray}
{\cal {PF}} & = & \frac{1}{(p_{1}^{2}-m_{1}^{2}+i\epsilon)(p_{2}^{2}-m_{2}^{2}+i\epsilon)(p_{k}^{2}-m_{k}^{2}+i\epsilon)}\left(\frac{\Lambda_{k}^{2}-m_{k}^{2}}{\Lambda_{k}^{2}-p_{k}^{2}}\right)^{2}.\label{eq:TF}
\end{eqnarray}
In the given equation, strong coupling constants, such as $g_{VPP}$, $f_{1{\cal {B}}_{c}{\cal {B}}_{c}V}$,
and $f_{2{\cal {B}}_{c}{\cal {B}}_{c}V}$ are calculated on-shell.
The reliability of the strong coupling constants is compromised due
to the exchange states $V$ being generally off-shell. 
The form factor
of the exchange particle $F(p_{j},m_{j})$ are introduced to account
for off-shell effects and self-consistency of the theoretical framework~\cite{Cheng:2004ru},
Furthermore, it is necessary to implement an appropriate regularization scheme to manage the inevitable divergence that occurs in the master integrals of the triangle diagram amplitude. The form factor introduced in Eq.\eqref{eq:TF} aligns with the Pauli-Villars regularization scheme.
\begin{align}
F(p_{k},m_{k})=\left(\frac{\Lambda_{k}^{2}-m_{k}^{2}}{\Lambda_{k}^{2}-p_{k}^{2}}\right)^{n}.\label{eq:Ffactor}
\end{align}
The cutoff $\Lambda_{k}$ can be given as 
\begin{align}
\Lambda_{k}=m_{k}+\eta\Lambda_{{\rm QCD}}.
\end{align}
with $\Lambda_{{\rm QCD}}=330\text{MeV}$ for the charm quark decays.
The phenomenological parameter $\eta$ 
is determined by experimental data {which can not be derived from the first-principle method}. In this work,
multiple strong vertices require extensive experimental data for individual
parameter calculation. In the next section, we will use the experimental data to determine the value of the parameter $\eta$. The form factor given by Eq.~(\ref{eq:Ffactor}) typically exhibits monopole or dipole behavior, with the exponential factor $n$ taking on values of $1$ or $2$. The branching ratios for $B$
meson decays in Ref.~\cite{Cheng:2004ru} are similar for both choices,
and then we select $n=2$.

After gathering all the fragments, the amplitude of decay ${\cal {B}}_{cc}\to{\cal {B}}_{c}P$
can be expressed as: 
\begin{align}
\mathcal{A}({\cal {B}}_{cc}\to{\cal {B}}_{c}P) & =\mathcal{M}_{SD}({\cal {B}}_{cc}\to{\cal {B}}_{c}P)+\mathcal{M}[P,{\cal {B}}_{c};V]+\mathcal{M}[V,{\cal {B}}_{c};P]+\mathcal{M}[V,{\cal {B}}_{c};V]\nonumber \\
 & +\mathcal{M}[{\cal {B}}_{c},P;{\cal {B}}_{c}]+\mathcal{M}[{\cal {B}}_{c},V;{\cal {B}}_{c}]+\mathcal{M}[{\cal {B}}_{c},P;{\cal {B}}_{c}^{*}]+\mathcal{M}[{\cal {B}}_{c},V;{\cal {B}}_{c}^{*}],
\end{align}
where $\mathcal{M}_{SD}$ labels its short-distance contributions {including the factorization contributions of $T$ and $C$ diagrams}. The amplitudes
for all channels can be found in Appendix~\ref{app:amp}.

\subsection{Decay asymmetry parameters and CP violation}
As the amplitude for the two body nonleptonic weak decay of doubly heavy baryon $\mathcal{B}_{cc}\to\mathcal{B}_{c}P$ can be deduced as 
\begin{align}
  \mathcal{M}({\cal B}_{cc}\to{\cal B}_{c}P) & =i\bar{u}_{\mathcal{B}_{c}}(A+B\gamma_{5})u_{\mathcal{B}_{cc}}=\mathcal{S} + \mathcal{P}\sigma\cdot\vec{p}_{\mathcal{B}_{c}}=i(H_{1/2}+H_{-1/2}),\label{eq:B2BPc}
\end{align} 
here $\vec{p}_{\mathcal{B}_{c}}$ is the three-momentum of singly charmed baryon $\mathcal{B}_{c}$ in the rest frame of the mother particle $\mathcal{B}_{cc}$. In this decay process, two helicity amplitudes can be obtained in the Pauli-Dirac representation, 
\begin{align}
  H_{\pm1/2}=\sqrt{2m_{\mathcal{B}_{cc}}(E_{\mathcal{B}_{c}} +m_{\mathcal{B}_{c}})}A\mp\sqrt{2m_{\mathcal{B}_{cc}}(E_{\mathcal{B}_{c}} -m_{\mathcal{B}_{c}})}B,
\end{align} 
here the subscripte $\pm 1/2$ denotes the helicity of douly charmed baryons. 
The amplitudes $\mathcal{S}$ and $\mathcal{P}$
denote the parity-violating (PV) S-wave and parity-conserving (PC) P-wave amplitudes, respectively, In the above Eq.~(\ref{eq:B2BPc}), $A$ and $B$ generally receive
both factorizable short distance (SD) from Eq.~(\ref{eq:B2BP}) and non-factorizable long distance (LD) contributions
\begin{align}
{A}=A_{SD}+A_{LD},~~{B}=B_{SD}+B_{LD}.\label{eq:AB}
\end{align} 
Then the decay width can be given as following,
\begin{align}
  \Gamma(\mathcal{B}_{cc}\to\mathcal{B}_{c}P)&=\frac{|\vec{p}_{\mathcal{B}_{c}}|}{8\pi}\Big[\frac{(m_{\mathcal{B}_{cc}}+m_{\mathcal{B}_{c}})^2-m_{P}^2}{ m_{\mathcal{B}_{cc}}^{2}}|A|^2+\frac{(m_{\mathcal{B}_{cc}}-m_{\mathcal{B}_{c}})^2-m_{P}^2}{ m_{\mathcal{B}_{cc}}^{2}}|B|^2\Big]\nonumber\\
  &=2|\vec{p}_{\mathcal{B}_{c}}|(|\mathcal{S}|^2 + |\mathcal{P}|^2)=\frac{|\vec{p}_{\mathcal{B}_{c}}|}{16\pi m_{\mathcal{B}_{cc}}^2}(|H_{1/2}|^2+|H_{-1/2}|^2),\label{eq:widthc}
  \end{align}
  where $|\vec{p}_{\mathcal{B}_{c}}|= \sqrt{E_{\mathcal{B}_{c}}^2-m_{\mathcal{B}_{c}}^2}$
  is size of the final baryon ${\mathcal{B}_{c}}$ momentum and the relations between $|A|^2,~|B|^2$ and $|\mathcal{S}|^2,~|\mathcal{P}|^2$ can be given as
  \begin{align}
    &|\mathcal{S}|^2=\frac{(m_{\mathcal{B}_{cc}}+m_{\mathcal{B}_{c}})^2-m_{P}^2}{16\pi m_{\mathcal{B}_{cc}}^{2}}|A|^2,~~|\mathcal{P}|^2=\frac{(m_{\mathcal{B}_{cc}}-m_{\mathcal{B}_{c}})^2-m_{P}^2}{16\pi m_{\mathcal{B}_{cc}}^{2}}|B|^2,\label{eq:relaspab}\\
   & |H_{1/2}|^2+|H_{-1/2}|^2=2(|A|^{2}+\kappa^2|B|^2),\label{eq:relahab}
    \end{align}
If the polarization of decay is measurable, there are additional experimental observables in the decay angular distribution. The asymmetry parameters can be used to to understand the angular correlations and dynamics of strong interactions in this decay process $\mathcal{B}_{cc}\to\mathcal{B}_{c}P$, which can be defined as following~\cite{He:2015fsa,Cheng:2020wmk,Zhang:2024rbl}
\begin{align}
  &\alpha  =\frac{2\text{Re}(\mathcal{S}^{*}\mathcal{P})}{|\mathcal{S}|^2+|\mathcal{P}|^2}=\frac{2\kappa \text{Re}(A^{*}B)}{|A|^{2}+\kappa^2|B|^2}=-\text{\ensuremath{\frac{|H_{{1/2}}|^{2}-|H_{-{1/2}}|^{2}}{|H_{{1/2}}|^{2}+|H_{-{1/2}}|^{2}}}},\label{eq:alpha}\\
  &\beta=\frac{2\text{Im}(\mathcal{S}^{*}\mathcal{P})}{|\mathcal{S}|^2+|\mathcal{P}|^2}=\frac{2\kappa \text{Im}(A^{*}B)}{|A|^{2}+\kappa^2|B|^2}=-\text{\ensuremath{\frac{2\text{Im}(H_{{1/2}}H_{-{1/2}}^{*})}{|H_{{1/2}}|^{2}+|H_{-{1/2}}|^{2}}}},\label{eq:beta}\\
  &\gamma=\frac{|\mathcal{S}|^2-|\mathcal{P}|^2}{|\mathcal{S}|^2+|\mathcal{P}|^2}=\frac{|A|^{2}-\kappa^2|B|^2}{|A|^{2}+\kappa^2|B|^2}=\text{\ensuremath{\frac{2\text{Re}(H_{{1/2}}H_{-{1/2}}^{*})}{|H_{{1/2}}|^{2}+|H_{-{1/2}}|^{2}}}}.\label{eq:gamma}
  \end{align}
Here $\kappa = |\vec{p}_{\mathcal{B}_{c}}|/(E_{\mathcal{B}_{c}} +m_{\mathcal{B}_{c}} ) =
\sqrt{(E_{\mathcal{B}_{c}}-m_{\mathcal{B}_{c}})/(E_{\mathcal{B}_{c}}+m_{\mathcal{B}_{c}})}$.
Only two of the variables are independent, subject to the constraint $\alpha^2+\beta^2+\gamma^2=1$. This constraint allows one to obtain
\begin{align}
  \beta=\sqrt{1-\alpha^2}\sin\phi,~\gamma=\sqrt{1-\alpha^2}\cos\phi,~\phi=\tan^{-1}(\beta/\gamma).
\end{align}
To experimentally measure the parameter $\alpha$, it is necessary to assess either the initial or final baryon polarization. While determining the initial polarization at LHCb might be challenging, the final baryon polarization can be ascertained from the decays of the final baryon. For measuring $\beta$ and $\gamma$, information on both the initial and final baryon polarization is essential. In the following, we will mainly discuss $\alpha$, which is more likely to be measured. 

As our calculation can give the strong phase of the decay amplitudes, the information of $CP$ violation can be derived.
The direct $CP$ asymmetry is defined as:
\begin{align}\label{eq:direct_CPV}
    A^\text{dir}_{CP}(\mathcal{B}_{cc}\to\mathcal{B}_{c}P) 
    & = \frac{\Gamma(\mathcal{B}_{cc}\to\mathcal{B}_{c}P)-\overline{\Gamma}(\mathcal{B}_{cc}\to\mathcal{B}_{c}P)}{\Gamma(\mathcal{B}_{cc}\to\mathcal{B}_{c}P) + \overline{\Gamma}(\mathcal{B}_{cc}\to\mathcal{B}_{c}P)} \nonumber\\
    & = \frac{2r\sin\Delta\delta\sin\Delta\phi}{1+r^2+2r\cos\Delta\delta\cos\Delta\phi} \,,   
\end{align}
where $r$ denotes the amplitude ratio. Furthermore, $\Delta\delta$ and $\Delta\phi$ symbolize the strong and weak phase differences respectively.

The violation of $CP$ arises from the parameters associated with decay asymmetry.
\begin{align} 
\alpha_{CP}  =\frac{\alpha-\bar{\alpha}}{\alpha+\bar{\alpha}},\qquad
\beta_{CP} & =\frac{\beta-\bar{\beta}}{\beta+\bar{\beta}},\qquad
\gamma_{CP}  =\frac{\gamma-\bar{\gamma}}{\gamma+\bar{\gamma}}.
\end{align}
\section{Numerical results and discussions}
\label{sec:results}
\subsection{Input parameters}
\begin{table}[b]
\centering \caption{{Decay constants (in units of MeV) of the light mesons~\cite{ParticleDataGroup:2024cfk,Choi:2015ywa,Feldmann:1998vh} and strong couplings~\cite{Cheng:2004ru,Nakayama:2006ps,Aliev:2010yx,Aliev:2010nh,Aliev:2010ev,Yan:1992gz,Casalbuoni:1996pg,Meissner:1987ge,Li:2012bt} used in this work.}}
\label{table:decayconstants} %
\begin{tabular}{ccccccccccc}
\hline 
$f_{\pi}$~  & ~$f_{\eta_{1}}$~  & ~$f_{\rho}$~  & ~$f_{\omega}$~  & ~$f_{\eta_{8}}$~  & ~$f_{\phi}$~  & ~$f_{K}$~  
& ~$f_{K^{*}}$\tabularnewline
\hline 
$130.2\pm1.2$  & $151\pm2.6$  & $216\pm5$  & $195\pm3$  & $169\pm2.6$  & $233\pm4.6$  & $155.7\pm3$  & $217\pm7$ \tabularnewline
\hline 
\end{tabular}\\
\begin{tabular}{ccccccccccccccc}
  \hline
   $g_{\rho\pi\pi}$~  &~$g_{\Xi_{c}^{+}\Xi_{c}^{+}\pi^{0}}$~  & ~$g_{\Xi_{c}^{\prime+}\Xi_{c}^{+}\pi^{0}}$~  & ~$g_{\Sigma_{c}^{+}\Sigma_{c}^{0}\pi^{+}}$~  & ~$g_{\Sigma_{c}^{*0}\Lambda_{c}^{+}\pi^{-}}$~&$g_{\Sigma_{c}^{*+}\Sigma_{c}^{0}\pi^{+}}$ \tabularnewline
\hline 
$6.05\pm0.02$& $0.70\pm0.22$  & $3.1\pm1.1$  & $8.0\pm2.8$  & $3.9\pm0.6$ & $4.3\pm0.4$ \tabularnewline
\hline 
 $g_{\omega \rho \pi}$&$g_{\Sigma_{c}^{*+}\Lambda_{c}^{+}\rho^{0}}$~  &~$g_{\Sigma_{c}^{*0}\Sigma_{c}^{+}\rho^{-}}$~  & ~$g_{\Xi_{c}^{0}\Lambda_{c}^{+}K^{*-}}$~  & $g_{\Sigma_{c}^{0}\Lambda_{c}^{+}\rho^{-}}$& $g_{\Sigma_{c}^{+}\Sigma_{c}^{0}\rho^{+}}$\tabularnewline\hline
  $-10\pm1$& $10\pm1.8$  &$5.77\pm0.5$  & $\{4.6\pm1.5,6\pm2\}$  & $\{2.6\pm0.9,16\pm5.3\}$&$\{4\pm1.3,27\pm9\}$ \tabularnewline\hline 
\end{tabular}
\end{table}

Calculation of this work requires inputs such as initial and final
state masses, decay constants of pseudoscalar and vector mesons, strong
couplings, transition form factors, and the
lifetimes of the doubly charmed baryons, as $\mathcal{BR}_{i}=\Gamma_{i}\cdot\tau$.
In the following, we will classify and discuss input parameter values in this work.
\begin{itemize}
  \item[(i)] 
The LHCb collaboration has successfully measured the mass and lifetime
of $\Xi_{cc}^{++}$: $m_{\Xi_{cc}^{++}}=3.621\rm{GeV}$~\cite{Aaij:2017ueg,Aaij:2018gfl}, $\tau_{\Xi_{cc}^{++}}=256\rm{fs}$~\cite{Aaij:2018wzf}. 
After these measurements, many theoretical works have studied the masses and lifetimes of doubly
charmed baryons~\cite{Workman:2022ynf,Chen:2016spr,Yu:2018com,Yu:2019lxw,Cheng:2018mwu,Berezhnoy:2018bde}.
The measurement of $\Xi_{cc}^{++}$ would benefit theoretical predictions on the other doubly charmed baryons. 
The results from Refs.~\cite{Yu:2018com,Berezhnoy:2018bde} are adopted in this work, $m_{\Xi_{cc}^{+}}=3.621\rm{GeV}$, $m_{\Omega_{cc}^{+}}=3.738\rm{GeV}$, $\tau_{\Xi_{cc}^{+}}=44\rm{fs}$ and $\tau_{\Omega_{cc}^{+}}=206\rm{fs}$. The masses of the final states including singly heavy baryons,
light pseudoscalar and vector mesons, can be easily found in Particle
Data Group~\cite{Workman:2022ynf}. 
\item[(ii)] 
The decay constants of pseudoscalar
and vector mesons are obtained from the literature~\cite{ParticleDataGroup:2024cfk,Choi:2015ywa,Feldmann:1998vh}, and summarized in Tab.~\ref{table:decayconstants}. The form factors
of the transition $\mathcal{B}_{cc}\to\mathcal{B}_{c}$ have been
calculated in many works~\cite{Wang:2017mqp}. In this work, we utilize
the theoretical results of form factors within the light-front quark
model, which have been successfully used to predict the discovery
channel of $\Xi_{cc}^{++}$ in Ref.~\cite{Yu:2017zst}.
\item[(iii)] 
In addition, the non-perturbative strong coupling constants serve a pivotal role in our calculation. 
Most of them can be taken from theoretical works. In this work, we use theoretical results~\cite{Cheng:2004ru,Nakayama:2006ps,Aliev:2010yx,Aliev:2010nh,Aliev:2010ev,Lin:2017mtz,Yan:1992gz,Casalbuoni:1996pg,Meissner:1987ge,Li:2012bt}, which have been listed in Tab.~\ref{table:decayconstants} and conmbine SU(3) symmetry to obtain.
For example, the strong couplings of singly heavy baryons and light mesons can be determined using the LCSRs method, as evidenced by $g_{\Xi_{c}^{+}\Xi_{c}^{+}\pi^{0}} = 0.7$ ~\cite{Aliev:2010yx}. The remaining couplings of this type can be inferred from $SU(3)$ flavor symmetry. 
\end{itemize}

\subsection{The determination of $\eta$}

\begin{figure}[t]
\centering {%
\begin{minipage}[c]{0.45\linewidth}%
\includegraphics[width=2.8in]{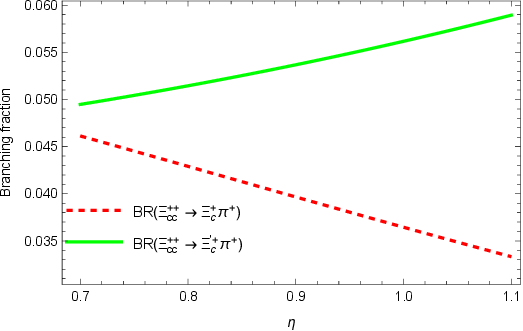} %
\end{minipage}} \hspace{1cm} { %
\begin{minipage}[c]{0.45\linewidth}%
\includegraphics[width=2.8in]{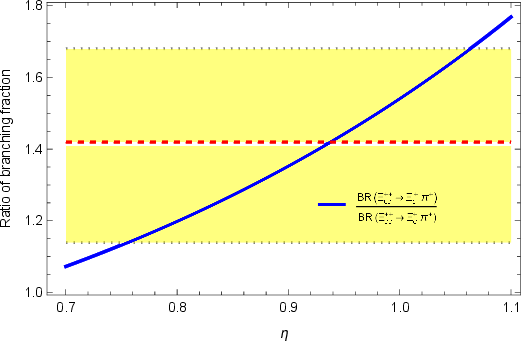} %
\end{minipage}} \caption{{(a): The theoretical predictions for the branching fractions of $\Xi_{cc}^{++}\to\Xi_{c}^{+}\pi^{+}$
and $\Xi_{cc}^{++}\to\Xi_{c}^{\prime+}\pi^{+}$ in logarithmic coordinates;
(b): Ratio of branching fractions with $\eta\sim[0.7,1.1]$.}
}
\label{fig:fig4} 
\end{figure}

In order to describe the off-shell effect of the exchanged particles,
one form factor is introduced as given in Eq.~(\ref{eq:Ffactor}).
In this form factor, the parameter $\eta$ is usually calculated from
the experimental data as in \cite{Cheng:2004ru}. In the case of doubly
charmed baryon decays, using the experiment data~\cite{LHCb:2019qed}
\begin{equation}
{\cal RB}=\frac{{\cal B}(\Xi_{cc}^{++}\to\Xi_{c}^{\prime+}\pi^{+})}{{\cal B}(\Xi_{cc}^{++}\to\Xi_{c}^{+}\pi^{+})}=1.41\pm0.17\pm0.10,
\end{equation}
this parameter can be determined as $\eta=0.9\pm0.2$.
With $\eta\sim[0.7,1.1]$, the dependence of the branching
fractions for $\Xi_{cc}^{++}\to\Xi_{c}^{+}\pi^{+}$ and $\Xi_{cc}^{++}\to\Xi_{c}^{\prime+}\pi^{+}$
can be shown as Fig.~\ref{fig:fig4} (a). {The long-distance contributions, primarily stemming from Fig.~\ref{fig:trianglesev} (a) and (c), coupled with the different form factors of the two processes, lead to an interesting outcome. In process  $\Xi_{cc}^{++}\to\Xi_{c}^{\prime+}\pi^{+}$, the contributions from Fig.~\ref{fig:trianglesev} (a) and (c) synergistically reinforce one another. Conversely, in process $\Xi_{cc}^{++}\to\Xi_{c}^{+}\pi^{+}$, these contributions negate each other. Consequently, as $\eta$ increases, there is a corresponding decrease in the branching ratio of process $\Xi_{cc}^{++}\to\Xi_{c}^{+}\pi^{+}$.}
It is said that the branching fractions are highly dependent on the value of the parameter $\eta$, the branching fractions could change about $20\%$ of magnitude. 
While as depicted
in Fig.~\ref{fig:fig4}~(b), the ratio of the two branching fractions
can also be given as,  
\begin{equation}
{\cal RB}=\frac{{\cal B}(\Xi_{cc}^{++}\to\Xi_{c}^{\prime+}\pi^{+})}{{\cal B}(\Xi_{cc}^{++}\to\Xi_{c}^{+}\pi^{+})}={1.35_{-0.66}^{+0.81}}.
\end{equation}
{the uncertainty of the ratio
is $60\%$ from the parameter $\eta$.} 
The lack of more experimental data for the calculation of $\eta$
will lead to large uncertainties in the theoretical predictions of
the branching fractions.
This is a well-known issue with final state interaction
effects, which have large theoretical uncertainties.
Nonetheless, the theoretical prediction using this value of $\eta$ remains highly reliable.
In this work, we will present the
numerical predictions of physical observation with $\eta=0.9\pm0.2$,
to show the absolute uncertainty of each mode. 
\subsection{Ratios of amplitudes and decay widths}
Upon determining $\eta$, the modulus of amplitude including $A$, $B$, and helicity amplitudes can be numerically computed using the amplitude for each nonleptonic decay mode, as detailed in Appendix \ref{app:amp}. 
Due to the limitations on the length of the article, we have not provided modulus of amplitude and widths for all processes. {Instead, we have only presented results for processes that are short-distance contributions dominated in the upper part of Tab.~\ref{tab:ampSD}, as well as those that are long-distance contributions dominated and Cabibbo-favored processes in the lower part. In addition to the error introduced by the parameter $\eta$, we also discuss errors arising from other input parameters. As illustrated in Tab.~\ref{tab:ampSD}, we present two distinct errors for each amplitude. The first source of uncertainty stems from the input parameters including decay constants and strong couplings. The second uncertainty is derived from the parameter $\eta=0.9\pm0.2$. And since the amplitudes ${A}_{SD}$ and ${B}_{SD}$ not involving strong couplings and $\eta$, there is only uncertainty from decay constants. Then we can found that the uncertainty from decay constants is much smaller than that from the strong coupling constants and $\eta$. This is because almost of these decay constants have been precisely determined using experimental data~\cite{ParticleDataGroup:2024cfk}. Nevertheless, there is a significant lack of experimental data available for determining the parameters: strong couplings and $\eta$.} In Tab.~\ref{Tab:amcomparison}, we conduct a comparative analysis with other theoretical works. In contrast to other studies, our work incorporates both real and imaginary contributions within the amplitudes, {which will provide the strong phases naturally to understand the $CP$ asymmetries and decay asymmetries.}
From {Tab.~\ref{tab:ampSD}}, it is evident that the imaginary component of the long-distance contribution closely approximates its real counterpart.

\begin{sidewaystable}\centering
\caption{{The amplitudes of ${\cal B}_{cc}\to{\cal B}_{c}P$ are in unit $10^{-2}G_F \rm{GeV}^2$. The processes shown in the upper part of this table are dominated by short-distance contributions, while those in the lower part are dominated by long-distance contributions and Cabibbo-favored. The first uncertainty comes from input parameters: decay constants and strong couplings, which have been listed in Tab.~\ref{table:decayconstants}. The second uncertainty is due to the variation of the phenomenological parameter $\eta=0.9\pm0.2$. While for the amplitudes ${A}_{SD}$ and ${B}_{SD}$, there is only uncertainty from decay constants.}} 
\label{tab:ampSD}
{\tiny{
\begin{tabular}{l|c|c|c|c|c|c}
\hline\hline
channels &${A}_{SD}$ &${B}_{SD}$
&${A}_{LD}$&${B}_{LD}$ 
&${A}_{tot}$&${B}_{tot}$ \\ \hline
$\Xi_{cc}^{++}\!\!\to\!\!\Xi_{c}^{+}\pi^{+} $ &
$9.92_{-0.09}^{+0.09}$ &
$14.80_{-0.14}^{+0.14}$ &
$-1.82_{-0.33-0.67}^{+0.32+0.61}+0.21_{-0.04-0.11}^{+0.02+0.16}i$ &
$-5.10_{-1.13-1.81}^{+1.01+1.69}+3.26_{-0.91-1.18}^{+1.11+1.37}i$ &
$8.10_{-0.29-0.67}^{+0.28+0.61}+0.21_{-0.04-0.11}^{+0.02+0.16}i$ &
$9.66_{-1.06-1.81}^{+0.94+1.69}+3.26_{-0.91-1.18}^{+1.11+1.37}i$ \\
$\Xi_{cc}^{++}\!\!\to\!\!\Xi_{c}^{\prime+}\pi^{+} $ &
$-5.27_{-0.05}^{+0.05}$ &
$-42.30_{-0.39}^{+0.39}$ &
$-1.02_{-0.40-0.36}^{+0.28+0.33}+1.58_{-0.46-0.58}^{+0.65+0.67}i$ &
$-2.61_{-0.56-0.85}^{+0.52+0.80}+2.48_{-0.94-1.02}^{+1.28+1.32}i$ &
$-6.29_{-0.42-0.36}^{+0.31+0.33}+1.58_{-0.46-0.58}^{+0.65+0.67}i$ &
$-44.90_{-0.75-0.85}^{+0.72+0.80}+2.48_{-0.94-1.02}^{+1.28+1.32}i$ \\
$\Xi_{cc}^{++}\!\!\to\!\!\Sigma_{c}^{+}\pi^{+} $ &
$1.18_{-0.01}^{+0.01}$ &
$8.22_{-0.08}^{+0.08}$ &
$-0.18_{-0.00-0.05}^{+0.03+0.05}-0.44_{-0.18-0.18}^{+0.12+0.16}i$ &
$-2.50_{-0.40-0.80}^{+0.40+0.74}+0.66_{-0.14-0.08}^{+0.03+0.04}i$ &
$1.01_{-0.01-0.05}^{+0.04+0.05}-0.44_{-0.18-0.18}^{+0.12+0.16}i$ &
$5.72_{-0.36-0.80}^{+0.36+0.74}+0.65_{-0.14-0.08}^{+0.03+0.04}i$ \\
$\Xi_{cc}^{++}\!\!\to\!\!\Lambda_{c}^{+}\pi^{+} $ &
$-2.29_{-0.02}^{+0.02}$ &
$-2.87_{-0.03}^{+0.03}$ &
$-0.24_{-0.03-0.08}^{+0.03+0.08}+0.36_{-0.07-0.13}^{+0.10+0.14}i$ &
$1.24_{-0.26-0.43}^{+0.29+0.49}-0.87_{-0.25-0.41}^{+0.22+0.34}i$ &
$-2.54_{-0.04-0.08}^{+0.04+0.08}+0.36_{-0.07-0.13}^{+0.10+0.14}i$ &
$-1.63_{-0.25-0.43}^{+0.28+0.49}-0.87_{-0.25-0.41}^{+0.22+0.34}i$ \\
$\Xi_{cc}^{++}\!\!\to\!\!\Xi_{c}^{\prime+}K^{+} $ &
$-1.54_{-0.03}^{+0.03}$ &
$-12.30_{-0.24}^{+0.24}$ &
$-0.02_{-0.02-0.03}^{+0.01+0.02}-0.06_{-0.01-0.02}^{+0.03+0.02}i$ &
$-1.46_{-0.33-0.48}^{+0.28+0.46}+0.58_{-0.17-0.13}^{+0.20+0.08}i$ &
$-1.56_{-0.03-0.03}^{+0.02+0.02}-0.06_{-0.01-0.02}^{+0.03+0.02}i$ &
$-13.70_{-0.45-0.48}^{+0.40+0.46}+0.58_{-0.17-0.13}^{+0.20+0.08}i$ \\
$\Xi_{cc}^{++}\!\!\to\!\!\Xi_{c}^{+}K^{+} $ &
$2.96_{+0.06}^{+0.06}$ &
$4.27_{-0.08}^{+0.08}$ &
$-0.23_{-0.01-0.04}^{+0.02+0.05}-0.34_{-0.09-0.16}^{+0.07+0.13}i$ &
$-1.84_{-0.29-0.61}^{+0.30+0.58}-0.94_{-0.25-0.33}^{+0.17+0.31}i$ &
$2.73_{-0.01-0.04}^{+0.03+0.05}-0.34_{-0.09-0.16}^{+0.07+0.13}i$ &
$2.43_{-0.25-0.61}^{+0.26+0.58}-0.94_{-0.25-0.33}^{+0.17+0.31}i$ \\
$\Xi_{cc}^{++}\!\!\to\!\!\Lambda_{c}^{+}K^{+} $ &
$-0.69_{-0.01}^{+0.01}$ &
$-0.84_{-0.02}^{+0.02}$ &
$0.10_{-0.01-0.03}^{+0.01+0.03}+0.07_{-0.01-0.02}^{+0.02+0.02}i$ &
$0.18_{-0.03-0.05}^{+0.02+0.04}+0.20_{-0.04-0.08}^{+0.05+0.10}i$ &
$-0.60_{-0.01-0.03}^{+0.00+0.03}+0.07_{-0.01-0.02}^{+0.02+0.02}i$ &
$-0.65_{-0.02-0.05}^{+0.01+0.04}+0.20_{-0.04-0.08}^{+0.05+0.10}i$ \\
$\Xi_{cc}^{++}\!\!\to\!\!\Sigma_{c}^{+}K^{+} $ &
$0.35_{-0.01}^{+0.01}$ &
$2.40_{0.05}^{+0.05}$ &
$-0.09_{-0.02-0.04}^{+0.02+0.03}+0.04_{-0.00-0.02}^{+0.00+0.03}i$ &
$0.46_{-0.09-0.13}^{+0.09+0.14}-0.28_{-0.09-0.09}^{+0.07+0.08}i$ &
$0.26_{-0.01-0.04}^{+0.01+0.03}+0.04_{-0.00-0.02}^{+0.00+0.03}i$ &
$2.86_{-0.11-0.13}^{+0.12+0.14}-0.28_{-0.09-0.09}^{+0.07+0.08}i$ \\
$\Xi_{cc}^{+}\!\!\to\!\!\Xi_{c}^{0}\pi^{+} $ &
$9.89_{-0.09}^{+0.09}$ &
$14.80_{-0.14}^{+0.14}$ &
$2.07_{-0.38-0.68}^{+0.41+0.73}-0.81_{-0.19-0.42}^{+0.16+0.33}i$ &
$5.55_{-1.01-1.85}^{+1.11+2.01}-5.18_{-1.16-2.49}^{+1.01+2.03}i$ &
$12.00_{-0.42-0.68}^{+0.46+0.73}-0.81_{-0.19-0.42}^{+0.16+0.33}i$ &
$20.30_{-1.08-1.85}^{+1.17+2.01}-5.18_{-1.16-2.49}^{+1.01+2.03}i$ \\
$\Xi_{cc}^{+}\!\!\to\!\!\Xi_{c}^{\prime0}\pi^{+} $ &
$-5.27_{-0.05}^{+0.05}$ &
$-42.30_{-0.39}^{+0.39}$ &
$0.54_{-0.11-0.18}^{+0.14+0.20}-0.62_{-0.16-0.26}^{+0.13+0.23}i$ &
$3.08_{-0.51-0.95}^{+0.52+1.00}-2.25_{-0.44-1.20}^{+0.40+0.94}i$ &
$-4.73_{-0.09-0.18}^{+0.12+0.20}-0.62_{-0.16-0.26}^{+0.13+0.23}i$ &
$-39.30_{-0.31-0.95}^{+0.32+1.00}-2.25_{-0.44-1.20}^{+0.40+0.94}i$ \\
$\Xi_{cc}^{+}\!\!\to\!\!\Sigma_{c}^{0}\pi^{+} $ &
$1.67_{-0.02}^{+0.02}$ &
$11.60_{0.11}^{+0.11}$ &
$-0.76_{-0.14-0.27}^{+0.13+0.25}-0.01_{-0.03-0.01}^{+0.02+0.03}i$ &
$-4.51_{-0.80-1.63}^{+0.77+1.50}+1.66_{-0.22-0.64}^{+0.17+0.78}i$ &
$0.92_{-0.13-0.27}^{+0.12+0.25}-0.01_{-0.03-0.01}^{+0.02+0.03}i$ &
$7.12_{-0.74-1.63}^{+0.72+1.50}+1.66_{-0.22-0.64}^{+0.17+0.78}i$ \\
$\Xi_{cc}^{+}\!\!\to\!\!\Xi_{c}^{\prime0}K^{+} $ &
$1.54_{0.03}^{+0.03}$ &
$12.30_{-0.24}^{+0.24}$ &
$-0.69_{-0.16-0.27}^{+0.14+0.24}+0.39_{-0.07-0.16}^{+0.07+0.20}i$ &
$1.26_{-0.18-0.41}^{+0.14+0.44}-0.44_{-0.11-0.37}^{+0.09+0.25}i$ &
$0.85_{-0.14-0.27}^{+0.12+0.24}+0.39_{-0.07-0.16}^{+0.07+0.20}i$ &
$13.50_{-0.30-0.41}^{+0.26+0.44}-0.45_{-0.11-0.37}^{+0.09+0.25}i$ \\
$\Xi_{cc}^{+}\!\!\to\!\!\Xi_{c}^{0}K^{+} $ &
$-2.96_{-0.06}^{+0.06}$ &
$-4.28_{-0.08}^{+0.08}$ &
$-0.00_{-0.02-0.01}^{+0.03+0.03}-0.23_{-0.05-0.11}^{+0.05+0.09}i$ &
$-0.92_{-0.08-0.30}^{+0.12+0.29}-0.36_{-0.16-0.03}^{+0.10+0.07}i$ &
$-2.96_{0.00-0.01}^{+0.01+0.03}-0.23_{-0.05-0.11}^{+0.05+0.09}i$ &
$-5.20_{-0.13-0.30}^{+0.17+0.29}-0.36_{-0.16-0.03}^{+0.10+0.07}i$ \\
$\Xi_{cc}^{+}\!\!\to\!\!\Sigma_{c}^{0}K^{+} $ &
$2.14_{-0.04}^{+0.04}$ &
$14.70_{-0.28}^{+0.28}$ &
$\cdots$ &
$\cdots$ &
$2.14_{-0.04}^{+0.04}$ &
$14.70_{-0.28}^{+0.28}$ \\
$\Omega_{cc}^{+}\!\!\to\!\!\Omega_{c}^{0}\pi^{+} $ &
$-7.39_{-0.07}^{+0.07}$ &
$-61.60_{-0.57}^{+0.57}$ &
$\cdots$ &
$\cdots$ &
$-7.39_{-0.07}^{+0.07}$ &
$-61.60_{-0.57}^{+0.57}$ \\
$\Omega_{cc}^{+}\!\!\to\!\!\Xi_{c}^{0}\pi^{+} $ &
$2.16_{-0.02}^{+0.02}$ &
$2.99_{-0.03}^{+0.03}$ &
$0.26_{-0.05-0.09}^{+0.06+0.10}-0.30_{-0.09-0.13}^{+0.07+0.11}i$ &
$-0.88_{-0.13-0.28}^{+0.14+0.28}+0.03_{-0.12-0.09}^{+0.05+0.04}i$ &
$2.42_{-0.06-0.09}^{+0.07+0.10}-0.30_{-0.09-0.13}^{+0.07+0.11}i$ &
$2.12_{-0.11-0.28}^{+0.13+0.28}+0.02_{-0.12-0.09}^{+0.05+0.04}i$ \\
$\Omega_{cc}^{+}\!\!\to\!\!\Xi_{c}^{\prime0}\pi^{+} $ &
$1.16_{-0.01}^{+0.01}$ &
$8.47_{-0.08}^{+0.08}$ &
$-0.16_{0.00-0.04}^{+0.01+0.05}-0.42_{-0.13-0.16}^{+0.10+0.15}i$ &
$-3.61_{-0.66-1.31}^{+0.63+1.21}+1.63_{-0.22-0.56}^{+0.16+0.63}i$ &
$1.00_{-0.01-0.04}^{+0.01+0.05}-0.42_{-0.13-0.16}^{+0.10+0.15}i$ &
$4.86_{-0.62-1.31}^{+0.59+1.21}+1.63_{-0.22-0.56}^{+0.16+0.63}i$ \\
$\Omega_{cc}^{+}\!\!\to\!\!\Omega_{c}^{0}K^{+} $ &
$-2.17_{-0.04}^{+0.04}$ &
$-17.90_{-0.34}^{+0.34}$ &
$-0.13_{-0.02-0.06}^{+0.02+0.05}-0.16_{-0.08-0.04}^{+0.05+0.05}i$ &
$-1.22_{-0.33-0.40}^{+0.26+0.39}+0.29_{-0.09-0.10}^{+0.02+0.00}i$ &
$-2.29_{-0.04-0.06}^{+0.04+0.05}-0.16_{-0.08-0.04}^{+0.05+0.05}i$ &
$-19.10_{-0.50-0.40}^{+0.43+0.39}+0.29_{-0.09-0.10}^{+0.02+0.00}i$ \\
$\Omega_{cc}^{+}\!\!\to\!\!\Xi_{c}^{0}K^{+} $ &
$0.65_{-0.01}^{+0.01}$ &
$0.87_{-0.02}^{+0.02}$ &
$-0.07_{-0.01-0.02}^{+0.01+0.02}-0.00_{-0.00-0.00}^{+0.00+0.00}i$ &
$-0.04_{-0.00-0.00}^{+0.01+0.01}-0.21_{-0.07-0.10}^{+0.05+0.08}i$ &
$0.59_{-0.00-0.02}^{+0.00+0.02}-0.00_{-0.00-0.00}^{+0.00+0.00}i$ &
$0.83_{-0.01-0.00}^{+0.02+0.01}-0.21_{-0.07-0.10}^{+0.05+0.08}i$ \\
$\Omega_{cc}^{+}\!\!\to\!\!\Xi_{c}^{\prime0}K^{+} $ &
$0.34_{-0.01}^{+0.01}$ &
$2.48_{-0.05}^{+0.05}$ &
$-0.06_{-0.02-0.02}^{+0.01+0.02}+0.06_{-0.01-0.02}^{+0.02+0.03}i$ &
$0.32_{-0.06-0.11}^{+0.06+0.12}-0.28_{-0.05-0.13}^{+0.05+0.11}i$ &
$0.28_{-0.01-0.02}^{+0.01+0.02}+0.06_{-0.01-0.02}^{+0.02+0.03}i$ &
$2.80_{-0.08-0.11}^{+0.09+0.12}-0.28_{-0.05-0.13}^{+0.05+0.11}i$ \\
\hline
$\Xi_{cc}^{++}\!\!\to\!\!\Sigma_{c}^{++}\bar{K}^{0} $ &
$-0.15_{-0.00}^{+0.00}$ &
$-0.23_{-0.00}^{+0.00}$ &
$-0.20_{-0.38-0.03}^{+0.22+0.11}-1.37_{-0.00-0.54}^{+0.20+0.46}i$ &
$-2.58_{-0.19-0.92}^{+0.30+0.85}-13.40_{-3.31-5.24}^{+2.43+4.66}i$ &
$-0.35_{-0.38-0.03}^{+0.22+0.11}-1.37_{-0.00-0.54}^{+0.20+0.46}i$ &
$-2.81_{-0.19-0.92}^{+0.30+0.85}-13.40_{-3.31-5.24}^{+2.43+4.66}i$ 
\\
$\Xi_{cc}^{+}\!\!\to\!\!\Omega_{c}^{0}K^{+} $ &
$\cdots$ &
$\cdots$ &
$2.20_{-0.45-0.75}^{+0.52+0.82}-2.18_{-0.52-1.02}^{+0.44+0.84}i$ &
$-7.29_{-1.21-2.37}^{+1.19+2.29}+1.80_{-0.19-0.82}^{+0.07+1.05}i$ &
$2.20_{-0.45-0.75}^{+0.52+0.82}-2.18_{-0.52-1.02}^{+0.44+0.84}i$ &
$-7.29_{-1.21-2.37}^{+1.19+2.29}+1.80_{-0.19-0.82}^{+0.07+1.05}i$ 
\\
$\Xi_{cc}^{+}\!\!\to\!\!\Sigma_{c}^{+}\bar{K}^{0} $ &
$-0.11_{-0.00}^{+0.00}$ &
$-0.16_{-0.00}^{+0.00}$ &
$-0.50_{-0.48-0.06}^{+0.29+0.10}-0.54_{-0.18-0.19}^{+0.41+0.17}i$ &
$-0.58_{-0.08-0.21}^{+0.49+0.19}-13.00_{-3.07-5.20}^{+2.26+4.58}i$ &
$-0.66_{-0.48-0.06}^{+0.29+0.10}-0.54_{-0.18-0.19}^{+0.41+0.17}i$ &
$-0.81_{-0.07-0.21}^{+0.49+0.19}-13.00_{-3.07-5.20}^{+2.26+4.58}i$ 
\\
$\Xi_{cc}^{+}\!\!\to\!\!\Lambda_{c}^{+}\bar{K}^{0} $ &
$-0.15_{-0.00}^{+0.00}$ &
$-0.23_{-0.00}^{+0.00}$ &
$2.23_{-0.23-0.76}^{+0.13+0.83}-1.41_{-0.03-0.68}^{+0.10+0.55}i$ &
$1.93_{-0.29-0.61}^{+0.46+0.63}-14.10_{-2.74-6.19}^{+2.16+5.22}i$ &
$2.07_{-0.23-0.76}^{+0.13+0.83}-1.41_{-0.03-0.68}^{+0.10+0.55}i$ &
$1.70_{-0.29-0.61}^{+0.45+0.63}-14.10_{-2.74-6.19}^{+2.16+5.22}i$ 
\\
$\Xi_{cc}^{+}\!\!\to\!\!\Sigma_{c}^{++}K^{-} $ &
$\cdots$ &
$\cdots$ &
$-0.49_{-0.30-0.18}^{+0.20+0.17}+0.48_{-0.24-0.19}^{+0.37+0.23}i$ &
$3.35_{-0.50-1.11}^{+0.64+1.20}-9.70_{-1.73-4.28}^{+1.36+3.61}i$ &
$-0.49_{-0.30-0.18}^{+0.20+0.17}+0.48_{-0.24-0.19}^{+0.37+0.23}i$ &
$3.35_{-0.50-1.11}^{+0.64+1.20}-9.70_{-1.73-4.28}^{+1.36+3.61}i$ 
\\
$\Xi_{cc}^{+}\!\!\to\!\!\Xi_{c}^{\prime+}\pi^{0} $ &
$\cdots$ &
$\cdots$ &
$-0.67_{-0.08-0.25}^{+0.09+0.23}+0.13_{-0.13-0.06}^{+0.06+0.09}i$ &
$3.42_{-0.64-1.18}^{+0.68+1.27}-2.00_{-0.84-0.76}^{+0.66+0.69}i$ &
$-0.67_{-0.08-0.25}^{+0.09+0.23}+0.13_{-0.13-0.06}^{+0.06+0.09}i$ &
$3.42_{-0.64-1.18}^{+0.68+1.27}-2.00_{-0.84-0.76}^{+0.66+0.69}i$ 
\\
$\Xi_{cc}^{+}\!\!\to\!\!\Xi_{c}^{\prime+}\eta_{1} $ &
$\cdots$ &
$\cdots$ &
$-0.52_{-0.28-0.18}^{+0.19+0.17}+0.78_{-0.29-0.29}^{+0.43+0.34}i$ &
$1.64_{-0.20-0.54}^{+0.24+0.57}-3.79_{-0.20-1.67}^{+0.23+1.41}i$ &
$-0.52_{-0.28-0.18}^{+0.19+0.17}+0.78_{-0.29-0.29}^{+0.43+0.34}i$ &
$1.64_{-0.20-0.54}^{+0.24+0.57}-3.79_{-0.20-1.67}^{+0.23+1.41}i$ 
\\
$\Xi_{cc}^{+}\!\!\to\!\!\Xi_{c}^{\prime+}\eta_{8} $ &
$\cdots$ &
$\cdots$ &
$-0.12_{-0.02-0.03}^{+0.04+0.04}+0.43_{-0.03-0.15}^{+0.03+0.17}i$ &
$-3.40_{-0.58-1.22}^{+0.46+1.13}+10.50_{-1.46-3.94}^{+1.85+4.69}i$ &
$-0.12_{-0.02-0.03}^{+0.04+0.04}+0.43_{-0.03-0.15}^{+0.03+0.17}i$ &
$-3.40_{-0.58-1.22}^{+0.46+1.13}+10.50_{-1.46-3.94}^{+1.85+4.69}i$ 
\\
$\Xi_{cc}^{+}\!\!\to\!\!\Xi_{c}^{+}\pi^{0} $ &
$\cdots$ &
$\cdots$ &
$0.74_{-0.08-0.21}^{+0.02+0.20}+1.04_{-0.23-0.35}^{+0.30+0.38}i$ &
$2.39_{-0.37-0.66}^{+0.31+0.61}+3.35_{-0.58-1.27}^{+0.75+1.54}i$ &
$0.74_{-0.08-0.21}^{+0.02+0.20}+1.04_{-0.23-0.35}^{+0.30+0.38}i$ &
$2.39_{-0.37-0.66}^{+0.31+0.61}+3.35_{-0.58-1.27}^{+0.75+1.54}i$ 
\\
$\Xi_{cc}^{+}\!\!\to\!\!\Xi_{c}^{+}\eta_{1} $ &
$\cdots$ &
$\cdots$ &
$-0.13_{-0.15-0.04}^{+0.09+0.04}+0.14_{-0.14-0.06}^{+0.23+0.07}i$ &
$1.85_{-0.30-0.61}^{+0.38+0.67}-6.49_{-1.10-2.86}^{+0.92+2.42}i$ &
$-0.13_{-0.15-0.04}^{+0.09+0.04}+0.14_{-0.14-0.06}^{+0.23+0.07}i$ &
$1.85_{-0.30-0.61}^{+0.38+0.67}-6.49_{-1.10-2.86}^{+0.92+2.42}i$ 
\\
$\Xi_{cc}^{+}\!\!\to\!\!\Xi_{c}^{+}\eta_{8} $ &
$\cdots$ &
$\cdots$ &
$0.42_{-0.21-0.14}^{+0.34+0.15}-0.53_{-0.48-0.25}^{+0.30+0.21}i$ &
$-3.79_{-0.85-1.38}^{+0.66+1.27}+12.80_{-1.96-4.74}^{+2.40+5.62}i$ &
$0.42_{-0.21-0.14}^{+0.34+0.15}-0.53_{-0.48-0.25}^{+0.30+0.21}i$ &
$-3.79_{-0.85-1.38}^{+0.66+1.27}+12.80_{-1.96-4.74}^{+2.40+5.62}i$ 
\\
$\Omega_{cc}^{+}\!\!\to\!\!\Xi_{c}^{\prime+}\bar{K}^{0} $ &
$0.08_{-0.00}^{+0.00}$ &
$0.68_{-0.01}^{+0.01}$ &
$0.99_{-0.12-0.42}^{+0.02+0.53}-1.63_{-0.01-0.73}^{+0.22+0.60}i$ &
$-5.86_{-0.60-1.96}^{+0.90+1.87}-8.35_{-1.44-3.07}^{+1.06+2.81}i$ &
$1.10_{-0.12-0.42}^{+0.02+0.53}-1.63_{-0.01-0.73}^{+0.22+0.60}i$ &
$-4.90_{-0.60-1.96}^{+0.90+1.87}-8.35_{-1.44-3.07}^{+1.06+2.81}i$ 
\\
$\Omega_{cc}^{+}\!\!\to\!\!\Xi_{c}^{+}\bar{K}^{0} $ &
$0.12_{-0.00}^{+0.00}$ &
$0.96_{-0.01}^{+0.01}$ &
$2.41_{-0.46-0.86}^{+0.56+0.98}-4.31_{-1.02-1.88}^{+0.83+1.59}i$ &
$-3.36_{-0.55-1.10}^{+0.64+1.06}-11.10_{-2.20-4.72}^{+1.71+4.04}i$ &
$2.53_{-0.46-0.86}^{+0.56+0.98}-4.31_{-1.02-1.88}^{+0.83+1.59}i$ &
$-2.40_{-0.55-1.10}^{+0.64+1.06}-11.10_{-2.20-4.72}^{+1.71+4.04}i$ 
\\
\hline\hline
\end{tabular} }}\end{sidewaystable}

\begin{table}[tb]
  \renewcommand{\arraystretch}{1.5}
  \addtolength{\arraycolsep}{5pt}
    \centering
    \caption{ Comparison of the decay amplitudes of $\Xi_{cc}^{++}\to\Xi_{c}^{(\prime)+}\pi^{+}$ from this work with those from the
literature. All the amplitudes below are in unit $10^{-2}G_F \rm{GeV}^2$.  
    In this table, we list the theoretical prediction from the pole
    approximation (PA) in tandem with the SU(3)F symmetry~\cite{Liu:2023dvg}, the non-relativistic quark model (NRQM)~\cite{Zeng:2022egh},
    the pole model (PM) with current algebra (CA)~\cite{Cheng:2020wmk}, the Ab initio three-loop calculation (AITLC)~\cite{Gutsche:2018msz}, 
    the heavy quark effective theory (HQET) with pole model (PM)~\cite{Sharma:2017txj}, light-cone sum rules (LCSR) with HQET~\cite{Shi:2022kfa}, and the bag model~\cite{Liu:2022igi}. 
    }
    \label{Tab:amcomparison}
    \begin{tabular}{l|c|c|c|c|c|c}
      \hline\hline
      $\Xi_{cc}^{++}\to\Xi_{c}^{+}\pi^{+}$&${A_{SD}}$ &${A_{LD}}$&${A_{tot}}$
      &${B_{SD}}$  &${B_{LD}}$ &${B_{tot}}$\\ \hline
        This work&$9.92$  &
$-1.82+0.21i$ &
$8.10+0.21i$ &
$14.80$ &
$-5.11+3.26i$ &
$9.66+3.26i$\\
PA+SU(3)~\cite{Liu:2023dvg}&$5.36$&$-1.12$&4.24&$7.59$&$-5.43$&$2.16$\\
        NRQM~\cite{Zeng:2022egh}&$7.4$ &$-12.4$ &$-5.0$ &$-13$ &$21.8$&$8.8$\\
        PM+CA~\cite{Cheng:2020wmk}& $7.40$&$-10.79$ &$-3.38$ &$-15.06$ &$18.91$&$3.85$\\
        AITLC~\cite{Gutsche:2018msz}& $-8.1$ &$11.5$ &3.4& $13.0$& $-18.5$& $-5.6$\\
        HQET+PM~\cite{Sharma:2017txj}& $9.52$& $ 0$& $ 9.52 $& $-19.45$& $-24.95 $& $-44.40$\\
        LCSR+HQET~\cite{Shi:2022kfa}& $9.52 $& $-16.67$& $ -7.18 $& $-19.45 $& $-20.47$&  $-39.92$\\
        Bag Model~\cite{Liu:2022igi}&$4.83$& $-9.99$& $-5.16$& $5.16$& $13.6$& $18.8$\\\hline
              $\Xi_{cc}^{++}\to\Xi_{c}^{\prime+}\pi^{+}$&${A_{SD}}$ &${A_{LD}}$&${A_{tot}}$
      &${B_{SD}}$  &${B_{LD}}$ &${B_{tot}}$\\ \hline
        This work&$-5.28$ &
$-1.02+1.58i$ &
$-6.30+1.58i$ &
$-42.40$ &
$-2.61+2.48i$ &
$-45.00+2.48i$\\
PA+SU(3)~\cite{Liu:2023dvg}&$-2.80$& $0$&$-2.80$& $-22.31$& $0$&$-22.31$\\
        NRQM~\cite{Zeng:2022egh}& $3.7$ &$0 $& $3.7$& $ -37.1$& $ 0$& $ -31.0$\\
        PM+CA~\cite{Cheng:2020wmk}& $4.5$ &$ -0.04 $ &$4.5 $ &$-48.5 $ &$-0.06 $ &$-48.4$\\
        AITLC~\cite{Gutsche:2018msz}& $-4.3$& $ -0.1$& $ -4.4$& $ 37.6 $& $1.4$& $ 39.0$\\
        HQET+PM~\cite{Sharma:2017txj}& $ 5.10$& $ 0 $& $5.10 $& $-62.37$& $ 0 $& $-62.37$\\
        LCSR+HQET~\cite{Shi:2022kfa}& $ 5.10$& $ -0.83$& $ 4.27$& $ -62.37$& $ -8.86$& $ -71.23 $\\
        Bag Model~\cite{Liu:2022igi}&$ 7.38$& $ -4.82$& $ 2.56 $& $-51.0$& $ 7.26 $& $-43.7$\\\hline\hline
      \end{tabular}
  \end{table}
Besides, the topological diagrams have the relations of $\frac{|C|}{|T|}\sim\frac{|C^{\prime}|}{|C|}\sim\frac{|E_{1}|}{|C|}\sim\frac{|E_{2}|}{|C|}\sim O(\frac{\Lambda_{\text{QCD}}^{h}}{m_{Q}})$
in the heavy baryon decays, manifested by the soft-colliner effective
theory \cite{Leibovich:2003tw,Mantry:2003uz}. These relations are
important in the phenomenological studies on the searches for the
double-heavy-flavor baryons, and give us more hints on the dynamics
of heavy baryon decays. From the above relations, all the tree-level
topological diagrams are at the same order in charmed baryon decays
due to $\Lambda_{\text{QCD}}^{h}/m_{c}\sim1$. It would be very useful
to numerically test these relations in our framework.

Based on the topological analysis in Tab.~\ref{tab:topologies}, $\mathcal{A}(\Xi_{cc}^{++}\to\Sigma_{c}^{++}\overline{K}^{0})\sim\tilde{C}$ and $\mathcal{A}(\Xi_{cc}^{++}\to\Xi_{c}^{\prime+}\pi^{+})\sim(\tilde{T}+\tilde{C}^{\prime})/\sqrt{2}$,
we obtain the ratios $\tilde{C}_{LD}/\tilde{T}_{SD}$ and $\tilde{C}_{LD}^{\prime}/\tilde{C}_{LD}$
as, 
\begin{align}
\frac{|\tilde{C}_{LD}|}{|\tilde{T}_{SD}|} & =\frac{|\mathcal{A}(\Xi_{cc}^{++}\to\Sigma_{c}^{++}\overline{K}^{0})|_{LD}}{\sqrt{2}|\mathcal{A}(\Xi_{cc}^{++}\to\Xi_{c}^{\prime+}\pi^{+})|_{SD}}={0.14\sim0.30},\label{eq:ct}\\
\frac{|\tilde{C}_{LD}^{\prime}|}{|\tilde{C}_{LD}|} & =\frac{\sqrt{2}|\mathcal{A}(\Xi_{cc}^{++}\to\Xi_{c}^{\prime+}\pi^{+})|_{LD}}{|\mathcal{A}(\Xi_{cc}^{++}\to\Sigma_{c}^{++}\overline{K}^{0})|_{LD}}={1.02\sim1.06}.\label{eq:cc}
\end{align}

In the following to consider the relations between $C$, $E_{1}$
and $E_{2}$, it would be more convenient to study the processes with
a pure topological diagram. The advantage of avoiding interference
between diagrams is that they could be directly determined by the
experimental data in the future.
With the help of $\mathcal{A}(\Xi_{cc}^{+}\to\Sigma_{c}^{++}K^{-})\sim\tilde{E}_{1}$, $\mathcal{A}(\Xi_{cc}^{+}\to\Omega_{c}^{0}K^{+})\sim\tilde{E}_{2}$ and $\mathcal{A}(\Xi_{cc}^{++}\to\Sigma_{c}^{++}\overline{K}^{0})\sim\tilde{C}$, the ratios $\tilde{E}_{1LD}/\tilde{C}_{LD}$ and $\tilde{E}_{2LD}/\tilde{C}_{ LD}$ can be easily carried out,
\begin{align}
\frac{|\tilde{E}_{1LD}|}{|\tilde{C}_{LD}|} & =\frac{|\mathcal{A}(\Xi_{cc}^{+}\to\Sigma_{c}^{++}K^{-})|_{LD}}{|\mathcal{A}(\Xi_{cc}^{++}\to\Sigma_{c}^{++}\overline{K}^{0})|_{LD}}={0.48\sim0.52},\label{eq:e1c}\\
\frac{|\tilde{E}_{2LD}|}{|\tilde{C}_{LD}|} & =\frac{|\mathcal{A}(\Xi_{cc}^{+}\to\Omega_{c}^{0}K^{+})|_{LD}}{|\mathcal{A}(\Xi_{cc}^{++}\to\Sigma_{c}^{++}\overline{K}^{0})|_{LD}}={0.70\sim0.77}.\label{eq:e2c}
\end{align}
From Eqs.~(\ref{eq:ct}-\ref{eq:e2c}), considering the relatively
large parameter uncertainty, all these results are consistent with
the relations found in \cite{Leibovich:2003tw,Mantry:2003uz}, 
\begin{align}
\frac{|C|}{|T|}\sim\frac{|C^{\prime}|}{|C|}\sim\frac{|E_{1}|}{|C|}\sim\frac{|E_{2}|}{|C|}\sim\mathcal{O}\left({\frac{\Lambda_{QCD}^{h}}{m_{c}}}\right)\sim\mathcal{O}(1).
\end{align}
The results shown with Eqs.~(\ref{eq:ct}-\ref{eq:e2c}) are different
with each other. This can be understood by the flavor $SU(3)$ breaking
effects. The flavor $SU(3)$ symmetry is of great significance in
the weak decays of heavy hadrons. In terms of a few $SU(3)$ irreducible
amplitudes, several relations between the widths of doubly charmed
baryon decays are obtained \cite{Wang:2017azm}. It is important to
numerically test the flavor $SU(3)$ symmetry and its breaking effects.

In the following, we show our numerical results on the ratios of decay
widths. In the $SU(3)$ limit, they should be unity. Any deviation
would indicate the $SU(3)$ breaking effects. 
\begin{align}
&\frac{\Gamma\big(\Xi_{cc}^{++}\to\Lambda_{c}^{+}\pi^{+}\big)}{\Gamma\big(\Xi_{cc}^{++}\to\Xi_{c}^{+}K^{+}\big)}
 =\frac{|\lambda_{d}(T+C^{\prime})|^{2}}{|\lambda_{s}(T+C^{\prime})|^{2}} ={1.01\sim1.13},\label{eq:SU3-1}\\
 & \frac{\Gamma\big(\Xi_{cc}^{+}\to\Xi_{c}^{+}K^{0}\big)}{\Gamma\big(\Omega_{cc}^{+}\to\Lambda_{c}^{+}\overline{K}^{0}\big)}=\frac{|\lambda_{s}C^{\prime}+\lambda_{d}E_{1}|^{2}}{|\lambda_{d}C^{\prime}+\lambda_{s}E_{1}|^{2}}
 ={0.53\sim0.56},\label{eq:SU3-2}\\
 & \frac{\Gamma\big(\Xi_{cc}^{+}\to\Xi_{c}^{0}K^{+}\big)}{\Gamma\big(\Omega_{cc}^{+}\to\Xi_{c}^{0}\pi^{+}\big)}= \frac{|\lambda_{s}T+\lambda_{d}E_{2}|^{2}}{|-\lambda_{d}T-\lambda_{s}E_{2}|^{2}}
 ={{1.02\sim1.19}},\label{eq:SU3-3}\\
 & \frac{\Gamma\big(\Xi_{cc}^{++}\to\Sigma_{c}^{++}\pi^{0}\big)}{\frac{1}{3}\Gamma\big(\Xi_{cc}^{++}\to\Sigma_{c}^{++}\eta_{8}\big)} = \frac{|-\frac{1}{\sqrt{2}}\tilde{C}|^{2}}{|-\frac{1}{\sqrt{6}}(\lambda_{d}-2\lambda_{s})\tilde{C}|^{2}}
  ={1.24\sim1.50},\label{eq:SU3-4}\\
  &\frac{\Gamma\big(\Xi_{cc}^{++}\to\Sigma_{c}^{+}\pi^{+}\big)}{\Gamma\big(\Xi_{cc}^{++}\to\Xi_{c}^{\prime+}K^{+}\big)} =\frac{|\frac{1}{\sqrt{2}}\lambda_{d}(\tilde{T}+\tilde{C}^{\prime})|^{2}}{|\frac{1}{\sqrt{2}}\lambda_{s}(\tilde{T}+\tilde{C}^{\prime})|^{2}}
   ={0.44\sim0.45},\label{eq:SU3-5}\\
   &\frac{\Gamma\big(\Xi_{cc}^{+}\to\Sigma_{c}^{++}\pi^{-}\big)}{\Gamma\big(\Omega_{cc}^{+}\to\Sigma_{c}^{++}K^{-}\big)} =\frac{|\lambda_{d}\tilde{E}_{1}|^{2}}{|\lambda_{s}\tilde{E}_{1}|^{2}}
  ={0.49\sim0.62},\label{eq:SU3-6}\\
  & \frac{\Gamma\big(\Omega_{cc}^{+}\to\Omega_{c}^{0}K^{+}\big)}{\Gamma\big(\Xi_{cc}^{+}\to\Sigma_{c}^{0}\pi^{+}\big)} =\frac{|\lambda_{s}(\tilde{T}+\tilde{E}_{2})|^{2}}{|\lambda_{d}(\tilde{T}+\tilde{E}_{2})|^{2}}
={3.54\sim6.36},\label{eq:SU3-7}\\
& \frac{\Gamma\big(\Xi_{cc}^{+}\to\Xi_{c}^{\prime+}K^{0}\big)}{\Gamma\big(\Omega_{cc}^{+}\to\Sigma_{c}^{+}\overline{K}^{0}\big)}=\frac{|\frac{1}{\sqrt{2}}(\lambda_{s}\tilde{C}^{\prime}+\lambda_{d}\tilde{E}_{1})|^{2}}{|\frac{1}{\sqrt{2}}(\lambda_{d}\tilde{C}^{\prime}+\lambda_{s}\tilde{E}_{1})|^{2}}
 ={0.69\sim0.72},\label{eq:SU3-8}\\
&\frac{\Gamma\big(\Omega_{cc}^{+}\to\Xi_{c}^{\prime0}\pi^{+}\big)}{\Gamma\big(\Xi_{cc}^{+}\to\Xi_{c}^{\prime0}K^{+}\big)} =\frac{|\frac{1}{\sqrt{2}}(\lambda_{d}\tilde{T}^{\prime}+\lambda_{s}\tilde{E}_{2})|^{2}}{|\frac{1}{\sqrt{2}}(\lambda_{s}\tilde{T}^{\prime}+\lambda_{d}\tilde{E}_{2})|^{2}} ={{0.50\sim0.51}}.\label{eq:SU3-9}
\end{align}

The numerical results presented above indicate that long-distance final-state interactions can significantly contribute to $SU(3)$ symmetry breaking. This effect originates from several factors, including the exchanged particles, hadronic strong coupling constants, transition form factors, and decay constants.
The interference between distinct diagrams is also considered. 

In our calculation of Eqs.~(\ref{eq:SU3-1})-(\ref{eq:SU3-9}), the large values mainly stem from the strong coupling constants. 
Taking ${\Gamma\big(\Omega_{cc}^{+}\to\Omega_{c}^{0}K^{+}\big)}/{\Gamma\big(\Xi_{cc}^{+}\to\Sigma_{c}^{0}\pi^{+}\big)}={{3.54\sim6.36}}$ as an example, the two decay modes are both dominated by the triangle diagram $\mathcal{M}(V,B;V)$ contribution, including $\mathcal{M}(K^{*+},\Omega_{c}^{0};\phi)$, $\mathcal{M}(\rho^+,\Xi_{c}^{(\prime)0};\overline{K}^{*0})$ for $\Omega_{cc}^{+}\to\Omega_{c}^{0}K^{+}$ and $\mathcal{M}(\rho^+,\Sigma_{c}^{0};\rho^0)$, $\mathcal{M}(K^{*+},\Xi_{c}^{(\prime)0};K^{*0})$ for $\Xi_{cc}^{+}\to\Sigma_{c}^{0}\pi^{+}$. Due to $g_{K^{*+}\phi K^{+}}/g_{\rho^{+}\rho^{0}\pi^{+}}= g_{\Omega_c^0\Omega_c^0\phi}/g_{\Sigma_c^0\Sigma_c^0\rho^{0}}=\sqrt{2}$, the ratio  $\mathcal{M}(K^{*+},\Omega_{c}^{0};\phi)/\mathcal{M}(\rho^+,\Sigma_{c}^{0};\rho^0)$ will be $2$. While the ratios of other two amplitudes $\mathcal{M}(\rho^+,\Xi_{c}^{(\prime)0};\overline{K}^{*0})/\mathcal{M}(K^{*+},\Xi_{c}^{(\prime)0};K^{*0})$ will be $1$, for $g_{\rho^{+}\bar{K}^{*0}K^{+}}/g_{K^{*+}K^{*0}\pi^{+}}=g_{\Omega_c^0\Xi_c^{\prime0}K^{*0}}/g_{\Sigma_c^0\Xi_c^{\prime0}K^{*0}}=1$.
Therefore, ${\Gamma\big(\Omega_{cc}^{+}\to\Omega_{c}^{0}K^{+}\big)}$ is slightly larger than ${\Gamma\big(\Xi_{cc}^{+}\to\Sigma_{c}^{0}\pi^{+}\big)}$, combining the effects from other input parameters. 

All strong couplings utilized in this study are sourced from existing literature and align with SU(3) flavor symmetry. These non-perturbative values carry significant theoretical uncertainties, warranting a comprehensive and meticulous examination in future research.

\subsection{Branching ratio and Decay asymmetry parameters}
\begin{table}
\caption{{Branching ratios and decay asymmetry parameters for the short-distance dynamics dominated modes. 
The ``CF\char`\"{}, ``SCS\char`\"{} and ``DCS\char`\"{} represent
CKM favored, singly CKM suppressed and doubly CKM suppressed processes,
respectively. And ${\cal B}_{SD}$, ${\cal B}_{LD}$ and ${\cal B}_{tot}$ denote the branching ratio of short-distance contribution, long-distance contribution and total one respectively. The first uncertainty comes from input parameters: decay constants and strong couplings, which have been listed in Tab.~\ref{table:decayconstants}. The second uncertainty is due to the variation of the phenomenological parameter $\eta=0.9\pm0.2$. While for the branching ratio ${\cal B}_{SD}$, there is only uncertainty from decay constants.}}
\label{tab:result1} %
\footnotesize
\begin{tabular}{l|c|c|c|c|c|c|c}
  \hline\hline
  channels &CKM&${\cal B}_{SD}[10^{-3}]$ &${\cal B}_{LD}[10^{-3}]$
  &${\cal B}_{tot}[10^{-3}]$ 
  &$\alpha$&$\beta$&$\gamma$\\ \hline
  $\Xi_{cc}^{++}\to\Xi_{c}^{+}\pi^{+} $ &CF
  &$60.70_{-1.11}^{+1.12}$ &
  $2.67_{-0.82-1.50}^{+1.30+2.37}$ &
  $39.70_{-2.73-6.33}^{+2.97+6.47}$ &
  $-0.43_{-0.02-0.03}^{+0.03+0.04}$ &
  $0.13_{-0.04-0.05}^{+0.06+0.07}$ &
  $-0.89_{-0.00-0.01}^{+0.01+0.01}$ \\
  $\Xi_{cc}^{++}\to\Xi_{c}^{\prime+}\pi^{+} $ &CF
  &$42.30_{-0.78}^{+0.78}$ &
  $2.13_{-0.87-1.24}^{+2.40+2.09}$ &
  $53.70_{-3.58-4.20}^{+5.86+5.26}$ &
  $-0.97_{0.00-0.00}^{+0.00+0.00}$ &
  $0.19_{-0.04-0.06}^{+0.04+0.05}$ &
  $0.14_{-0.07-0.06}^{+0.04+0.05}$ \\
  $\Xi_{cc}^{++}\to\Sigma_{c}^{+}\pi^{+} $ &SCS
  &$2.23_{-0.04}^{+0.04}$ &
  $0.27_{-0.08-0.14}^{+0.17+0.21}$ &
  $1.39_{-0.04-0.12}^{+0.09+0.18}$ &
  $-0.87_{-0.04-0.07}^{+0.07+0.12}$ &
  $0.50_{-0.11-0.16}^{+0.06+0.14}$ &
  $0.00_{-0.17-0.18}^{+0.10+0.12}$ \\
  $\Xi_{cc}^{++}\to\Lambda_{c}^{+}\pi^{+} $ &SCS
  &$3.44_{-0.06}^{+0.06}$ &
  $0.18_{-0.06-0.11}^{+0.11+0.17}$ &
  $4.08_{-0.13-0.25}^{+0.17+0.32}$ &
  $-0.26_{-0.05-0.10}^{+0.06+0.11}$ &
  $0.19_{-0.04-0.06}^{+0.04+0.06}$ &
  $-0.95_{-0.00-0.01}^{+0.01+0.02}$ \\
  $\Xi_{cc}^{++}\to\Xi_{c}^{\prime+}K^{+} $ &SCS
  &$2.74_{-0.10}^{+0.11}$ &
  $0.03_{-0.01-0.01}^{+0.02+0.02}$ &
  $3.18_{-0.15-0.16}^{+0.19+0.18}$ &
  $-0.96_{-0.00-0.01}^{+0.00+0.01}$ &
  $-0.08_{-0.00-0.01}^{+0.01+0.02}$ &
  $0.26_{-0.01-0.02}^{+0.01+0.02}$ \\
  $\Xi_{cc}^{++}\to\Xi_{c}^{+}K^{+} $ &SCS
  &$4.82_{-0.18}^{+0.19}$ &
  $0.15_{-0.04-0.08}^{+0.07+0.13}$ &
  $4.02_{-0.04-0.08}^{+0.13+0.16}$ &
  $-0.31_{-0.03-0.05}^{+0.03+0.06}$ &
  $-0.08_{-0.02-0.04}^{+0.02+0.03}$ &
  $-0.95_{-0.01-0.01}^{+0.01+0.02}$ \\
  $\Xi_{cc}^{++}\to\Lambda_{c}^{+}K^{+} $ &DCS
  &$0.29_{-0.01}^{+0.01}$ &
  $0.01_{-0.00-0.00}^{+0.00+0.01}$ &
  $0.21_{-0.00-0.02}^{+0.00+0.02}$ &
  $-0.44_{-0.01-0.01}^{+0.00+-0.00}$ &
  $-0.09_{-0.02-0.06}^{+0.01+0.04}$ &
  $-0.89_{0.00-0.00}^{+0.00+0.01}$ \\
  $\Xi_{cc}^{++}\to\Sigma_{c}^{+}K^{+} $ &DCS
  &$0.16_{-0.01}^{+0.01}$ &
  $0.01_{-0.00-0.01}^{+0.00+0.01}$ &
  $0.17_{-0.01-0.00}^{+0.01+0.01}$ &
  $-0.79_{-0.04-0.09}^{+0.05+0.12}$ &
  $-0.21_{-0.00-0.09}^{+0.02+0.08}$ &
  $0.58_{-0.06-0.12}^{+0.06+0.11}$ \\
  $\Xi_{cc}^{+}\to\Xi_{c}^{0}\pi^{+} $ &CF
  &$10.40_{-0.19}^{+0.19}$ &
  $0.68_{-0.21-0.39}^{+0.34+0.63}$ &
  $15.60_{-1.15-1.90}^{+1.38+2.26}$ &
  $-0.58_{-0.01-0.02}^{+0.01+0.02}$ &
  $-0.11_{-0.02-0.04}^{+0.02+0.04}$ &
  $-0.81_{-0.01-0.02}^{+0.01+0.02}$ \\
  $\Xi_{cc}^{+}\to\Xi_{c}^{\prime0}\pi^{+} $ &CF
  &$7.26_{-0.13}^{+0.13}$ &
  $0.10_{-0.03-0.06}^{+0.06+0.10}$ &
  $6.15_{-0.14-0.32}^{+0.13+0.33}$ &
  $-0.95_{-0.00-0.00}^{+0.01+0.01}$ &
  $-0.07_{-0.02-0.03}^{+0.02+0.02}$ &
  $0.31_{-0.01-0.01}^{+0.01+0.01}$ \\
  $\Xi_{cc}^{+}\to\Sigma_{c}^{0}\pi^{+} $ &SCS
  &$0.76_{-0.01}^{+0.01}$ &
  $0.14_{-0.04-0.08}^{+0.06+0.12}$ &
  $0.28_{-0.05-0.10}^{+0.06+0.13}$ &
  $-0.89_{-0.02-0.04}^{+0.06+0.10}$ &
  $0.22_{-0.05-0.09}^{+0.08+0.11}$ &
  $0.40_{-0.03-0.06}^{+0.05+0.12}$ \\
  $\Xi_{cc}^{+}\to\Xi_{c}^{\prime0}K^{+} $ &SCS
  &$0.47_{-0.02}^{+0.02}$ &
  $0.05_{-0.02-0.03}^{+0.03+0.06}$ &
  $0.41_{-0.00-0.01}^{+0.00+0.01}$ &
  $-0.69_{-0.08-0.16}^{+0.12+0.23}$ &
  $-0.34_{-0.07-0.18}^{+0.07+0.15}$ &
  $0.64_{-0.08-0.14}^{+0.06+0.08}$ \\
  $\Xi_{cc}^{+}\to\Xi_{c}^{0}K^{+} $ &SCS
  &$0.82_{-0.03}^{+0.03}$ &
  $0.01_{-0.00-0.00}^{+0.00+0.01}$ &
  $0.85_{-0.01-0.00}^{+0.01+0.00}$ &
  $-0.55_{-0.01-0.03}^{+0.01+0.03}$ &
  $-0.00_{-0.00-0.02}^{+0.00+0.01}$ &
  $-0.83_{-0.01-0.02}^{+0.01+0.02}$ \\
  $\Xi_{cc}^{+}\to\Sigma_{c}^{0}K^{+} $ &DCS
  &$1.00_{-0.04}^{+0.04}$ &
  $\cdots$ &$1.00_{-0.04}^{+0.04}$ &
  $-0.98$ &
  $\cdots$ &
  $0.18$ \\
  $\Omega_{cc}^{+}\to\Omega_{c}^{0}\pi^{+} $ &CF
  &$68.10_{-1.25}^{+1.26}$ &
  $\cdots$ &$68.10_{-1.25}^{+1.26}$ &
  $-0.96$ &
  $\cdots$ &
  $0.28$ \\
  $\Omega_{cc}^{+}\to\Xi_{c}^{0}\pi^{+} $ &SCS
  &$2.46_{-0.05}^{+0.05}$ &
  $0.09_{-0.03-0.05}^{+0.06+0.09}$ &
  $2.99_{-0.15-0.21}^{+0.21+0.27}$ &
  $-0.34_{-0.03-0.06}^{+0.03+0.06}$ &
  $0.05_{-0.01-0.01}^{+0.00+-0.00}$ &
  $-0.94_{-0.01-0.02}^{+0.01+0.02}$ \\
  $\Omega_{cc}^{+}\to\Xi_{c}^{\prime0}\pi^{+} $ &SCS
  &$1.75_{-0.03}^{+0.03}$ &
  $0.34_{-0.10-0.19}^{+0.17+0.30}$ &
  $0.96_{-0.00-0.10}^{+0.07+0.18}$ &
  $-0.74_{-0.09-0.17}^{+0.17+0.33}$ &
  $0.66_{-0.13-0.24}^{+0.07+0.18}$ &
  $-0.15_{-0.15-0.21}^{+0.11+0.19}$ \\
  $\Omega_{cc}^{+}\to\Omega_{c}^{0}K^{+} $ &SCS
  &$4.43_{-0.17}^{+0.17}$ &
  $0.03_{-0.01-0.02}^{+0.02+0.02}$ &
  $5.03_{-0.20-0.21}^{+0.24+0.25}$ &
  $-0.98_{-0.00-0.00}^{+0.00+-0.00}$ &
  $-0.08_{-0.03-0.01}^{+0.02+0.02}$ &
  $0.17_{-0.00-0.01}^{+0.01+0.00}$ \\
  $\Omega_{cc}^{+}\to\Xi_{c}^{0}K^{+} $ &DCS
  &$0.20_{-0.01}^{+0.01}$ &
  $0.00_{-0.00-0.00}^{+0.00+0.00}$ &
  $0.17_{-0.00-0.01}^{+0.00+0.01}$ &
  $-0.50_{-0.01-0.01}^{+0.01+0.01}$ &
  $-0.13_{-0.04-0.06}^{+0.03+0.05}$ &
  $-0.86_{-0.01-0.01}^{+0.01+0.02}$ \\
  $\Omega_{cc}^{+}\to\Xi_{c}^{\prime0}K^{+} $ &DCS
  &$0.12_{-0.00}^{+0.00}$ &
  $0.01_{-0.00-0.00}^{+0.00+0.01}$ &
  $0.13_{-0.00-0.00}^{+0.00+0.01}$ &
  $-0.85_{-0.04-0.07}^{+0.05+0.09}$ &
  $-0.28_{-0.06-0.11}^{+0.05+0.10}$ &
  $0.45_{-0.05-0.08}^{+0.04+0.08}$ \\
  \hline\hline
  \end{tabular}
\end{table}

\begin{table}
  \caption{{Branching ratios and decay asymmetry parameters for the long-distance dominated Cabibbo-favored ($\lambda_{sd}$)
  modes. The first uncertainty comes from input parameters: decay constants and strong couplings, which have been listed in Tab.~\ref{table:decayconstants}. The second uncertainty is due to the variation of the phenomenological parameter $\eta=0.9\pm0.2$. While for the branching ratio ${\cal B}_{SD}$, there is only uncertainty from decay constants.}}\label{tab:result2}
  \footnotesize
  \begin{tabular}{l|c|c|c|c|c|c}
    \hline\hline
    channels &${\cal B}_{SD}[10^{-5}]$ &${\cal B}_{LD}[10^{-3}]$
    &${\cal B}_{tot}[10^{-3}]$ 
    &$\alpha$&$\beta$&$\gamma$\\ \hline
    $\Xi_{cc}^{++}\to\Sigma_{c}^{++}\bar{K}^{0} $
    &$1.32_{-0.02}^{+0.02}$ &
    $3.92_{-0.86-2.22}^{+1.71+3.63}$ &
    $3.98_{-0.89-2.22}^{+1.77+3.62}$ &
    $-0.88_{-0.03-0.01}^{+0.17+0.01}$ &
    $0.04_{-0.18-0.10}^{+0.22+0.12}$ &
    $0.48_{-0.18-0.05}^{+0.11+0.02}$ \\
    $\Xi_{cc}^{+}\to\Omega_{c}^{0}K^{+} $
    &$\cdots$ &
    $0.78_{-0.25-0.46}^{+0.45+0.78}$ &
    $0.78_{-0.25-0.46}^{+0.45+0.78}$ &
    $0.47_{-0.03-0.02}^{+0.03+0.02}$ &
    $-0.28_{-0.01-0.02}^{+0.00+0.02}$ &
    $-0.83_{-0.02-0.02}^{+0.02+0.02}$ \\
    $\Xi_{cc}^{+}\to\Sigma_{c}^{+}\bar{K}^{0} $
    &$0.11_{-0.00}^{+0.00}$ &
    $0.51_{-0.12-0.29}^{+0.34+0.47}$ &
    $0.52_{-0.13-0.29}^{+0.36+0.47}$ &
    $-0.45_{-0.22-0.02}^{+0.30+0.01}$ &
    $0.48_{-0.25-0.10}^{+0.15+0.11}$ &
    $0.75_{-0.14-0.10}^{+0.07+0.06}$ \\
    $\Xi_{cc}^{+}\to\Lambda_{c}^{+}\bar{K}^{0} $
    &$0.25_{-0.00}^{+0.00}$ &
    $1.53_{-0.32-0.91}^{+0.44+1.57}$ &
    $1.46_{-0.31-0.89}^{+0.43+1.55}$ &
    $-0.65_{-0.04-0.01}^{+0.06+0.01}$ &
    $-0.74_{-0.00-0.01}^{+0.03+0.02}$ &
    $0.17_{-0.09-0.00}^{+0.13+0.01}$ \\
    $\Xi_{cc}^{+}\to\Sigma_{c}^{++}K^{-} $
    &$\cdots$ &
    $0.33_{-0.09-0.20}^{+0.22+0.35}$ &
    $0.33_{-0.09-0.20}^{+0.22+0.35}$ &
    $0.60_{-0.28-0.00}^{+0.13+0.00}$ &
    $0.30_{-0.08-0.01}^{+0.03+0.01}$ &
    $0.74_{-0.18-0.00}^{+0.12+0.00}$ \\
    $\Xi_{cc}^{+}\to\Xi_{c}^{\prime+}\pi^{0} $
    &$\cdots$ &
    $0.08_{-0.02-0.05}^{+0.04+0.08}$ &
    $0.08_{-0.02-0.05}^{+0.04+0.08}$ &
    $0.94_{-0.14-0.02}^{+0.01+0.01}$ &
    $0.34_{-0.25-0.03}^{+0.19+0.04}$ &
    $-0.03_{-0.16-0.01}^{+0.11+0.00}$ \\
    $\Xi_{cc}^{+}\to\Xi_{c}^{\prime+}\eta_{1} $
    &$\cdots$ &
    $0.10_{-0.04-0.06}^{+0.12+0.10}$ &
    $0.10_{-0.04-0.06}^{+0.12+0.10}$ &
    $0.89_{-0.14-0.01}^{+0.07+0.00}$ &
    $0.16_{-0.05-0.01}^{+0.02+0.01}$ &
    $-0.43_{-0.18-0.01}^{+0.53+0.01}$ \\
    $\Xi_{cc}^{+}\to\Xi_{c}^{\prime+}\eta_{8} $
    &$\cdots$ &
    $0.03_{-0.01-0.02}^{+0.01+0.03}$ &
    $0.03_{-0.01-0.02}^{+0.01+0.03}$ &
    $-0.89_{-0.04-0.02}^{+0.06+0.02}$ &
    $0.03_{-0.07-0.00}^{+0.08+0.00}$ &
    $0.46_{-0.10-0.04}^{+0.07+0.03}$ \\
    $\Xi_{cc}^{+}\to\Xi_{c}^{+}\pi^{0} $
    &$\cdots$ &
    $0.22_{-0.06-0.12}^{+0.11+0.18}$ &
    $0.22_{-0.06-0.12}^{+0.11+0.18}$ &
    $-0.89_{-0.00-0.02}^{+0.01+0.02}$ &
    $-0.00_{-0.05-0.04}^{+0.06+0.03}$ &
    $-0.46_{-0.01-0.03}^{+0.01+0.04}$ \\
    $\Xi_{cc}^{+}\to\Xi_{c}^{+}\eta_{1} $
    &$\cdots$ &
    $0.11_{-0.03-0.07}^{+0.07+0.12}$ &
    $0.11_{-0.03-0.07}^{+0.07+0.12}$ &
    $0.29_{-0.33-0.01}^{+0.23+0.01}$ &
    $0.15_{-0.10-0.01}^{+0.05+0.01}$ &
    $0.95_{-0.16-0.00}^{+0.01+0.00}$ \\
    $\Xi_{cc}^{+}\to\Xi_{c}^{+}\eta_{8} $
    &$\cdots$ &
    $0.13_{-0.04-0.08}^{+0.13+0.14}$ &
    $0.13_{-0.04-0.08}^{+0.13+0.14}$ &
    $0.72_{-0.44-0.01}^{+0.11+0.01}$ &
    $0.29_{-0.13-0.02}^{+0.02+0.03}$ &
    $0.63_{-0.29-0.00}^{+0.18+0.00}$ \\
    $\Omega_{cc}^{+}\to\Xi_{c}^{\prime+}\bar{K}^{0} $
    &$0.82_{-0.02}^{+0.02}$ &
    $2.76_{-0.31-1.63}^{+0.07+2.82}$ &
    $2.74_{-0.30-1.63}^{+0.05+2.83}$ &
    $-0.42_{-0.01-0.04}^{+0.01+0.02}$ &
    $-0.88_{-0.02-0.00}^{+0.06+0.01}$ &
    $-0.20_{-0.14-0.05}^{+0.28+0.04}$ \\
    $\Omega_{cc}^{+}\to\Xi_{c}^{+}\bar{K}^{0} $
    &$2.07_{-0.04}^{+0.04}$ &
    $13.10_{-3.97-7.83}^{+7.37+13.60}$ &
    $13.30_{-4.00-7.89}^{+7.42+13.70}$ &
    $-0.54_{-0.03-0.00}^{+0.01+-0.00}$ &
    $-0.49_{-0.00-0.01}^{+0.01+0.01}$ &
    $-0.69_{-0.01-0.00}^{+0.03+0.00}$ \\
    \hline\hline
    \end{tabular}
  \end{table}

\begin{table}
\caption{{Same as Table.\ref{tab:result2} but for the long-distance dominated singly Cabibbo-suppressed modes.}}
\label{tab:result3} 
\footnotesize
\begin{tabular}{l|c|c|c|c|c|c}
  \hline\hline
  channels &${\cal B}_{SD}[10^{-7}]$ &${\cal B}_{LD}[10^{-4}]$
  &${\cal B}_{tot}[10^{-4}]$ 
  &$\alpha$&$\beta$&$\gamma$\\ \hline
  $\Xi_{cc}^{++}\to\Sigma_{c}^{++}\pi^{0} $
  &$3.89_{-0.07}^{+0.07}$ &
  $4.05_{-1.31-2.25}^{+2.95+3.73}$ &
  $3.92_{-1.27-2.21}^{+2.90+3.70}$ &
  $-0.47_{-0.15-0.06}^{+0.15+0.07}$ &
  $0.58_{-0.16-0.03}^{+0.07+0.01}$ &
  $0.66_{-0.04-0.06}^{+0.01+0.07}$ \\
  $\Xi_{cc}^{++}\to\Sigma_{c}^{++}\eta_{1} $
  &$3.81_{-0.15}^{+0.15}$ &
  $0.09_{-0.03-0.05}^{+0.11+0.07}$ &
  $0.09_{-0.03-0.04}^{+0.10+0.07}$ &
  $0.57_{-0.46-0.25}^{+0.05+0.13}$ &
  $0.82_{-0.11-0.11}^{+-0.01+0.13}$ &
  $0.04_{-0.25-0.05}^{+0.23+0.02}$ \\
  $\Xi_{cc}^{++}\to\Sigma_{c}^{++}\eta_{8} $
  &$1.16_{-0.04}^{+0.05}$ &
  $8.22_{-3.13-4.74}^{+7.84+7.78}$ &
  $7.84_{-3.04-4.61}^{+7.71+7.65}$ &
  $-0.50_{-0.06-0.01}^{+0.04+0.01}$ &
  $0.20_{-0.01-0.01}^{+0.03+0.01}$ &
  $-0.84_{-0.03-0.01}^{+0.05+0.01}$ \\
  $\Xi_{cc}^{+}\to\Sigma_{c}^{+}\pi^{0} $
  &$0.47_{-0.01}^{+0.01}$ &
  $0.27_{-0.06-0.16}^{+0.11+0.29}$ &
  $0.28_{-0.06-0.16}^{+0.12+0.29}$ &
  $-0.56_{-0.06-0.06}^{+0.13+0.05}$ &
  $-0.12_{-0.18-0.06}^{+0.21+0.07}$ &
  $0.82_{-0.13-0.04}^{+0.05+0.02}$ \\
  $\Xi_{cc}^{+}\to\Sigma_{c}^{+}\eta_{1} $
  &$0.33_{-0.01}^{+0.01}$ &
  $0.04_{-0.01-0.03}^{+0.05+0.04}$ &
  $0.04_{-0.01-0.02}^{+0.04+0.04}$ &
  $0.98_{-0.42-0.06}^{+-0.06+0.01}$ &
  $-0.11_{-0.13-0.10}^{+0.01+0.04}$ &
  $0.14_{-0.32-0.11}^{+0.39+0.20}$ \\
  $\Xi_{cc}^{+}\to\Sigma_{c}^{+}\eta_{8} $
  &$0.10_{-0.00}^{+0.00}$ &
  $0.74_{-0.30-0.43}^{+0.83+0.70}$ &
  $0.70_{-0.28-0.41}^{+0.81+0.68}$ &
  $-0.21_{-0.01-0.02}^{+0.01+0.02}$ &
  $0.40_{-0.06-0.01}^{+0.15+0.01}$ &
  $-0.89_{-0.03-0.01}^{+0.10+0.01}$ \\
  $\Xi_{cc}^{+}\to\Lambda_{c}^{+}\pi^{0} $
  &$1.22_{-0.02}^{+0.02}$ &
  $0.96_{-0.24-0.58}^{+0.38+1.01}$ &
  $0.93_{-0.24-0.56}^{+0.38+1.00}$ &
  $-0.62_{-0.02-0.05}^{+0.01+0.06}$ &
  $-0.69_{-0.01-0.04}^{+0.01+0.03}$ &
  $-0.38_{0.00-0.03}^{+0.02+0.04}$ \\
  $\Xi_{cc}^{+}\to\Lambda_{c}^{+}\eta_{1} $
  &$0.73_{-0.03}^{+0.03}$ &
  $0.06_{-0.02-0.04}^{+0.03+0.07}$ &
  $0.07_{-0.01-0.04}^{+0.03+0.07}$ &
  $0.07_{-0.27-0.14}^{+0.24+0.08}$ &
  $-0.35_{-0.09-0.12}^{+0.10+0.07}$ &
  $0.93_{-0.11-0.05}^{+0.06+0.01}$ \\
  $\Xi_{cc}^{+}\to\Lambda_{c}^{+}\eta_{8} $
  &$0.26_{-0.01}^{+0.01}$ &
  $0.65_{-0.14-0.39}^{+0.20+0.68}$ &
  $0.69_{-0.15-0.40}^{+0.20+0.69}$ &
  $-0.71_{-0.00-0.03}^{+-0.00+0.04}$ &
  $-0.70_{0.00-0.04}^{+0.01+0.04}$ &
  $-0.01_{-0.05-0.06}^{+0.07+0.04}$ \\
  $\Xi_{cc}^{+}\to\Sigma_{c}^{++}\pi^{-} $
  &$\cdots$ &
  $0.18_{-0.05-0.10}^{+0.11+0.18}$ &
  $0.18_{-0.05-0.10}^{+0.11+0.18}$ &
  $0.73_{-0.24-0.00}^{+0.10+0.01}$ &
  $0.13_{-0.09-0.01}^{+0.05+0.00}$ &
  $0.67_{-0.18-0.00}^{+0.15+0.00}$ \\
  $\Xi_{cc}^{+}\to\Xi_{c}^{\prime+}K^{0} $
  &$\cdots$ &
  $0.32_{-0.10-0.18}^{+0.18+0.30}$ &
  $0.32_{-0.10-0.18}^{+0.18+0.30}$ &
  $0.96_{-0.02-0.01}^{+0.01+0.00}$ &
  $0.23_{-0.09-0.02}^{+0.07+0.03}$ &
  $0.13_{-0.01-0.01}^{+0.03+0.00}$ \\
  $\Xi_{cc}^{+}\to\Xi_{c}^{+}K^{0} $
  &$\cdots$ &
  $0.44_{-0.13-0.24}^{+0.29+0.37}$ &
  $0.44_{-0.13-0.24}^{+0.29+0.37}$ &
  $-0.19_{-0.15-0.05}^{+0.13+0.05}$ &
  $-0.90_{0.030.00}^{+0.05+0.00}$ &
  $-0.40_{-0.06-0.02}^{+0.02+0.02}$ \\
  $\Omega_{cc}^{+}\to\Sigma_{c}^{+}\bar{K}^{0} $
  &$\cdots$ &
  $2.09_{-0.62-1.24}^{+1.15+2.14}$ &
  $2.09_{-0.62-1.24}^{+1.15+2.14}$ &
  $-0.37_{-0.07-0.03}^{+0.03+0.03}$ &
  $-0.79_{-0.02-0.02}^{+0.03+0.02}$ &
  $-0.49_{-0.06-0.01}^{+0.12+0.01}$ \\
  $\Omega_{cc}^{+}\to\Lambda_{c}^{+}\bar{K}^{0} $
  &$\cdots$ &
  $3.93_{-1.12-2.31}^{+2.33+3.95}$ &
  $3.93_{-1.12-2.31}^{+2.33+3.95}$ &
  $0.34_{-0.20-0.04}^{+0.15+0.03}$ &
  $0.67_{-0.04-0.06}^{+-0.00+0.06}$ &
  $0.67_{-0.08-0.05}^{+0.04+0.04}$ \\
  $\Omega_{cc}^{+}\to\Sigma_{c}^{++}K^{-} $
  &$\cdots$ &
  $1.67_{-0.35-1.01}^{+0.71+1.77}$ &
  $1.67_{-0.35-1.01}^{+0.71+1.77}$ &
  $0.33_{-0.24-0.01}^{+0.18+0.01}$ &
  $0.32_{-0.07-0.02}^{+0.03+0.02}$ &
  $0.89_{-0.14-0.00}^{+0.04+0.00}$ \\
  $\Omega_{cc}^{+}\to\Xi_{c}^{\prime+}\pi^{0} $
  &$\cdots$ &
  $3.01_{-1.01-1.78}^{+2.15+3.03}$ &
  $3.01_{-1.01-1.78}^{+2.15+3.03}$ &
  $-0.24_{-0.02-0.07}^{+0.00+0.07}$ &
  $-0.76_{-0.04-0.05}^{+0.04+0.04}$ &
  $-0.60_{-0.03-0.02}^{+0.05+0.04}$ \\
  $\Omega_{cc}^{+}\to\Xi_{c}^{\prime+}\eta_{1} $
  &$\cdots$ &
  $0.48_{-0.15-0.28}^{+0.55+0.49}$ &
  $0.48_{-0.15-0.28}^{+0.55+0.49}$ &
  $0.96_{-0.25-0.00}^{+-0.12+0.00}$ &
  $0.24_{-0.04-0.01}^{+-0.03+0.00}$ &
  $-0.11_{-0.28-0.03}^{+0.49+0.02}$ \\
  $\Omega_{cc}^{+}\to\Xi_{c}^{\prime+}\eta_{8} $
  &$\cdots$ &
  $8.95_{-3.31-5.22}^{+8.17+8.72}$ &
  $8.95_{-3.31-5.22}^{+8.17+8.72}$ &
  $-0.62_{-0.06-0.00}^{+0.05+-0.00}$ &
  $0.11_{-0.01-0.01}^{+0.00+0.01}$ &
  $-0.78_{-0.03-0.00}^{+0.06+0.00}$ \\
  $\Omega_{cc}^{+}\to\Xi_{c}^{+}\pi^{0} $
  &$\cdots$ &
  $2.65_{-0.82-1.55}^{+1.53+2.62}$ &
  $2.65_{-0.82-1.55}^{+1.53+2.62}$ &
  $0.28_{-0.02-0.07}^{+-0.01+0.08}$ &
  $-0.50_{-0.01-0.03}^{+-0.00+0.04}$ &
  $-0.82_{-0.00-0.00}^{+0.00+0.01}$ \\
  $\Omega_{cc}^{+}\to\Xi_{c}^{+}\eta_{1} $
  &$\cdots$ &
  $0.72_{-0.15-0.44}^{+0.29+0.78}$ &
  $0.72_{-0.15-0.44}^{+0.29+0.78}$ &
  $-0.04_{-0.24-0.00}^{+0.24+0.00}$ &
  $0.12_{-0.06-0.01}^{+0.05+0.01}$ &
  $0.99_{-0.07-0.00}^{+-0.06+0.00}$ \\
  $\Omega_{cc}^{+}\to\Xi_{c}^{+}\eta_{8} $
  &$\cdots$ &
  $2.55_{-0.89-1.47}^{+1.98+2.42}$ &
  $2.55_{-0.89-1.47}^{+1.98+2.42}$ &
  $-0.80_{-0.04-0.00}^{+0.04+0.00}$ &
  $0.15_{-0.00-0.01}^{+-0.00+0.02}$ &
  $-0.58_{-0.05-0.01}^{+0.06+0.01}$ \\
  \hline\hline
  \end{tabular}
\end{table}

\begin{table}
\caption{{Same as Tab.~\ref{tab:result2} but for the long-distance dominated doubly Cabibbo-suppressed modes.}}
\label{tab:result4} 
\footnotesize
\begin{tabular}{l|c|c|c|c|c|c}
  \hline\hline
  channels &${\cal B}_{SD}[10^{-8}]$ &${\cal B}_{LD}[10^{-5}]$
  &${\cal B}_{tot}[10^{-5}]$ 
  &$\alpha$&$\beta$&$\gamma$\\ \hline
  $\Xi_{cc}^{++}\to\Sigma_{c}^{++}K^{0} $
  &$6.32_{-0.24}^{+0.25}$ &
  $1.59_{-0.48-0.86}^{+1.09+1.41}$ &
  $1.68_{-0.50-0.89}^{+1.12+1.43}$ &
  $-0.09_{-0.22-0.07}^{+0.19+0.08}$ &
  $0.85_{-0.09-0.05}^{+0.00+0.04}$ &
  $0.51_{-0.04-0.10}^{+0.00+0.09}$ \\
  $\Xi_{cc}^{+}\to\Sigma_{c}^{+}K^{0} $
  &$0.64_{-0.02}^{+0.03}$ &
  $0.10_{-0.02-0.05}^{+0.04+0.08}$ &
  $0.09_{-0.01-0.05}^{+0.03+0.08}$ &
  $0.05_{-0.15-0.02}^{+0.19+0.04}$ &
  $-0.50_{-0.21-0.03}^{+0.40+0.03}$ &
  $0.87_{-0.28-0.02}^{+-0.00+0.01}$ \\
  $\Xi_{cc}^{+}\to\Lambda_{c}^{+}K^{0} $
  &$1.76_{-0.07}^{+0.07}$ &
  $0.49_{-0.15-0.30}^{+0.30+0.53}$ &
  $0.50_{-0.15-0.30}^{+0.30+0.53}$ &
  $-0.55_{-0.03-0.05}^{+0.01+0.05}$ &
  $-0.66_{0.00-0.01}^{+0.01+0.02}$ &
  $-0.51_{-0.02-0.04}^{+0.04+0.03}$ \\
  $\Omega_{cc}^{+}\to\Sigma_{c}^{+}\pi^{0} $
  &$\cdots$ &
  $0.88_{-0.27-0.52}^{+0.54+0.90}$ &
  $0.88_{-0.27-0.52}^{+0.54+0.90}$ &
  $0.93_{-0.09-0.00}^{+0.02+0.00}$ &
  $0.26_{-0.02-0.00}^{+0.01+0.00}$ &
  $0.27_{-0.12-0.01}^{+0.16+0.01}$ \\
  $\Omega_{cc}^{+}\to\Sigma_{c}^{+}\eta_{1} $
  &$\cdots$ &
  $0.15_{-0.05-0.09}^{+0.18+0.16}$ &
  $0.15_{-0.05-0.09}^{+0.18+0.16}$ &
  $0.96_{-0.25-0.00}^{+0.07+0.00}$ &
  $0.27_{-0.05-0.01}^{+0.03+0.01}$ &
  $-0.04_{-0.25-0.02}^{+0.45+0.01}$ \\
  $\Omega_{cc}^{+}\to\Sigma_{c}^{+}\eta_{8} $
  &$\cdots$ &
  $0.10_{-0.04-0.06}^{+0.12+0.09}$ &
  $0.10_{-0.04-0.06}^{+0.12+0.09}$ &
  $-0.77_{-0.02-0.02}^{+0.12+0.02}$ &
  $0.62_{-0.07-0.03}^{+-0.00+0.03}$ &
  $-0.16_{-0.16-0.01}^{+0.38+0.02}$ \\
  $\Omega_{cc}^{+}\to\Lambda_{c}^{+}\pi^{0} $
  &$\cdots$ &
  $0.43_{-0.10-0.25}^{+0.23+0.40}$ &
  $0.43_{-0.10-0.25}^{+0.23+0.40}$ &
  $0.17_{-0.29-0.08}^{+0.25+0.07}$ &
  $-0.85_{0.03-0.03}^{+0.07+0.03}$ &
  $0.50_{-0.14-0.03}^{+-0.02+0.02}$ \\
  $\Omega_{cc}^{+}\to\Lambda_{c}^{+}\eta_{1} $
  &$\cdots$ &
  $0.17_{-0.04-0.11}^{+0.09+0.19}$ &
  $0.17_{-0.04-0.11}^{+0.09+0.19}$ &
  $-0.05_{-0.28-0.00}^{+0.27+0.00}$ &
  $0.06_{-0.09-0.01}^{+0.07+0.01}$ &
  $1.00_{-0.11-0.00}^{+-0.08+0.00}$ \\
  $\Omega_{cc}^{+}\to\Lambda_{c}^{+}\eta_{8} $
  &$\cdots$ &
  $1.21_{-0.37-0.67}^{+0.62+1.02}$ &
  $1.21_{-0.37-0.67}^{+0.62+1.02}$ &
  $-0.41_{-0.05-0.02}^{+0.07+0.03}$ &
  $-0.90_{-0.03-0.02}^{+0.04+0.02}$ &
  $-0.15_{-0.05-0.05}^{+0.03+0.05}$ \\
  $\Omega_{cc}^{+}\to\Sigma_{c}^{0}\pi^{+} $
  &$\cdots$ &
  $1.34_{-0.43-0.79}^{+0.79+1.33}$ &
  $1.34_{-0.43-0.79}^{+0.79+1.33}$ &
  $0.99_{-0.01-0.00}^{+0.01+0.00}$ &
  $0.01_{-0.01-0.00}^{+0.00+0.00}$ &
  $-0.16_{-0.05-0.00}^{+0.06+0.01}$ \\
  $\Omega_{cc}^{+}\to\Sigma_{c}^{++}\pi^{-} $
  &$\cdots$ &
  $0.69_{-0.17-0.42}^{+0.37+0.74}$ &
  $0.69_{-0.17-0.42}^{+0.37+0.74}$ &
  $0.49_{-0.24-0.00}^{+0.15+0.00}$ &
  $0.26_{-0.08-0.01}^{+0.04+0.01}$ &
  $0.83_{-0.15-0.00}^{+0.08+0.00}$ \\
  $\Omega_{cc}^{+}\to\Xi_{c}^{\prime+}K^{0} $
  &$2.31_{-0.09}^{+0.09}$ &
  $1.38_{-0.41-0.82}^{+0.96+1.44}$ &
  $1.32_{-0.39-0.80}^{+0.94+1.42}$ &
  $0.25_{-0.25-0.05}^{+0.16+0.05}$ &
  $0.67_{-0.14-0.05}^{+0.02+0.04}$ &
  $0.70_{-0.14-0.02}^{+0.10+0.03}$ \\
  $\Omega_{cc}^{+}\to\Xi_{c}^{+}K^{0} $
  &$4.95_{-0.19}^{+0.19}$ &
  $1.96_{-0.41-1.14}^{+0.52+1.93}$ &
  $2.10_{-0.43-1.19}^{+0.54+1.98}$ &
  $-0.32_{-0.06-0.04}^{+0.07+0.04}$ &
  $-0.93_{-0.030.00}^{+0.04+0.01}$ &
  $-0.19_{-0.05-0.11}^{+0.09+0.09}$ \\
  \hline\hline
  \end{tabular}
\end{table}

The branching ratio and three asymmetry parameters ($\alpha$, $\beta$, $\gamma$) of each nonleptonic decay $\mathcal{B}_{cc}\to\mathcal{B}_{c}P$
can be predicted. The branching
ratios and decay symmetry parameters of the short-distance contribution dominated channels (with
$T$ topology) are listed into Tab.~\ref{tab:result1}. For the long-distance
contribution-dominated processes, the numerical results are given in Tabs.~\ref{tab:result2}-\ref{tab:result4}. {As do in Tab.~\ref{tab:ampSD}, we also consider two uncertainty sources for branching ratio and three asymmetry parameters. The first source of uncertainty stems from decay constants and strong couplings. The second one is from the parameter $\eta=0.9\pm0.2$. Also for the branching ratio $\mathcal{B}_{SD}$ not involving input parameters: strong couplings and $\eta$, there is only uncertainty from decay constants of light mesons. Then we can found that the uncertainty from decay constants is also much smaller than that from the strong coupling constants and $\eta$. And compared with the branching ratios, the decay asymmetry parameters are less affected by the two parameters. Therefore, the decay asymmetry parameter has a weaker dependence on the input parameter: strong coupling constants and $\eta$, compared to the branching ratios.} In Tab.~\ref{Tab:amcomparison}, we conduct a comparative analysis with other theoretical works. In contrast to other studies, our work incorporates both real and imaginary contributions within the amplitudes, which will provide the strong phases naturally to understand the $CP$ asymmetries and decay asymmetries.In Table \ref{Tab:comparison}, we present a comparison of our findings with both the experimental data and results from existing theoretical literature.
The findings indicate that the diverse predictions do not entirely align with each other.

It can be found that the factorizable contributions of diagram amplitude
$T$ are dominant relative to the long-distance contributions of $C^{\prime}$
and $E_{2}$, as shown in Tab.~\ref{tab:result1}. On the other
hand, from Tabs.~\ref{tab:result2}-\ref{tab:result4}, the long-distance
contributions are dominated, since the short-distance amplitude $C_{SD}$
is deeply suppressed by the effective Wilson coefficient $a_{2}(m_{c})=-0.017<a_{1}(m_{c})=1.07$
at the charm mass scale. In this work, $a_{1}(m_{c})=1.07$ has been
used at the calculation of the weak decay vertex in the triangle diagram.
Currently, there is no absolute branching fraction of $\mathcal{B}_{cc}\to\mathcal{B}_{c}P$ been directly measured, but the branching fraction of $\Xi_{cc}^{++}\to \Xi_c^+\pi^+$ relative to $\Xi_{cc}^{++}\to \Lambda_c^+ K^-\pi^+\pi^+$ was measured to be \cite{Aaij:2018gfl}
\begin{equation}
\frac{\mathcal{B}(\Xi_{cc}^{++}\to\Xi_c^+\pi^+)\times \mathcal{B}(\Xi_c^+\to p K^-\pi^+)}
{\mathcal{B}(\Xi_{cc}^{++}\to\Lambda_c^+ K^- \pi^+\pi^+)\times
\mathcal{B}(\Lambda_c^+\to p K^- \pi^+)}=0.035\pm 0.009 ({\rm{stat.}})
\pm 0.003({\rm{syst.}}).\label{eq:Xicpi}
\end{equation}
As pointed out in Ref.~\cite{Zeng:2022egh}, using the measurements ${\cal B}(\Lambda_c^+\to p K^-\pi^+)=(6.24\pm0.28)\%$ and ${\cal B}(\Xi_c^+\to p K^-\pi^+)=(0.62\pm0.30)\%$~\cite{ParticleDataGroup:2024cfk},
and combining the theoretical predictions ${\cal B}(\Xi_{cc}^{++}\to\Lambda_c^+ K^- \pi^+\pi^+)\approx {2\over 3}{\cal B}(
\Xi_{cc}^{++}\to\Sigma_c^{++}\overline{K}^{*0})$~\cite{Cheng:2020wmk} and ${\cal B}(\Xi_{cc}^{++}\to\Sigma_c^{++}\overline{K}^{*0})=5.61\%$~\cite{Gutsche:2019iac}, 
the branching fraction of $\Xi_{cc}^{++}\to\Xi_c^+\pi^+$ can be obtained 
\begin{equation} \label{eq:BRexpt}
\mathcal{B}(\Xi_{cc}^{++}\to\Xi_c^+\pi^+)_{\rm expt}\approx (1.33\pm0.74)\%.
\end{equation}
Nevertheless, the calculated branching fraction {$\mathcal{B}(\Xi_{cc}^{++}\to\Xi_c^+\pi^+)_{\rm our}= (3.98_{-0.27-0.63}^{+0.30+0.65}) \%$ slightly larger than the expectation given by the above Eq.~\eqref{eq:BRexpt}}.

The decay asymmetry parameters, defined as ratios of the amplitudes $(\mathcal{S},~\mathcal{P})$, $(A,~B)$ or helicity amplitudes $H_{\pm1/2}$ in different form as given by Eqs.~(\eqref{eq:alpha}-\eqref{eq:gamma}), exhibit less sensitivity to model and input parameters. The sensitivity to the $\eta$ parameter is diminished for all three asymmetry parameters in comparison to branching ratios, regardless of whether the process is predominantly influenced by short-distance or long-distance contributions.

\begin{figure}[t]
  \centering {%
  \begin{minipage}[c]{0.45\linewidth}%
  \includegraphics[width=2.8in]{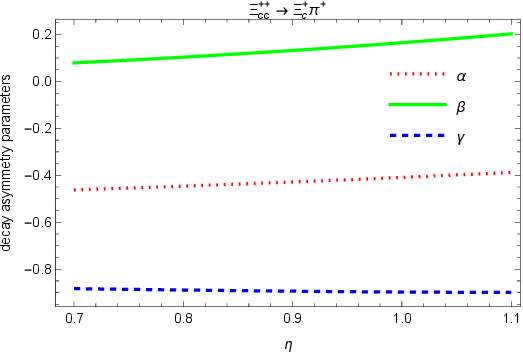} %
  \end{minipage}} \hspace{1cm} { %
  \begin{minipage}[c]{0.45\linewidth}%
  \includegraphics[width=2.8in]{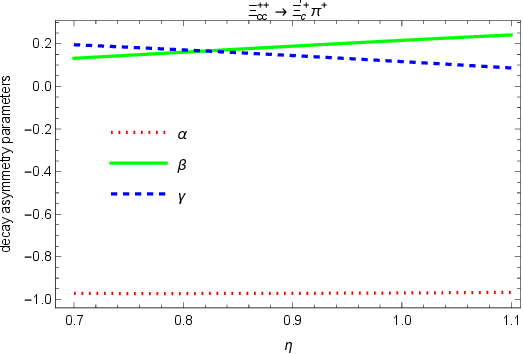} %
  \end{minipage}} \caption{{The dependence of decay asymmetry parameters on $\eta$ for decays (a): $\Xi_{cc}^{++}\to\Xi_{c}^{+}\pi^{+}$
  and (b):$\Xi_{cc}^{++}\to\Xi_{c}^{\prime+}\pi^{+}$ 
  with $\eta\sim[0.7,1.1]$.}}
  \label{fig:symmeteroneta} 
  \end{figure}

The dependence of the asymmetry parameters on the $\eta$ parameter is shown in Fig.\ref{fig:symmeteroneta}. We can see that these parameters have a small dependence on the parameter $\eta$. The theoretical predictions for the decay asymmetry parameter $\alpha$ for $\Xi_{cc}^{++}\to\Xi_{c}^{\prime+}\pi^{+}$ given in this paper are consistent with those in Refs.~\cite{Liu:2023dvg,Zeng:2022egh,Cheng:2020wmk,Gutsche:2018msz,Sharma:2017txj,Shi:2022kfa,Ke:2022gxm,Liu:2022igi}, which indicates that the calculations of relative magnitudes and phases of helicity amplitudes are reasonable.

\begin{table}[tb]
  \renewcommand{\arraystretch}{1.5}
  \addtolength{\arraycolsep}{5pt}
    \centering
    \caption{Comparison of the branching fractions, the decay asymmetry
    parameter $\alpha$ and the ratio of branching fractions $R_{{\cal B}^{\prime}/{\cal B}}$ in various approaches.  
    In this table, we list the experimental data from the LHCb collaboration~\cite{LHCb:2019qed}and the theoretical prediction from the pole
    approximation (PA) in tandem with the SU(3)F symmetry~\cite{Liu:2023dvg}, the non-relativistic quark model (NRQM)~\cite{Zeng:2022egh},
    the pole model (PM) with current algebra (CA)~\cite{Cheng:2020wmk}, the Ab initio three-loop calculation (AITLC)~\cite{Gutsche:2018msz}, 
    the heavy quark effective theory (HQET) with pole model (PM)~\cite{Sharma:2017txj}, light-cone sum rules (LCSR) with HQET~\cite{Shi:2022kfa}, the light front quark model (LFQM) with mixing mechanism~\cite{Ke:2022gxm}, the rescattering mechanism (RM) with Cutkosky cutting rule (CCR)~\cite{Han:2021azw} and the bag model~\cite{Liu:2022igi}. 
    }
    \label{Tab:comparison}
    \begin{tabular}{l|c|c|c|c|c}
      \hline\hline
      &${\cal B}(\Xi_{c}^{+}\pi^{+})[\%]$ & ${\cal B}(\Xi_{c}^{\prime+}\pi^{+})[\%]$&$\alpha(\Xi_{c}^{+}\pi^{+})$
        &$\alpha(\Xi_{c}^{\prime+}\pi^{+})$&$R_{{\cal B}^{\prime}/{\cal B}}$\\ \hline
        This work&$3.97_{-0.64}^{+0.63}$ & $5.38_{-0.42}^{+0.53}$& $-0.43_{-0.04}^{+0.03}$&$-0.97_{-0.00}^{+0.00}$ &$1.35_{-0.61}^{+0.81}$\\
PA+SU(3)~\cite{Liu:2023dvg}&$0.99\pm0.21$&$1.18\pm0.18$&$-0.19\pm0.07$&$-0.96\pm0.00$&$1.19\pm0.09$\\
        NRQM~\cite{Zeng:2022egh}&1.83 &2.86 &-0.78 &-0.89 &1.56\\
        PM+CA~\cite{Cheng:2020wmk}& 0.69&4.65 &-0.41 &-0.84 &6.74\\
        AITLC~\cite{Gutsche:2018msz}& $0.71\pm0.11$&$3.39\pm0.51$ &$-0.57\pm0.09$ &$-0.93\pm0.14$ &$4.33\pm0.65$\\
        HQET+PM~\cite{Sharma:2017txj}&9.30 &7.5 &-0.99 &-0.78 &-0.81\\
        LCSR+HQET~\cite{Shi:2022kfa}&$6.22\pm1.94$ &$8.85\pm0.62$ & $+0.99$&$-0.64$ &$1.42\pm0.78$\\
        LFQM+Mixing~\cite{Ke:2022gxm}&$2.14\pm0.18$ &$3.02\pm0.1$ &$-0.99\pm0.01$ &$-0.09\pm0.07$ &$1.41\pm0.20$\\
RM+CCR~\cite{Han:2021azw}&$8.48_{-1.37}^{+2.27}$&$4.72_{-0.00}^{+0.02}$&-&-&$0.56_{-0.12}^{+0.10}$\\
        Bag Model~\cite{Liu:2022igi}&$2.24$ & $3.25$&$-0.93$ &$-0.63$ &1.45\\\hline
        LHCb~\cite{LHCb:2019qed}&- & -&- &- &$1.41\pm0.17\pm0.10$\\
        \hline\hline
      \end{tabular}
  \end{table}

\subsection{$CP$ violations}
In this section, we provide our numerical calculation for direct charge parity ($CP$) violation in the decays of doubly charmed baryons. 
Using the parameter $\eta=0.9\pm0.2$, we give our predictions for $CP$ violations in singly CKM suppressed decays in Tabs.~\ref{Tab:CPV}-\ref{tab:CPVs}.

\begin{table}[tb]
\renewcommand{\arraystretch}{1.5}
\addtolength{\arraycolsep}{5pt}
	\centering
	\caption{{The $CP$ violation of the short-distance dynamics dominated modes. The first uncertainty comes from input parameters: decay constants and strong couplings, which have been listed in Tab.~\ref{table:decayconstants}. The second uncertainty is due to the variation of the phenomenological parameter $\eta=0.9\pm0.2$.}}\label{Tab:CPV}
  \begin{tabular}{l|c|c|c|c}
    \hline\hline
    channels & ${\cal A}_{CP}^{dir}[10^{-3}]$&$\alpha_{CP}[10^{-3}]$
    &$\beta_{CP}[10^{-3}]$&$\gamma_{CP}[10^{-3}]$\\ \hline
    $\Xi_{cc}^{++}\to\Sigma_{c}^{+}\pi^{+} $ &
    $0.13_{-0.02-0.07}^{+0.05+0.11}$ &
    $0.12_{-0.04-0.07}^{+0.08+0.15}$ &
    $-0.35_{-0.02-0.05}^{+0.04+0.06}$ &
    $-36.70_{-18.20-36.20}^{+18.40+37.10}$ \\
    $\Xi_{cc}^{++}\to\Lambda_{c}^{+}\pi^{+} $ &
    $-0.52_{-0.02-0.18}^{+0.07+0.22}$ &
    $-0.26_{0.03-0.14}^{+0.03+0.09}$ &
    $-0.53_{-0.03-0.02}^{+0.02+0.03}$ &
    $0.04_{-0.00-0.01}^{+0.00+0.00}$ \\
    $\Xi_{cc}^{++}\to\Xi_{c}^{\prime+}K^{+} $ &
    $-0.09_{-0.01-0.03}^{+0.01+0.03}$ &
    $-0.02_{-0.01-0.01}^{+0.00+0.01}$ &
    $-2.11_{-0.72-0.37}^{+0.35+0.26}$ &
    $0.44_{-0.06-0.15}^{+0.04+0.17}$ \\
    $\Xi_{cc}^{++}\to\Xi_{c}^{+}K^{+} $ &
    $0.10_{-0.15-0.13}^{+0.04+0.00}$ &
    $0.21_{-0.04-0.10}^{+0.04+0.16}$ &
    $-1.62_{-0.12-0.15}^{+0.20+0.18}$ &
    $-0.01_{-0.00-0.00}^{+0.01+0.01}$ \\
    $\Xi_{cc}^{+}\to\Sigma_{c}^{0}\pi^{+} $ &
    $0.06_{-0.02-0.03}^{+0.05+0.01}$ &
    $0.13_{-0.05-0.08}^{+0.17+0.23}$ &
    $-0.44_{-0.03-0.02}^{+0.04+0.16}$ &
    $-0.53_{-0.15-0.23}^{+0.12+0.21}$ \\
    $\Xi_{cc}^{+}\to\Xi_{c}^{\prime0}K^{+} $ &
    $0.10_{-0.02-0.04}^{+0.03+0.04}$ &
    $0.11_{-0.04-0.08}^{+0.14+0.22}$ &
    $-0.01_{-0.00-0.03}^{+0.01+0.05}$ &
    $-0.13_{-0.00-0.01}^{+0.01+0.02}$ \\
    $\Xi_{cc}^{+}\to\Xi_{c}^{0}K^{+} $ &
    $-0.04_{-0.01-0.01}^{+0.01+0.01}$ &
    $-0.06_{-0.01-0.03}^{+0.01+0.02}$ &
    $18.40_{-17.20-25.30}^{+6.08+14.50}$ &
    $0.03_{-0.01-0.01}^{+0.01+0.02}$ \\
    $\Omega_{cc}^{+}\to\Xi_{c}^{0}\pi^{+} $ &
    $-0.18_{-0.01-0.00}^{+0.08+0.05}$ &
    $-0.11_{-0.01-0.00}^{+0.01+0.02}$ &
    $-2.86_{-2.48-1.41}^{+0.55+0.67}$ &
    $0.02_{-0.01-0.00}^{+0.00+0.00}$ \\
    $\Omega_{cc}^{+}\to\Xi_{c}^{\prime0}\pi^{+} $ &
    $-0.22_{-0.08-0.26}^{+0.05+0.12}$ &
    $0.36_{-0.12-0.23}^{+0.48+0.89}$ &
    $-0.58_{-0.01-0.02}^{+0.04+0.05}$ &
    $2.43_{-2.84-8.87}^{+0.81+1.14}$ \\
    $\Omega_{cc}^{+}\to\Omega_{c}^{0}K^{+} $ &
    $-0.08_{-0.02-0.02}^{+0.02+0.02}$ &
    $-0.01_{0.00-0.00}^{+0.00+0.00}$ &
    $-1.69_{-0.22-0.33}^{+0.18+0.22}$ &
    $0.71_{-0.12-0.28}^{+0.13+0.37}$ \\
    \hline\hline
    \end{tabular} 
\end{table}

\begin{table}
  \caption{{CP violation of the nonleptonic weak decay of doubly charmed baryons  ${\cal B}_{cc}\to{\cal B}_{c}P$ which is the long-distance dominated singly Cabibbo-suppressed. The first uncertainty comes from input parameters: decay constants and strong couplings, which have been listed in Tab.~\ref{table:decayconstants}. The second uncertainty is due to the variation of the phenomenological parameter $\eta=0.9\pm0.2$.}}\label{tab:CPVs}
  \begin{tabular}{l|c|c|c|c}
    \hline\hline
    channels & ${\cal A}_{CP}^{dir}[10^{-4}]$&$\alpha_{CP}[10^{-4}]$
    &$\beta_{CP}[10^{-4}]$&$\gamma_{CP}[10^{-4}]$\\ \hline
    $\Xi_{cc}^{++}\to\Sigma_{c}^{++}\pi^{0} $ &
    $2.03_{-0.06-0.13}^{+0.03+0.20}$ &
    $-0.48_{-0.53-0.21}^{+0.06+0.30}$ &
    $-2.19_{-1.41-0.32}^{+0.25+0.16}$ &
    $1.93_{-0.13-0.30}^{+0.02+0.19}$ \\
    $\Xi_{cc}^{++}\to\Sigma_{c}^{++}\eta_{1} $ &
    $-32.30_{-0.92-1.31}^{+0.89+2.21}$ &
    $-20.30_{-5.44-25.90}^{+25.50+7.51}$ &
    $9.79_{-1.61-4.62}^{+1.14+2.53}$ &
    $-57.20_{-13.30-51.60}^{+28.10+9.86}$ \\
    $\Xi_{cc}^{++}\to\Sigma_{c}^{++}\eta_{8} $ &
    $2.31_{-0.26-0.04}^{+0.62+0.04}$ &
    $0.20_{-0.14-0.04}^{+0.20+0.02}$ &
    $6.60_{-0.34-0.40}^{+0.00+0.40}$ &
    $-0.42_{-0.34-0.02}^{+0.11+0.02}$ \\
    $\Xi_{cc}^{+}\to\Sigma_{c}^{+}\pi^{0} $ &
    $2.61_{-0.31-0.21}^{+0.40+0.14}$ &
    $-3.03_{-2.55-0.02}^{+0.71+0.16}$ &
    $29.20_{-23.50-9.20}^{+13.00+34.40}$ &
    $0.77_{-0.31-0.61}^{+0.92+0.88}$ \\
    $\Xi_{cc}^{+}\to\Sigma_{c}^{+}\eta_{1} $ &
    $-11.50_{-0.57-0.05}^{+0.60+0.07}$ &
    $-0.35_{-0.58-0.61}^{+13.80+0.08}$ &
    $69.90_{-22.00-14.40}^{+11.40+9.96}$ &
    $-22.70_{-8.97-64.30}^{+12.80+9.94}$ \\
    $\Xi_{cc}^{+}\to\Sigma_{c}^{+}\eta_{8} $ &
    $7.28_{-0.53-0.40}^{+0.26+0.43}$ &
    $-0.29_{-0.52-0.11}^{+1.78+0.04}$ &
    $4.57_{-0.85-0.12}^{+0.27+0.11}$ &
    $-0.90_{-1.20-0.05}^{+0.25+0.04}$ \\
    $\Xi_{cc}^{+}\to\Lambda_{c}^{+}\pi^{0} $ &
    $1.49_{-0.08-0.23}^{+0.02+0.34}$ &
    $-5.66_{-0.04-1.12}^{+0.28+0.69}$ &
    $5.28_{-0.33-0.47}^{+0.49+0.33}$ &
    $-2.15_{-0.42-0.33}^{+0.31+0.17}$ \\
    $\Xi_{cc}^{+}\to\Lambda_{c}^{+}\eta_{1} $ &
    $-6.57_{-0.25-0.49}^{+0.20+0.33}$ &
    $68.40_{-39.30-15.10}^{+31.30+39.90}$ &
    $2.36_{-0.52-3.52}^{+0.97+3.57}$ &
    $-0.75_{-1.10-0.42}^{+1.32+0.11}$ \\
    $\Xi_{cc}^{+}\to\Lambda_{c}^{+}\eta_{8} $ &
    $0.93_{-0.06-0.03}^{+0.03+0.01}$ &
    $-2.79_{-0.41-0.30}^{+0.44+0.24}$ &
    $2.87_{-0.36-0.27}^{+0.55+0.27}$ &
    $-51.40_{-22.70-4.47}^{+26.70+6.46}$ \\
    $\Xi_{cc}^{+}\to\Sigma_{c}^{++}\pi^{-} $ &
    $-0.62_{-0.35-0.02}^{+0.26+0.02}$ &
    $-1.15_{-0.16-0.01}^{+0.26+0.03}$ &
    $17.10_{-4.20-0.77}^{+7.15+1.45}$ &
    $0.72_{-0.24-0.05}^{+1.41+0.07}$ \\
    $\Xi_{cc}^{+}\to\Xi_{c}^{\prime+}K^{0} $ &
    $-2.27_{-0.62-0.09}^{+0.36+0.10}$ &
    $-0.34_{-0.09-0.05}^{+0.24+0.04}$ &
    $6.83_{-3.22-0.79}^{+4.62+0.87}$ &
    $-2.78_{-2.99-0.28}^{+4.79+0.25}$ \\
    $\Xi_{cc}^{+}\to\Xi_{c}^{+}K^{0} $ &
    $-8.58_{-0.58-0.67}^{+0.32+0.66}$ &
    $-13.20_{-1.46-2.81}^{+1.84+1.96}$ &
    $-2.98_{-0.72-0.48}^{+1.68+0.50}$ &
    $18.10_{-3.24-0.76}^{+0.68+0.66}$ \\
    $\Omega_{cc}^{+}\to\Sigma_{c}^{+}\bar{K}^{0} $ &
    $0.87_{-0.41-0.01}^{+0.37+0.01}$ &
    $3.48_{-1.57-0.32}^{+1.38+0.40}$ &
    $-0.01_{-0.02-0.06}^{+0.17+0.06}$ &
    $-1.96_{-0.02-0.09}^{+0.08+0.07}$ \\
    $\Omega_{cc}^{+}\to\Lambda_{c}^{+}\bar{K}^{0} $ &
    $0.48_{-0.06-0.02}^{+0.02+0.02}$ &
    $0.67_{-8.40-0.04}^{+0.73+0.06}$ &
    $-0.49_{-0.81-0.06}^{+0.89+0.06}$ &
    $0.32_{-0.41-0.09}^{+0.71+0.11}$ \\
    $\Omega_{cc}^{+}\to\Sigma_{c}^{++}K^{-} $ &
    $-0.34_{-0.10-0.01}^{+0.12+0.01}$ &
    $-3.78_{-1.43-0.48}^{+6.13+0.35}$ &
    $1.14_{-1.10-0.18}^{+2.65+0.14}$ &
    $0.36_{-0.39-0.03}^{+0.60+0.04}$ \\
    $\Omega_{cc}^{+}\to\Xi_{c}^{\prime+}\pi^{0} $ &
    $6.79_{-0.30-0.26}^{+0.37+0.05}$ &
    $-1.58_{-0.03-0.45}^{+0.19+0.21}$ &
    $2.35_{-0.12-0.36}^{+0.10+0.33}$ &
    $-3.57_{-1.08-0.46}^{+0.54+0.29}$ \\
    $\Omega_{cc}^{+}\to\Xi_{c}^{\prime+}\eta_{1} $ &
    $-3.66_{-0.60-0.22}^{+0.45+0.20}$ &
    $-0.16_{-0.98-0.03}^{+0.01+0.04}$ &
    $3.47_{-1.84-0.53}^{+1.81+0.50}$ &
    $-4.24_{-2.23-1.02}^{+2.32+0.72}$ \\
    $\Omega_{cc}^{+}\to\Xi_{c}^{\prime+}\eta_{8} $ &
    $-0.19_{-0.08-0.02}^{+0.03+0.02}$ &
    $-0.37_{-0.03-0.02}^{+0.04+0.02}$ &
    $5.32_{-0.19-0.61}^{+0.21+0.71}$ &
    $0.13_{-0.02-0.00}^{+0.04+0.00}$ \\
    $\Omega_{cc}^{+}\to\Xi_{c}^{+}\pi^{0} $ &
    $5.32_{-0.58-0.17}^{+1.00+0.01}$ &
    $-2.03_{-0.08-0.26}^{+0.14+0.17}$ &
    $2.34_{-0.39-0.42}^{+0.41+0.62}$ &
    $-0.64_{-0.21-0.02}^{+0.12+0.04}$ \\
    $\Omega_{cc}^{+}\to\Xi_{c}^{+}\eta_{1} $ &
    $-1.49_{-0.00-0.09}^{+0.00+0.08}$ &
    $-5.81_{-2.39-0.37}^{+2.76+0.28}$ &
    $-1.07_{-0.86-0.30}^{+0.24+0.27}$ &
    $0.03_{-0.00-0.01}^{+0.17+0.01}$ \\
    $\Omega_{cc}^{+}\to\Xi_{c}^{+}\eta_{8} $ &
    $-0.33_{-0.19-0.00}^{+0.10+0.00}$ &
    $-0.25_{-0.02-0.01}^{+0.03+0.01}$ &
    $-2.95_{-0.89-0.40}^{+2.09+0.35}$ &
    $0.68_{-0.11-0.02}^{+0.11+0.02}$ \\
    \hline\hline
    \end{tabular}
\end{table}

In this study, we exclusively examine the $CP$ violation in singly Cabibbo suppressed processes, which encompasses two CKM matrix elements: $V_{cd} V_{ud}^*$ and $V_{cs} V_{us}^*$. Within the rescattering mechanism, these elements contribute to an identical process, resulting in a non-zero weak phase difference. Furthermore, long-distance final state interaction effect offer significant phase information for the decay process. Consequently, our theoretical framework is capable of directly determining the magnitude of $CP$ violation and currently serves as the sole method to measure $CP$ violation in charmed baryon decays.
The decay amplitude, denoted by \(\mathcal{A}=\lambda_d\mathcal{A}_d+\lambda_s\mathcal{A}_s\), where \(\lambda_q=V_{cq}V_{uq}^*\) and \(q=d,s\), aids in the expression of \(CP\) violation,
\begin{align}
A_{CP}^{\rm dir}\approx -2\frac{{\rm Im}(\lambda_d^*\lambda_s)}{|\lambda_d|^2} \frac{{\rm Im}(\mathcal{A}_d^*\mathcal{A}_s)}{|\mathcal{A}_d-\mathcal{A}_s|^2}.
\end{align} 

\begin{figure}[t]
  \centering {%
  \begin{minipage}[c]{0.45\linewidth}%
  \includegraphics[width=2.8in]{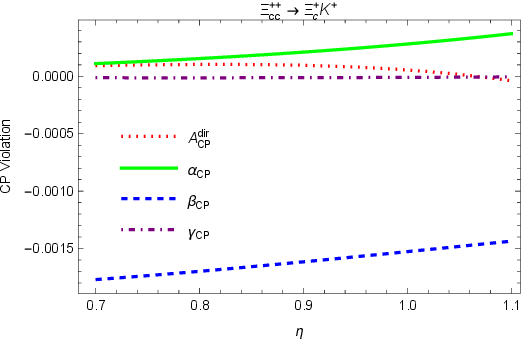} %
  \end{minipage}} \hspace{0.8cm} { %
  \begin{minipage}[c]{0.45\linewidth}%
  \includegraphics[width=2.8in]{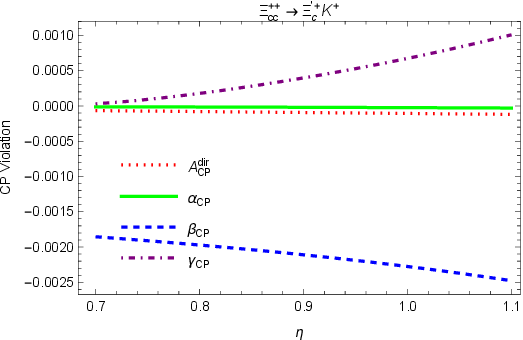} %
  \end{minipage}} 
  \caption{{The dependence of the CP violation on $\eta$  for decays (a): $\Xi_{cc}^{++}\to\Xi_{c}^{+}K^{+}$
  and (b): $\Xi_{cc}^{++}\to\Xi_{c}^{\prime+}K^{+}$ with $\eta\sim[0.7,1.1]$.}} \label{fig:CPV56} 
  \end{figure}

From Eq. \eqref{eq:direct_CPV}, we can see that the direct $CP$ violation is proportional to $\sin \Delta \phi $, where $\Delta \phi $ is the weak phase difference for tree-level operators in charmed baryon decays, which is about $6\times10^{-4}$ ($\Delta \phi =\arctan [{\text {Im}(V_{cd})}/{\text {Re}(V_{cd})}]$). Therefore, the expected size of $CP$ violation in doubly charmed baryon decays should be at this level, which is consistent with the theoretical expectations in Table \ref{Tab:CPV}.
For example, the decay channels $\Xi_{cc}^{++} \to \Xi_{c}^{+}K^+$ and $\Xi_{cc}^{++} \to \Xi_{c}^{\prime+}K^+$ are characterized by $CP$ violation induced by tree-level operators mediated by CKM matrix elements $V_{cd}V_{ud}$ and $V_{cs} V_{us}$. 
These elements contribute to the amplitudes in the triangle diagram, as depicted in Fig.~\ref{fig:trianglesev}. This influences the decay amplitude and induces a weak phase difference, denoted as $\Delta\phi$. The understanding and prediction of $CP$ violation in charmed baryon decays heavily relies on this weak phase difference, in conjunction with the strong phases derived from the triangle integrals.
To validate the reliability of the theoretical method for predicting the magnitude of $CP$ violation, it is crucial to investigate the relationship between $CP$ violation parameters and model parameters, such as $\eta$.
For the decays $\Xi_{cc}^{++} \to \Xi_{c}^{+}K^+$ and $\Xi_{cc}^{++} \to \Xi_{c}^{\prime+}K^+$, the relationship between their $CP$ violation parameters and the $\eta$ parameter is depicted in Fig.~\ref{fig:CPV56}.
This research offers a quantitative evaluation of the susceptibility of $CP$ violation effects to alterations in the model parameter $\eta$.
The study indicates that both the direct $CP$ violation and the asymmetry parameter-induced $CP$ violation exhibit minimal sensitivity to model parameters.
The results show that the predictions for $CP$ violation observables are independent of theoretical input parameters or model assumptions. This is important to check the robustness of the theoretical framework and to make sure that the predicted size of $CP$ violation indeed corresponds to the physical effects driving these decays.

\section{Summary}

\label{sec:summary}

In this work, we introduced the whole theoretical framework of the
rescattering mechanism by investigating the two-body nonleptonic weak decays $\mathcal{B}_{cc}\to\mathcal{B}_{c}P$, where $\mathcal{B}_{cc}=(\Xi_{cc}^{++},\Xi_{cc}^{+},\Omega_{cc}^{+})$
are the doubly charmed baryons, $\mathcal{B}_{c}=(\mathcal{B}_{\bar{3}},\mathcal{B}_{6})$
are the singly charmed baryons and $P=(\pi,K,\eta_{1,8})$ are the
light pseudoscalar mesons. It was interpreted in detail for the physical
foundation of the rescattering mechanism at the hadron level. On the
other hand, as a self-consistent test to the rescattering mechanism,
the relations of topological diagrams and flavor $SU(3)$ symmetry
have been discussed. The main points are the following: 
\begin{enumerate}
\item[(1)] We present a topological
analysis of the two-body nonleptonic weak decays $\mathcal{B}_{cc}\to\mathcal{B}_{c}P$ and give the amplitudes of each decay mode within the rescattering mechanism at the hadron level.
\item[(2)]
Using the experiment data~\cite{LHCb:2019qed}
\begin{equation}
{\cal RB}=\frac{{\cal B}(\Xi_{cc}^{++}\to\Xi_{c}^{\prime+}\pi^{+})}{{\cal B}(\Xi_{cc}^{++}\to\Xi_{c}^{+}\pi^{+})}=1.41\pm0.17\pm0.10,
\end{equation}
this parameter can be determined as $\eta=0.9\pm0.2$.
\item[(3)]
After the calculation of the amplitude of each decay mode, we conduct a comparative analysis with other theoretical works. In contrast to other studies, our work include both real and imaginary contributions of the amplitudes. {
  As illustrated in Tab.~\ref{tab:ampSD}, the uncertainty from decay constants is much smaller than that from the strong coupling constants and $\eta$. This is because almost of these decay constants have been precisely determined using experimental data~\cite{ParticleDataGroup:2024cfk}. Nevertheless, there is a significant lack of experimental data available for determining the parameters: strong couplings and $\eta$.}
\item[(4)] 
Also, we obtain the same counting rules as the analysis in SCET for the
topological amplitudes in charm decays, that is $\frac{|C|}{|T|}\sim\frac{|C^{\prime}|}{|C|}\sim\frac{|E_{1}|}{|C|}\sim\frac{|E_{2}|}{|C|}\sim O(\frac{\Lambda_{\text{QCD}}^{h}}{m_{c}})\sim O(1)$,
which will be significant guidance for the following studies to charmed
baryon decays. After the calculation of the decay width of each decay mode, we find that large $SU(3)$ symmetry breaking effects exist in our work.
It requires more studies on the $SU(3)$ breaking effects of the doubly
charmed baryon decays in the future. 
\item[(5)]With $\eta=0.9\pm0.2$, we theoretically predict
the branching ratios, decay asymmetry parameters and CP violation of all considered $\mathcal{B}_{cc}\to\mathcal{B}_{c}P$
and some discussions on the dependence of the parameter $\eta$. As depicted
in Fig.\ref{fig:fig4}(a), the branching fraction {$\mathcal{B}(\Xi_{cc}^{++}\to\Xi_c^+\pi^+)_{\rm our}= (3.98_{-0.27-0.63}^{+0.30+0.65}) \%$} is slightly larger than the expectation in Ref.~\cite{Zeng:2022egh}. The numerical results of the branching ratios exhibited the same
conclusion with the charm meson decays: the non-factorizable long-distance
contributions play an important role in doubly charmed baryon decays. 
\item[(6)] {From Tabs.~\ref{tab:result1}-\ref{tab:result4}, we can found that the decay asymmetry parameters are less affected by the two parameters: strong coupling constants and $\eta$,compared with the branching ratios. Therefore, the decay asymmetry parameter has a weaker dependence on the input parameter, compared to the branching ratios.}
\item[(7)] The study indicates that both the direct $CP$ violation and the asymmetry parameter-induced $CP$ violation exhibit minimal sensitivity to model parameters.

\end{enumerate}
\begin{acknowledgments}
The authors are very grateful to Jia-Jie Han and Hua-Yu Jiang for useful discussions.
This work is supported in part by the National Natural Science Foundation of China under Grant No. 12005294 and No. 12335003, 
the youth Foundation of China University of mining and technology No. JN210003, 
the opening Issues Foundation of Lanzhou Center for Theoretical Physics No.12247101, 
and the Fundamental Research Funds for the Central Universities under No. lzujbky-2024-oy02.
\end{acknowledgments}

\textbf{Data Availability Statement:}
No Data associated in the manuscript.

\appendix

\section{Amplitudes of each mode}

\label{app:amp}

The expressions of amplitudes for all the forty-seven $\mathcal{B}_{cc}\to\mathcal{B}_{c}P$
decays considered in this paper are collected in this Appendix. Twenty
amplitudes for the short-distance dominated modes are given as follows.
{\scriptsize{}{}{}{ 
\begin{align}
\mathcal{A}(\Xi_{cc}^{++}\to\Xi_{c}^{+}\pi^{+}) & =\mathcal{T}_{SD}(\Xi_{cc}^{++}\to\Xi_{c}^{+}\pi^{+})+\mathcal{C}_{SD}(\Xi_{cc}^{++}\to\Xi_{c}^{+}\pi^{+})+\mathcal{M}(\pi^{+},\Xi_{c}^{(\prime)+};\rho^{0})+\mathcal{M}(\rho^{+},\Xi_{c}^{(\prime)+};\pi^{0})+\mathcal{M}(\rho^{+},\Xi_{c}^{(\prime)+};\rho^{0})\nonumber \\
 & +\mathcal{M}(\Xi_{c}^{(\prime)+},\pi^{+};\Xi_{c}^{(\prime)0})+\mathcal{M}(\Xi_{c}^{(\prime)+},\rho^{+};\Xi_{c}^{(\prime)0})+\mathcal{M}(\Xi_{c}^{(\prime)+},\pi^{+};\Xi_{c}^{\prime*0})+\mathcal{M}(\Xi_{c}^{(\prime)+},\rho^{+};\Xi_{c}^{\prime*0}),\\
\mathcal{A}(\Xi_{cc}^{++}\to\Xi_{c}^{\prime+}\pi^{+}) & =\mathcal{T}_{SD}(\Xi_{cc}^{++}\to\Xi_{c}^{\prime+}\pi^{+})+\mathcal{C}_{SD}(\Xi_{cc}^{++}\to\Xi_{c}^{\prime+}\pi^{+})+\mathcal{M}(\pi^{+},\Xi_{c}^{(\prime)+};\rho^{0})+\mathcal{M}(\rho^{+},\Xi_{c}^{(\prime)+};\pi^{0})+\mathcal{M}(\rho^{+},\Xi_{c}^{(\prime)+};\rho^{0})\nonumber \\
 & +\mathcal{M}(\Xi_{c}^{(\prime)+},\pi^{+};\Xi_{c}^{(\prime)0})+\mathcal{M}(\Xi_{c}^{(\prime)+},\rho^{+};\Xi_{c}^{(\prime)0})+\mathcal{M}(\Xi_{c}^{(\prime)+},\pi^{+};\Xi_{c}^{\prime*0})+\mathcal{M}(\Xi_{c}^{(\prime)+},\rho^{+};\Xi_{c}^{\prime*0}),\\
\mathcal{A}(\Xi_{cc}^{++}\to\Sigma_{c}^{+}\pi^{+}) & =\mathcal{T}_{SD}(\Xi_{cc}^{++}\to\Sigma_{c}^{+}\pi^{+})+\mathcal{C}_{SD}(\Xi_{cc}^{++}\to\Sigma_{c}^{+}\pi^{+})+\mathcal{M}(\pi^{+},\Lambda_{c}^{+};\rho^{0})+\mathcal{M}(\rho^{+},\Lambda_{c}^{+};\pi^{0})+\mathcal{M}(K^{+},\Xi_{c}^{(\prime)+};K^{\ast0})\nonumber \\
 & +\mathcal{M}(K^{\ast+},\Xi_{c}^{(\prime)+};K^{0})+\mathcal{M}(\rho^{+},\Xi_{c}^{(\prime)+};\rho^{0})+\mathcal{M}(K^{\ast+},\Xi_{c}^{(\prime)+};K^{\ast0})\nonumber \\
 & +\mathcal{M}(\Lambda_{c}^{+}/\Sigma_{c}^{+},\pi^{+};\Sigma_{c}^{0})+\mathcal{M}(\Lambda_{c}^{+}/\Sigma_{c}^{+},\rho^{+};\Sigma_{c}^{0})+\mathcal{M}(\Xi_{c}^{(\prime)+},K^{+};\Xi_{c}^{(\prime)0})+\mathcal{M}(\Xi_{c}^{(\prime)+},K^{\ast+};\Xi_{c}^{(\prime)0})\nonumber \\
 & +\mathcal{M}(\Lambda_{c}^{+}/\Sigma_{c}^{+},\pi^{+};\Sigma_{c}^{*0})+\mathcal{M}(\Lambda_{c}^{+}/\Sigma_{c}^{+},\rho^{+};\Sigma_{c}^{*0})+\mathcal{M}(\Xi_{c}^{(\prime)+},K^{+};\Xi_{c}^{\prime*0})+\mathcal{M}(\Xi_{c}^{(\prime)+},K^{\ast+};\Xi_{c}^{\prime*0}),\\
\mathcal{A}(\Xi_{cc}^{++}\to\Lambda_{c}^{+}\pi^{+}) & =\mathcal{T}_{SD}(\Xi_{cc}^{++}\to\Lambda_{c}^{+}\pi^{+})+\mathcal{C}_{SD}(\Xi_{cc}^{++}\to\Lambda_{c}^{+}\pi^{+})+\mathcal{M}(\pi^{+},\Sigma_{c}^{+};\rho^{0})+\mathcal{M}(\rho^{+},\Sigma_{c}^{+};\pi^{0})+\mathcal{M}(K^{+},\Xi_{c}^{(\prime)+};K^{\ast0})\nonumber \\
 & +\mathcal{M}(K^{\ast+},\Xi_{c}^{(\prime)+};K^{0})+\mathcal{M}(\rho^{+},\Xi_{c}^{(\prime)+};\rho^{0})+\mathcal{M}(K^{\ast+},\Xi_{c}^{(\prime)+};K^{\ast0})\nonumber \\
 & +\mathcal{M}(\Lambda_{c}^{+}/\Sigma_{c}^{+},\pi^{+};\Sigma_{c}^{0})+\mathcal{M}(\Lambda_{c}^{+}/\Sigma_{c}^{+},\rho^{+};\Sigma_{c}^{0})+\mathcal{M}(\Xi_{c}^{(\prime)+},K^{+};\Xi_{c}^{(\prime)0})+\mathcal{M}(\Xi_{c}^{(\prime)+},K^{\ast+};\Xi_{c}^{(\prime)0})\nonumber \\
 & +\mathcal{M}(\Lambda_{c}^{+}/\Sigma_{c}^{+},\pi^{+};\Sigma_{c}^{*0})+\mathcal{M}(\Lambda_{c}^{+}/\Sigma_{c}^{+},\rho^{+};\Sigma_{c}^{*0})+\mathcal{M}(\Xi_{c}^{(\prime)+},K^{+};\Xi_{c}^{\prime*0})+\mathcal{M}(\Xi_{c}^{(\prime)+},K^{\ast+};\Xi_{c}^{\prime*0}),
\\
\mathcal{A}(\Xi_{cc}^{++}\to\Xi_{c}^{+}K^{+}) & =\mathcal{T}_{SD}(\Xi_{cc}^{++}\to\Xi_{c}^{+}K^{+})+\mathcal{C}_{SD}(\Xi_{cc}^{++}\to\Xi_{c}^{+}K^{+})+\mathcal{M}(K^{+},\Xi_{c}^{(\prime)+};\rho^{0})+\mathcal{M}(K^{+},\Xi_{c}^{(\prime)+};\omega)\nonumber \\
 &+\mathcal{M}(K^{\ast+},\Xi_{c}^{(\prime)+};\pi^{0})
 +\mathcal{M}(K^{\ast+},\Xi_{c}^{(\prime)+};\eta_{8})+\mathcal{M}(\pi^{+},\Lambda_{c}^{+};\overline{K}^{\ast0})+\mathcal{M}(\pi^{+},\Sigma_{c}^{+};\overline{K}^{\ast0})+\mathcal{M}(\rho^{+},\Lambda_{c}^{+};\overline{K}^{0})\nonumber \\
 & +\mathcal{M}(\rho^{+},\Sigma_{c}^{+};\overline{K}^{0})+\mathcal{M}(K^{+},\Xi_{c}^{(\prime)+};\phi)+\mathcal{M}(K^{\ast+},\Xi_{c}^{(\prime)+};\rho^{0})+\mathcal{M}(K^{\ast+},\Xi_{c}^{(\prime)+};\omega)\nonumber \\
 & +\mathcal{M}(\rho^{+},\Lambda_{c}^{+};\overline{K}^{\ast0})+\mathcal{M}(\rho^{+},\Sigma_{c}^{+};\overline{K}^{\ast0})+\mathcal{M}(K^{\ast+},\Xi_{c}^{(\prime)+};\phi)\nonumber \\
 & +\mathcal{M}(\Xi_{c}^{(\prime)+},K^{+};\Omega_{c}^{0})+\mathcal{M}(\Xi_{c}^{(\prime)+},K^{\ast+};\Omega_{c}^{0})+\mathcal{M}(\Lambda_{c}^{+}/\Sigma_{c}^{+},\pi^{+};\Xi_{c}^{(\prime)0})+\mathcal{M}(\Lambda_{c}^{+}/\Sigma_{c}^{+},\rho^{+};\Xi_{c}^{(\prime)0})\nonumber \\
 & +\mathcal{M}(\Xi_{c}^{(\prime)+},K^{+};\Omega_{c}^{*0})+\mathcal{M}(\Xi_{c}^{(\prime)+},K^{\ast+};\Omega_{c}^{*0})+\mathcal{M}(\Lambda_{c}^{+}/\Sigma_{c}^{+},\pi^{+};\Xi_{c}^{\prime*0})+\mathcal{M}(\Lambda_{c}^{+}/\Sigma_{c}^{+},\rho^{+};\Xi_{c}^{\prime*0}),\\
\mathcal{A}(\Xi_{cc}^{++}\to\Xi_{c}^{\prime+}K^{+}) & =\mathcal{T}_{SD}(\Xi_{cc}^{++}\to\Xi_{c}^{\prime+}K^{+})+\mathcal{C}_{SD}(\Xi_{cc}^{++}\to\Xi_{c}^{\prime+}K^{+})+\mathcal{M}(K^{+},\Xi_{c}^{(\prime)+};\rho^{0})+\mathcal{M}(K^{+},\Xi_{c}^{(\prime)+};\omega)\nonumber \\
 & +\mathcal{M}(K^{\ast+},\Xi_{c}^{(\prime)+};\pi^{0})
 +\mathcal{M}(K^{\ast+},\Xi_{c}^{(\prime)+};\eta_{8})+\mathcal{M}(\pi^{+},\Lambda_{c}^{+};\overline{K}^{\ast0})+\mathcal{M}(\pi^{+},\Sigma_{c}^{+};\overline{K}^{\ast0})+\mathcal{M}(\rho^{+},\Lambda_{c}^{+};\overline{K}^{0})\nonumber \\
 & +\mathcal{M}(\rho^{+},\Sigma_{c}^{+};\overline{K}^{0})+\mathcal{M}(K^{+},\Xi_{c}^{(\prime)+};\phi)+\mathcal{M}(K^{\ast+},\Xi_{c}^{(\prime)+};\rho^{0})+\mathcal{M}(K^{\ast+},\Xi_{c}^{(\prime)+};\omega)\nonumber \\
 & +\mathcal{M}(\rho^{+},\Lambda_{c}^{+};\overline{K}^{\ast0})+\mathcal{M}(\rho^{+},\Sigma_{c}^{+};\overline{K}^{\ast0})+\mathcal{M}(K^{\ast+},\Xi_{c}^{(\prime)+};\phi)\nonumber \\
 & +\mathcal{M}(\Xi_{c}^{(\prime)+},K^{+};\Omega_{c}^{0})+\mathcal{M}(\Xi_{c}^{(\prime)+},K^{\ast+};\Omega_{c}^{0})+\mathcal{M}(\Lambda_{c}^{+}/\Sigma_{c}^{+},\pi^{+};\Xi_{c}^{(\prime)0})+\mathcal{M}(\Lambda_{c}^{+}/\Sigma_{c}^{+},\rho^{+};\Xi_{c}^{(\prime)0})\nonumber \\
 & +\mathcal{M}(\Xi_{c}^{(\prime)+},K^{+};\Omega_{c}^{*0})+\mathcal{M}(\Xi_{c}^{(\prime)+},K^{\ast+};\Omega_{c}^{*0})+\mathcal{M}(\Lambda_{c}^{+}/\Sigma_{c}^{+},\pi^{+};\Xi_{c}^{\prime*0})+\mathcal{M}(\Lambda_{c}^{+}/\Sigma_{c}^{+},\rho^{+};\Xi_{c}^{\prime*0}),\\
\mathcal{A}(\Xi_{cc}^{++}\to\Lambda_{c}^{+}K^{+}) & =\mathcal{T}_{SD}(\Xi_{cc}^{++}\to\Lambda_{c}^{+}K^{+})+\mathcal{C}_{SD}(\Xi_{cc}^{++}\to\Lambda_{c}^{+}K^{+})+\mathcal{M}(K^{+},\Lambda_{c}^{+};\omega)+\mathcal{M}(K^{+},\Sigma_{c}^{+};\rho^{0})\nonumber \\
 & +\mathcal{M}(K^{\ast+},\Lambda_{c}^{+};\eta_{8})
 +\mathcal{M}(K^{\ast+},\Sigma_{c}^{+};\pi^{0})+\mathcal{M}(K^{\ast+},\Lambda_{c}^{+};\omega)+\mathcal{M}(K^{\ast+},\Sigma_{c}^{+};\rho^{0})\nonumber \\
 & +\mathcal{M}(\Lambda_{c}^{+}/\Sigma_{c}^{+},K^{+};\Xi_{c}^{(\prime)0})+\mathcal{M}(\Lambda_{c}^{+}/\Sigma_{c}^{+},K^{\ast+};\Xi_{c}^{(\prime)0})\nonumber \\
 & +\mathcal{M}(\Lambda_{c}^{+}/\Sigma_{c}^{+},K^{+};\Xi_{c}^{\prime*0})+\mathcal{M}(\Lambda_{c}^{+}/\Sigma_{c}^{+},K^{\ast+};\Xi_{c}^{\prime*0}),\\
\mathcal{A}(\Xi_{cc}^{++}\to\Sigma_{c}^{+}K^{+}) & =\mathcal{T}_{SD}(\Xi_{cc}^{++}\to\Sigma_{c}^{+}K^{+})+\mathcal{C}_{SD}(\Xi_{cc}^{++}\to\Sigma_{c}^{+}K^{+})+\mathcal{M}(K^{+},\Lambda_{c}^{+};\rho^{0})+\mathcal{M}(K^{+},\Sigma_{c}^{+};\omega)\nonumber \\
 &+\mathcal{M}(K^{\ast+},\Lambda_{c}^{+};\pi^{0})
 +\mathcal{M}(K^{\ast+},\Sigma_{c}^{+};\eta_{8})+\mathcal{M}(K^{\ast+},\Lambda_{c}^{+};\rho^{0})+\mathcal{M}(K^{\ast+},\Sigma_{c}^{+};\omega)\nonumber \\
 & +\mathcal{M}(\Lambda_{c}^{+}/\Sigma_{c}^{+},K^{+};\Xi_{c}^{(\prime)0})+\mathcal{M}(\Lambda_{c}^{+}/\Sigma_{c}^{+},K^{\ast+};\Xi_{c}^{(\prime)0})\nonumber \\
 & +\mathcal{M}(\Lambda_{c}^{+}/\Sigma_{c}^{+},K^{+};\Xi_{c}^{\prime*0})+\mathcal{M}(\Lambda_{c}^{+}/\Sigma_{c}^{+},K^{\ast+};\Xi_{c}^{\prime*0}),
\\
\mathcal{A}(\Xi_{cc}^{+}\to\Xi_{c}^{0}\pi^{+}) & =\mathcal{T}_{SD}(\Xi_{cc}^{+}\to\Xi_{c}^{0}\pi^{+})+\mathcal{M}(\pi^{+},\Xi_{c}^{(\prime)0};\rho^{0})+\mathcal{M}(\rho^{+},\Xi_{c}^{(\prime)0};\pi^{0})+\mathcal{M}(\rho^{+},\Xi_{c}^{(\prime)0};\rho^{0}),\\
\mathcal{A}(\Xi_{cc}^{+}\to\Xi_{c}^{\prime0}\pi^{+}) & =\mathcal{T}_{SD}(\Xi_{cc}^{+}\to\Xi_{c}^{(\prime)0}\pi^{+})+\mathcal{M}(\pi^{+},\Xi_{c}^{(\prime)0};\rho^{0})+\mathcal{M}(\rho^{+},\Xi_{c}^{(\prime)0};\pi^{0})+\mathcal{M}(\rho^{+},\Xi_{c}^{(\prime)0};\rho^{0}),\\
\mathcal{A}(\Xi_{cc}^{+}\to\Xi_{c}^{0}K^{+}) & =\mathcal{T}_{SD}(\Xi_{cc}^{+}\to\Xi_{c}^{0}K^{+})+\mathcal{M}(\pi^{+},\Sigma_{c}^{0};\overline{K}^{\ast0})+\mathcal{M}(\rho^{+},\Sigma_{c}^{0};\overline{K}^{0})+\mathcal{M}(K^{+},\Xi_{c}^{(\prime)0};\phi)\nonumber \\
 & +\mathcal{M}(K^{\ast+},\Xi_{c}^{(\prime)0};\eta_{8})+\mathcal{M}(\rho^{+},\Sigma_{c}^{0};\overline{K}^{\ast0})+\mathcal{M}(K^{\ast+},\Xi_{c}^{(\prime)0};\phi),\\
\mathcal{A}(\Xi_{cc}^{+}\to\Xi_{c}^{\prime0}K^{+}) & =\mathcal{T}_{SD}(\Xi_{cc}^{+}\to\Xi_{c}^{\prime0}K^{+})+\mathcal{M}(\pi^{+},\Sigma_{c}^{0};\overline{K}^{\ast0})+\mathcal{M}(\rho^{+},\Sigma_{c}^{0};\overline{K}^{0})+\mathcal{M}(K^{+},\Xi_{c}^{(\prime)0};\phi)\nonumber\\
 & +\mathcal{M}(K^{\ast+},\Xi_{c}^{(\prime)0};\eta_{8})+\mathcal{M}(\rho^{+},\Sigma_{c}^{0};\overline{K}^{\ast0})+\mathcal{M}(K^{\ast+},\Xi_{c}^{(\prime)0};\phi),\\
\mathcal{A}(\Xi_{cc}^{+}\to\Sigma_{c}^{0}K^{+}) & =\mathcal{T}_{SD}(\Xi_{cc}^{+}\to\Sigma_{c}^{0}K^{+}),\\
\mathcal{A}(\Xi_{cc}^{+}\to\Sigma_{c}^{0}\pi^{+}) & =\mathcal{T}_{SD}(\Xi_{cc}^{+}\to\Sigma_{c}^{0}\pi^{+})+\mathcal{M}(\pi^{+},\Sigma_{c}^{0};\rho^{0})+\mathcal{M}(\rho^{+},\Sigma_{c}^{0};\pi^{0})+\mathcal{M}(K^{+},\Xi_{c}^{(\prime)0};K^{\ast0})\nonumber \\
 & +\mathcal{M}(K^{\ast+},\Xi_{c}^{(\prime)0};K^{0})+\mathcal{M}(\rho^{+},\Sigma_{c}^{0};\rho^{0})+\mathcal{M}(K^{\ast+},\Xi_{c}^{(\prime)0};K^{\ast0}),
\end{align}} 
\begin{align}
\mathcal{A}(\Omega_{cc}^{+}\to\Omega_{c}^{0}\pi^{+}) & =\mathcal{T}_{SD}(\Omega_{cc}^{+}\to\Omega_{c}^{0}\pi^{+}),\\
\mathcal{A}(\Omega_{cc}^{+}\to\Xi_{c}^{0}\pi^{+}) & =\mathcal{T}_{SD}(\Omega_{cc}^{+}\to\Xi_{c}^{0}\pi^{+})+\mathcal{M}(\pi^{+},\Xi_{c}^{(\prime)0};\rho^{0})+\mathcal{M}(\rho^{+},\Xi_{c}^{(\prime)0};\pi^{0})+\mathcal{M}(K^{+},\Omega_{c}^{0};K^{\ast0})\nonumber \\
 & +\mathcal{M}(K^{\ast+},\Omega_{c}^{0};K^{0})+\mathcal{M}(\rho^{+},\Xi_{c}^{(\prime)0};\rho^{0})+\mathcal{M}(K^{\ast+},\Omega_{c}^{0};K^{\ast0}),\\
\mathcal{A}(\Omega_{cc}^{+}\to\Xi_{c}^{\prime0}\pi^{+}) & =\mathcal{T}_{SD}(\Omega_{cc}^{+}\to\Xi_{c}^{\prime0}\pi^{+})+\mathcal{M}(\pi^{+},\Xi_{c}^{(\prime)0};\rho^{0})+\mathcal{M}(\rho^{+},\Xi_{c}^{(\prime)0};\pi^{0})+\mathcal{M}(K^{+},\Omega_{c}^{0};K^{\ast0})\nonumber \\
 & +\mathcal{M}(K^{\ast+},\Omega_{c}^{0};K^{0})+\mathcal{M}(\rho^{+},\Xi_{c}^{(\prime)0};\rho^{0})+\mathcal{M}(K^{\ast+},\Omega_{c}^{0};K^{\ast0}),\\
\mathcal{A}(\Omega_{cc}^{+}\to\Omega_{c}^{0}K^{+}) & =\mathcal{T}_{SD}(\Omega_{cc}^{+}\to\Omega_{c}^{0}K^{+})+\mathcal{M}(\pi^{+},\Xi_{c}^{(\prime)0};\overline{K}^{\ast0})+\mathcal{M}(\rho^{+},\Xi_{c}^{(\prime)0};\overline{K}^{0})\nonumber \\
 & +\mathcal{M}(K^{+},\Omega_{c}^{0};\phi)+\mathcal{M}(K^{\ast+},\Omega_{c}^{0};\eta_{8})+\mathcal{M}(\rho^{+},\Xi_{c}^{(\prime)0};\overline{K}^{\ast0})+\mathcal{M}(K^{\ast+},\Omega_{c}^{0};\phi),\\
\mathcal{A}(\Omega_{cc}^{+}\to\Xi_{c}^{0}K^{+}) & =\mathcal{T}_{SD}(\Omega_{cc}^{+}\to\Xi_{c}^{0}K^{+})+\mathcal{M}(K^{+},\Xi_{c}^{(\prime)0};\phi)+\mathcal{M}(K^{\ast+},\Xi_{c}^{(\prime)0};\eta_{8})+\mathcal{M}(K^{\ast+},\Xi_{c}^{(\prime)0};\phi),\nonumber \\
\mathcal{A}(\Omega_{cc}^{+}\to\Xi_{c}^{(\prime)0}K^{+}) & =\mathcal{T}_{SD}(\Omega_{cc}^{+}\to\Xi_{c}^{(\prime)0}K^{+})+\mathcal{M}(K^{+},\Xi_{c}^{(\prime)0};\phi)+\mathcal{M}(K^{\ast+},\Xi_{c}^{(\prime)0};\eta_{8})+\mathcal{M}(K^{\ast+},\Xi_{c}^{(\prime)0};\phi).
\end{align}
 The amplitudes for long-distance dominated and Cabibbo flavored
modes can be given as follows. { 
\begin{align}
\mathcal{A}(\Xi_{cc}^{++}\to\Sigma_{c}^{++}\overline{K}^{0}) & =\mathcal{C}_{SD}(\Xi_{cc}^{++}\to\Sigma_{c}^{++}\overline{K}^{0})+\mathcal{M}(\pi^{+},\Xi_{c}^{(\prime)+};K^{\ast+})+\mathcal{M}(\rho^{+},\Xi_{c}^{(\prime)+};K^{+})+\mathcal{M}(\rho^{+},\Xi_{c}^{(\prime)+};K^{\ast+})\nonumber \\
 & +\mathcal{M}(\Xi_{c}^{(\prime)+},\pi^{+};\Lambda_{c}^{+}/\Sigma_{c}^{+})+\mathcal{M}(\Xi_{c}^{(\prime)+},\rho^{+};\Lambda_{c}^{+}/\Sigma_{c}^{+})+\mathcal{M}(\Xi_{c}^{(\prime)+},\pi^{+};\Sigma_{c}^{*+})+\mathcal{M}(\Xi_{c}^{(\prime)+},\rho^{+};\Sigma_{c}^{*+}),\\
\mathcal{A}(\Xi_{cc}^{+}\to\Omega_{c}^{0}K^{+}) & =\mathcal{M}(\pi^{+},\Xi_{c}^{(\prime)0};\overline{K}^{\ast0})+\mathcal{M}(\rho^{+},\Xi_{c}^{(\prime)0};\overline{K}^{0})+\mathcal{M}(\rho^{+},\Xi_{c}^{(\prime)0};\overline{K}^{\ast0}),\\
\mathcal{A}(\Xi_{cc}^{+}\to\Sigma_{c}^{+}\overline{K}^{0}) & =\mathcal{C}_{SD}(\Xi_{cc}^{+}\to\Sigma_{c}^{+}\overline{K}^{0})+\mathcal{M}(\pi^{+},\Xi_{c}^{(\prime)0};K^{\ast+})+\mathcal{M}(\rho^{+},\Xi_{c}^{(\prime)0};K^{+})+\mathcal{M}(\rho^{+},\Xi_{c}^{(\prime)0};K^{\ast+})\nonumber \\
 & +\mathcal{M}(\Xi_{c}^{(\prime)0},\pi^{+};\Sigma_{c}^{0})+\mathcal{M}(\Xi_{c}^{(\prime)0},\rho^{+};\Sigma_{c}^{0})+\mathcal{M}(\Xi_{c}^{(\prime)0},\pi^{+};\Sigma_{c}^{*0})+\mathcal{M}(\Xi_{c}^{(\prime)0},\rho^{+};\Sigma_{c}^{*0}),\\
\mathcal{A}(\Xi_{cc}^{+}\to\Lambda_{c}^{+}\overline{K}^{0}) & =\mathcal{C}_{SD}(\Xi_{cc}^{+}\to\Lambda_{c}^{+}\overline{K}^{0})+\mathcal{M}(\pi^{+},\Xi_{c}^{(\prime)0};K^{\ast+})+\mathcal{M}(\rho^{+},\Xi_{c}^{(\prime)0};K^{+})+\mathcal{M}(\rho^{+},\Xi_{c}^{(\prime)0};K^{\ast+})\nonumber \\
 & +\mathcal{M}(\Xi_{c}^{(\prime)0},\pi^{+};\Sigma_{c}^{0})+\mathcal{M}(\Xi_{c}^{(\prime)0},\rho^{+};\Sigma_{c}^{0})+\mathcal{M}(\Xi_{c}^{(\prime)0},\pi^{+};\Sigma_{c}^{*0})+\mathcal{M}(\Xi_{c}^{(\prime)0},\rho^{+};\Sigma_{c}^{*0}),\\
\mathcal{A}(\Xi_{cc}^{+}\to\Sigma_{c}^{++}K^{-}) & =\mathcal{M}(\Xi_{c}^{(\prime)0},\pi^{+};\Lambda_{c}^{+}/\Sigma_{c}^{+})+\mathcal{M}(\Xi_{c}^{(\prime)0},\rho^{+};\Lambda_{c}^{+}/\Sigma_{c}^{+})+\mathcal{M}(\Xi_{c}^{(\prime)0},\pi^{+};\Sigma_{c}^{*+})+\mathcal{M}(\Xi_{c}^{(\prime)0},\rho^{+};\Sigma_{c}^{*+}),\\
\mathcal{A}(\Xi_{cc}^{+}\to\Xi_{c}^{+}\pi^{0}) & =\mathcal{C}_{SD}(\Xi_{cc}^{+}\to\Xi_{c}^{+}\pi^{0})+\mathcal{M}(\pi^{+},\Xi_{c}^{(\prime)0};\rho^{+})+\mathcal{M}(\rho^{+},\Xi_{c}^{(\prime)0};\pi^{+})+\mathcal{M}(\rho^{+},\Xi_{c}^{(\prime)0};\rho^{+})\nonumber \\
 & +\mathcal{M}(\Xi_{c}^{(\prime)0},\pi^{+};\Xi_{c}^{(\prime)0})+\mathcal{M}(\Xi_{c}^{(\prime)0},\rho^{+};\Xi_{c}^{(\prime)0})+\mathcal{M}(\Xi_{c}^{(\prime)0},\pi^{+};\Xi_{c}^{\prime*0})+\mathcal{M}(\Xi_{c}^{(\prime)0},\rho^{+};\Xi_{c}^{\prime*0}),\\
\mathcal{A}(\Xi_{cc}^{+}\to\Xi_{c}^{+}\eta_{1}) & =\mathcal{M}(\Xi_{c}^{0},\pi^{+};\Xi_{c}^{0})+\mathcal{M}(\Xi_{c}^{\prime0},\pi^{+};\Xi_{c}^{\prime0})+\mathcal{M}(\Xi_{c}^{0},\rho^{+};\Xi_{c}^{0})+\mathcal{M}(\Xi_{c}^{\prime0},\rho^{+};\Xi_{c}^{\prime0})\nonumber \\
 & +\mathcal{M}(\Xi_{c}^{\prime0},\pi^{+};\Xi_{c}^{\prime*0})+\mathcal{M}(\Xi_{c}^{\prime0},\rho^{+};\Xi_{c}^{\prime*0}),\\
\mathcal{A}(\Xi_{cc}^{+}\to\Xi_{c}^{+}\eta_{8}) & =\mathcal{M}(\Xi_{c}^{(\prime)0},\pi^{+};\Xi_{c}^{(\prime)0})+\mathcal{M}(\Xi_{c}^{(\prime)0},\rho^{+};\Xi_{c}^{(\prime)0})+\mathcal{M}(\Xi_{c}^{(\prime)0},\pi^{+};\Xi_{c}^{\prime*0})+\mathcal{M}(\Xi_{c}^{(\prime)0},\rho^{+};\Xi_{c}^{\prime*0}),\\
\mathcal{A}(\Xi_{cc}^{+}\to\Xi_{c}^{\prime+}\pi^{0}) & =\mathcal{C}_{SD}(\Xi_{cc}^{+}\to\Xi_{c}^{(\prime)+}\pi^{0})+\mathcal{M}(\pi^{+},\Xi_{c}^{(\prime)0};\rho^{+})+\mathcal{M}(\rho^{+},\Xi_{c}^{(\prime)0};\pi^{+})+\mathcal{M}(\rho^{+},\Xi_{c}^{(\prime)0};\rho^{+})\nonumber \\
 & +\mathcal{M}(\Xi_{c}^{(\prime)0},\pi^{+};\Xi_{c}^{(\prime)0})+\mathcal{M}(\Xi_{c}^{(\prime)0},\rho^{+};\Xi_{c}^{(\prime)0})+\mathcal{M}(\Xi_{c}^{(\prime)0},\pi^{+};\Xi_{c}^{\prime*0})+\mathcal{M}(\Xi_{c}^{(\prime)0},\rho^{+};\Xi_{c}^{\prime*0}),\\
\mathcal{A}(\Xi_{cc}^{+}\to\Xi_{c}^{\prime+}\eta_{1}) & =\mathcal{M}(\Xi_{c}^{0},\pi^{+},;\Xi_{c}^{0})+\mathcal{M}(\Xi_{c}^{\prime0},\pi^{+},;\Xi_{c}^{\prime0})+\mathcal{M}(\Xi_{c}^{0},\rho^{+};\Xi_{c}^{0})+\mathcal{M}(\Xi_{c}^{\prime0},\rho^{+};\Xi_{c}^{\prime0})\nonumber \\
 & +\mathcal{M}(\Xi_{c}^{\prime0},\pi^{+},;\Xi_{c}^{\prime*0})+\mathcal{M}(\Xi_{c}^{\prime0},\rho^{+};\Xi_{c}^{\prime*0}),\\
\mathcal{A}(\Xi_{cc}^{+}\to\Xi_{c}^{\prime+}\eta_{8}) & =\mathcal{M}(\Xi_{c}^{(\prime)0},\pi^{+};\Xi_{c}^{(\prime)0})+\mathcal{M}(\Xi_{c}^{(\prime)0},\rho^{+};\Xi_{c}^{(\prime)0})+\mathcal{M}(\Xi_{c}^{(\prime)0},\pi^{+};\Xi_{c}^{\prime*0})+\mathcal{M}(\Xi_{c}^{(\prime)0},\rho^{+};\Xi_{c}^{\prime*0}),\\
\mathcal{A}(\Omega_{cc}^{+}\to\Xi_{c}^{+}\overline{K}^{0}) & =\mathcal{C}_{SD}(\Omega_{cc}^{+}\to\Xi_{c}^{+}\overline{K}^{0})+\mathcal{M}(\pi^{+},\Omega_{c}^{0};K^{\ast+})+\mathcal{M}(\rho^{+},\Omega_{c}^{0};K^{+})+\mathcal{M}(\rho^{+},\Omega_{c}^{0};K^{\ast+})\nonumber \\
 & +\mathcal{M}(\Omega_{c}^{0},\pi^{+};\Xi_{c}^{(\prime)0})+\mathcal{M}(\Omega_{c}^{0},\rho^{+};\Xi_{c}^{(\prime)0})+\mathcal{M}(\Omega_{c}^{0},\pi^{+};\Xi_{c}^{\prime*0})+\mathcal{M}(\Omega_{c}^{0},\rho^{+};\Xi_{c}^{\prime*0}),\\
\mathcal{A}(\Omega_{cc}^{+}\to\Xi_{c}^{\prime+}\overline{K}^{0}) & =\mathcal{C}_{SD}(\Omega_{cc}^{+}\to\Xi_{c}^{(\prime)+}\overline{K}^{0})+\mathcal{M}(\pi^{+},\Omega_{c}^{0};K^{\ast+})+\mathcal{M}(\rho^{+},\Omega_{c}^{0};K^{+})+\mathcal{M}(\rho^{+},\Omega_{c}^{0};K^{\ast+})\nonumber \\
 & +\mathcal{M}(\Omega_{c}^{0},\pi^{+};\Xi_{c}^{(\prime)0})+\mathcal{M}(\Omega_{c}^{0},\rho^{+};\Xi_{c}^{(\prime)0})+\mathcal{M}(\Omega_{c}^{0},\pi^{+};\Xi_{c}^{\prime*0})+\mathcal{M}(\Omega_{c}^{0},\rho^{+};\Xi_{c}^{\prime*0}).
\end{align}
} The amplitudes for long-distance dominated singly Cabibbo suppressed
modes can be given as follows. {
\begin{align}
\mathcal{A}(\Xi_{cc}^{++}\to\Sigma_{c}^{++}\pi^{0}) & =\mathcal{C}_{SD}(\Xi_{cc}^{++}\to\Sigma_{c}^{++}\pi^{0})+\mathcal{M}(\pi^{+},\Lambda_{c}^{+}/\Sigma_{c}^{+};\rho^{+})+\mathcal{M}(\rho^{+},\Lambda_{c}^{+}/\Sigma_{c}^{+};\pi^{+})+\mathcal{M}(\rho^{+},\Lambda_{c}^{+}/\Sigma_{c}^{+};\rho^{+})\nonumber \\
 & +\mathcal{M}(K^{+},\Xi_{c}^{(\prime)+};K^{\ast+})+\mathcal{M}(K^{\ast+},\Xi_{c}^{(\prime)+};K^{+})+\mathcal{M}(\Xi_{c}^{(\prime)+},K^{+};\Xi_{c}^{(\prime)+})+\mathcal{M}(\Xi_{c}^{(\prime)+},K^{\ast+};\Xi_{c}^{(\prime)+})\nonumber \\
 & +\mathcal{M}(\Lambda_{c}^{+},\pi^{+};\Sigma^{+})+\mathcal{M}(\Sigma_{c}^{+},\pi^{+};\Lambda^{+})+\mathcal{M}(\Lambda_{c}^{+},\rho^{+};\Sigma^{+})+\mathcal{M}(\Sigma_{c}^{+},\rho^{+};\Lambda^{+})\nonumber\\
 & +\mathcal{M}(\Xi_{c}^{(\prime)+},K^{+};\Xi_{c}^{\prime*+})+\mathcal{M}(\Xi_{c}^{(\prime)+},K^{\ast+};\Xi_{c}^{\prime*+})+\mathcal{M}(\Lambda_{c}^{+},\pi^{+};\Sigma^{*+})+\mathcal{M}(\Lambda_{c}^{+},\rho^{+};\Sigma^{*+}),\\
\mathcal{A}(\Xi_{cc}^{++}\to\Sigma_{c}^{++}\eta_{1}) & =\mathcal{C}_{SD}(\Xi_{cc}^{++}\to\Sigma_{c}^{++}\eta_{1})\nonumber \\
 & +\mathcal{M}(\Sigma_{c}^{+},\pi^{+};\Sigma_{c}^{+})+\mathcal{M}(\Lambda_{c}^{+},\pi^{+};\Lambda_{c}^{+})+\mathcal{M}(\Sigma_{c}^{+},\rho^{+};\Sigma_{c}^{+})+\mathcal{M}(\Lambda_{c}^{+},\rho^{+};\Lambda_{c}^{+})\nonumber \\
 & +\mathcal{M}(\Xi_{c}^{+},K^{+};\Xi_{c}^{+})+\mathcal{M}(\Xi_{c}^{\prime+},K^{+};\Xi_{c}^{\prime+})+\mathcal{M}(\Xi_{c}^{+},K^{\ast+};\Xi_{c}^{+})+\mathcal{M}(\Xi_{c}^{\prime+},K^{\ast+};\Xi_{c}^{\prime+})\nonumber \\
 & +\mathcal{M}(\Sigma_{c}^{+},\pi^{+};\Sigma_{c}^{*+})+\mathcal{M}(\Sigma_{c}^{+},\rho^{+};\Sigma_{c}^{*+})+\mathcal{M}(\Xi_{c}^{\prime+},K^{+};\Xi_{c}^{\prime*+})+\mathcal{M}(\Xi_{c}^{\prime+},K^{\ast+};\Xi_{c}^{\prime*+}),\end{align}}{
  \begin{align}
\mathcal{A}(\Xi_{cc}^{++}\to\Sigma_{c}^{++}\eta_{8}) & =\mathcal{C}_{SD}(\Xi_{cc}^{++}\to\Sigma_{c}^{++}\eta_{8})\nonumber \\
 & +\mathcal{M}(K^{+},\Xi_{c}^{(\prime)+};K^{\ast+})+\mathcal{M}(K^{\ast+},\Xi_{c}^{(\prime)+};K^{+})+\mathcal{M}(K^{\ast+},\Xi_{c}^{(\prime)+};K^{*+})\nonumber \\
 & +\mathcal{M}(\Sigma_{c}^{+},\pi^{+};\Sigma_{c}^{+})+\mathcal{M}(\Lambda_{c}^{+},\pi^{+};\Lambda_{c}^{+})+\mathcal{M}(\Sigma_{c}^{+},\rho^{+};\Sigma_{c}^{+})+\mathcal{M}(\Lambda_{c}^{+},\rho^{+};\Lambda_{c}^{+})\nonumber \\
 & +\mathcal{M}(\Xi_{c}^{(\prime)+},K^{+};\Xi_{c}^{(\prime)+})+\mathcal{M}(\Xi_{c}^{(\prime)+},K^{\ast+};\Xi_{c}^{(\prime)+})+\mathcal{M}(\Xi_{c}^{(\prime)+},K^{+};\Xi_{c}^{\prime*+})+\mathcal{M}(\Xi_{c}^{(\prime)+},K^{\ast+};\Xi_{c}^{\prime*+}),  \\
\mathcal{A}(\Xi_{cc}^{+}\to\Sigma_{c}^{+}\pi^{0}) & =\mathcal{C}_{SD}(\Xi_{cc}^{+}\to\Sigma_{c}^{+}\pi^{0})+\mathcal{M}(\pi^{+},\Sigma_{c}^{0};\rho^{+})+\mathcal{M}(\rho^{+},\Sigma_{c}^{0};\pi^{+})+\mathcal{M}(K^{+},\Xi_{c}^{(\prime)0};K^{\ast+})\nonumber \\
 & +\mathcal{M}(K^{\ast+},\Xi_{c}^{(\prime)0};K^{+})+\mathcal{M}(\rho^{+},\Sigma_{c}^{0};\rho^{+})+\mathcal{M}(K^{\ast+},\Xi_{c}^{(\prime)0};K^{\ast+})\nonumber \\
 & +\mathcal{M}(\Sigma_{c}^{0},\pi^{+};\Sigma_{c}^{0})+\mathcal{M}(\Sigma_{c}^{0},\rho^{+};\Sigma_{c}^{0})+\mathcal{M}(\Xi_{c}^{(\prime)0},K^{+};\Xi_{c}^{(\prime)0})+\mathcal{M}(\Xi_{c}^{(\prime)0},K^{\ast+};\Xi_{c}^{(\prime)0})\nonumber \\
 & +\mathcal{M}(\Sigma_{c}^{0},\pi^{+};\Sigma_{c}^{*0})+\mathcal{M}(\Sigma_{c}^{0},\rho^{+};\Sigma_{c}^{*0})+\mathcal{M}(\Xi_{c}^{(\prime)0},K^{+};\Xi_{c}^{\prime*0})+\mathcal{M}(\Xi_{c}^{(\prime)0},K^{\ast+};\Xi_{c}^{\prime*0}),\\
\mathcal{A}(\Xi_{cc}^{+}\to\Sigma_{c}^{+}\eta_{1}) & =\mathcal{C}_{SD}(\Xi_{cc}^{+}\to\Sigma_{c}^{+}\eta_{1})+\mathcal{M}(\Xi_{c}^{0},K^{+};\Xi_{c}^{0})+\mathcal{M}(\Xi_{c}^{\prime0},K^{+};\Xi_{c}^{\prime0})+\mathcal{M}(\Xi_{c}^{0},K^{\ast+};\Xi_{c}^{0})\nonumber \\
 & +\mathcal{M}(\Xi_{c}^{\prime0},K^{\ast+};\Xi_{c}^{\prime0})+\mathcal{M}(\Sigma_{c}^{0},\pi^{+};\Sigma_{c}^{0})+\mathcal{M}(\Sigma_{c}^{0},\rho^{+};\Sigma_{c}^{0})\nonumber\\
 & +\mathcal{M}(\Xi_{c}^{\prime0},K^{+};\Xi_{c}^{\prime*0})+\mathcal{M}(\Xi_{c}^{\prime0},K^{\ast+};\Xi_{c}^{\prime*0})+\mathcal{M}(\Sigma_{c}^{0},\pi^{+};\Sigma_{c}^{*0})+\mathcal{M}(\Sigma_{c}^{0},\rho^{+};\Sigma_{c}^{*0}),
\\
\mathcal{A}(\Xi_{cc}^{+}\to\Sigma_{c}^{+}\eta_{8}) & =\mathcal{C}_{SD}(\Xi_{cc}^{+}\to\Sigma_{c}^{+}\eta_{8})+\mathcal{M}(K^{+},\Xi_{c}^{(\prime)0};K^{\ast+})+\mathcal{M}(K^{\ast+},\Xi_{c}^{(\prime)0};K^{+})+\mathcal{M}(K^{\ast+},\Xi_{c}^{(\prime)0};K^{\ast+})\nonumber \\
 & +\mathcal{M}(\Sigma_{c}^{0},\pi^{+};\Sigma_{c}^{0})+\mathcal{M}(\Sigma_{c}^{0},\rho^{+};\Sigma_{c}^{0})+\mathcal{M}(\Xi_{c}^{(\prime)0},K^{+};\Xi_{c}^{(\prime)0})+\mathcal{M}(\Xi_{c}^{(\prime)0},K^{\ast+};\Xi_{c}^{(\prime)0})\nonumber \\
 & +\mathcal{M}(\Sigma_{c}^{0},\pi^{+};\Sigma_{c}^{*0})+\mathcal{M}(\Sigma_{c}^{0},\rho^{+};\Sigma_{c}^{*0})+\mathcal{M}(\Xi_{c}^{(\prime)0},K^{+};\Xi_{c}^{\prime*0})+\mathcal{M}(\Xi_{c}^{(\prime)0},K^{\ast+};\Xi_{c}^{\prime*0}),\\
\mathcal{A}(\Xi_{cc}^{+}\to\Lambda_{c}^{+}\pi^{0}) & =\mathcal{C}_{SD}(\Xi_{cc}^{+}\to\Lambda_{c}^{+}\pi^{0})+\mathcal{M}(\pi^{+},\Sigma_{c}^{0};\rho^{+})+\mathcal{M}(\rho^{+},\Sigma_{c}^{0};\pi^{+})+\mathcal{M}(K^{+},\Xi_{c}^{(\prime)0};K^{\ast+})\nonumber \\
 & +\mathcal{M}(K^{\ast+},\Xi_{c}^{(\prime)0};K^{+})+\mathcal{M}(\rho^{+},\Sigma_{c}^{0};\rho^{+})+\mathcal{M}(K^{\ast+},\Xi_{c}^{(\prime)0};K^{\ast+})\nonumber \\
 & +\mathcal{M}(\Sigma_{c}^{0},\pi^{+};\Sigma_{c}^{0})+\mathcal{M}(\Sigma_{c}^{0},\rho^{+};\Sigma_{c}^{0})+\mathcal{M}(\Xi_{c}^{(\prime)0},K^{+};\Xi_{c}^{(\prime)0})+\mathcal{M}(\Xi_{c}^{(\prime)0},K^{\ast+};\Xi_{c}^{(\prime)0})\nonumber \\
 & +\mathcal{M}(\Sigma_{c}^{0},\pi^{+};\Sigma_{c}^{*0})+\mathcal{M}(\Sigma_{c}^{0},\rho^{+};\Sigma_{c}^{*0})+\mathcal{M}(\Xi_{c}^{(\prime)0},K^{+};\Xi_{c}^{\prime*0})+\mathcal{M}(\Xi_{c}^{(\prime)0},K^{\ast+};\Xi_{c}^{\prime*0}),\\
\mathcal{A}(\Xi_{cc}^{+}\to\Lambda_{c}^{+}\eta_{1}) & =\mathcal{C}_{SD}(\Xi_{cc}^{+}\to\Lambda_{c}^{+}\eta_{1})+\mathcal{M}(\Xi_{c}^{0},K^{+};\Xi_{c}^{0})+\mathcal{M}(\Xi_{c}^{\prime0},K^{+};\Xi_{c}^{\prime0})+\mathcal{M}(\Xi_{c}^{0},K^{\ast+};\Xi_{c}^{0})\nonumber \\
 & +\mathcal{M}(\Xi_{c}^{\prime0},K^{\ast+};\Xi_{c}^{\prime0})+\mathcal{M}(\Sigma_{c}^{0},\pi^{+};\Sigma_{c}^{0})+\mathcal{M}(\Sigma_{c}^{0},\rho^{+};\Sigma_{c}^{0})\nonumber\\
 & +\mathcal{M}(\Xi_{c}^{\prime0},K^{+};\Xi_{c}^{\prime*0})+\mathcal{M}(\Xi_{c}^{\prime0},K^{\ast+};\Xi_{c}^{\prime*0})+\mathcal{M}(\Sigma_{c}^{0},\pi^{+};\Sigma_{c}^{*0})+\mathcal{M}(\Sigma_{c}^{0},\rho^{+};\Sigma_{c}^{*0}),\\
\mathcal{A}(\Xi_{cc}^{+}\to\Lambda_{c}^{+}\eta_{8}) & =\mathcal{C}_{SD}(\Xi_{cc}^{+}\to\Lambda_{c}^{+}\eta_{8})+\mathcal{M}(K^{+},\Xi_{c}^{(\prime)0};K^{\ast+})+\mathcal{M}(K^{\ast+},\Xi_{c}^{(\prime)0};K^{+})+\mathcal{M}(K^{\ast+},\Xi_{c}^{(\prime)0};K^{\ast+})\nonumber \\
 & +\mathcal{M}(\Sigma_{c}^{0},\pi^{+};\Sigma_{c}^{0})+\mathcal{M}(\Sigma_{c}^{0},\rho^{+};\Sigma_{c}^{0})+\mathcal{M}(\Xi_{c}^{(\prime)0},K^{+};\Xi_{c}^{(\prime)0})+\mathcal{M}(\Xi_{c}^{(\prime)0},K^{\ast+};\Xi_{c}^{(\prime)0})\nonumber \\
 & +\mathcal{M}(\Sigma_{c}^{0},\pi^{+};\Sigma_{c}^{*0})+\mathcal{M}(\Sigma_{c}^{0},\rho^{+};\Sigma_{c}^{*0})+\mathcal{M}(\Xi_{c}^{(\prime)0},K^{+};\Xi_{c}^{\prime*0})+\mathcal{M}(\Xi_{c}^{(\prime)0},K^{\ast+};\Xi_{c}^{\prime*0}),\\
\mathcal{A}(\Xi_{cc}^{+}\to\Sigma_{c}^{++}\pi^{-}) & =\mathcal{M}(\Sigma_{c}^{0},\pi^{+};\Lambda_{c}^{+}/\Sigma_{c}^{+})+\mathcal{M}(\Sigma_{c}^{0},\rho^{+};\Lambda_{c}^{+}/\Sigma_{c}^{+})+\mathcal{M}(\Xi_{c}^{(\prime)0},K^{+};\Xi_{c}^{(\prime)+})+\mathcal{M}(\Xi_{c}^{(\prime)0},K^{\ast+};\Xi_{c}^{(\prime)+})\nonumber \\
 & +\mathcal{M}(\Sigma_{c}^{0},\pi^{+};\Sigma_{c}^{*+})+\mathcal{M}(\Sigma_{c}^{0},\rho^{+};\Sigma_{c}^{*+})+\mathcal{M}(\Xi_{c}^{(\prime)0},K^{+};\Xi_{c}^{\prime*+})+\mathcal{M}(\Xi_{c}^{(\prime)0},K^{\ast+};\Xi_{c}^{\prime*+}),\\
\mathcal{A}(\Xi_{cc}^{+}\to\Xi_{c}^{+}K^{0}) & =\mathcal{C}_{SD}(\Xi_{cc}^{+}\to\Xi_{c}^{+}K^{0})+\mathcal{M}(K^{+},\Xi_{c}^{(\prime)0};\rho^{+})+\mathcal{M}(K^{\ast+},\Xi_{c}^{(\prime)0};\rho^{+})+\mathcal{M}(K^{\ast+},\Xi_{c}^{(\prime)0};\rho^{+})\nonumber \\
 & +\mathcal{M}(\Xi_{c}^{(\prime)0},K^{+};\Omega_{c}^{0})+\mathcal{M}(\Xi_{c}^{(\prime)0},K^{\ast+};\Omega_{c}^{0})+\mathcal{M}(\Sigma_{c}^{0},\pi^{+};\Xi_{c}^{(\prime)0})+\mathcal{M}(\Sigma_{c}^{0},\rho^{+};\Xi_{c}^{(\prime)0})\nonumber \\
 & +\mathcal{M}(\Xi_{c}^{(\prime)0},K^{+};\Omega_{c}^{*0})+\mathcal{M}(\Xi_{c}^{(\prime)0},K^{\ast+};\Omega_{c}^{*0})+\mathcal{M}(\Sigma_{c}^{0},\pi^{+};\Xi_{c}^{\prime*0})+\mathcal{M}(\Sigma_{c}^{0},\rho^{+};\Xi_{c}^{\prime*0}),\\
\mathcal{A}(\Xi_{cc}^{+}\to\Xi_{c}^{\prime+}K^{0}) & =\mathcal{C}_{SD}(\Xi_{cc}^{+}\to\Xi_{c}^{(\prime)+}K^{0})+\mathcal{M}(K^{+},\Xi_{c}^{(\prime)0};\rho^{+})+\mathcal{M}(K^{\ast+},\Xi_{c}^{(\prime)0};\pi^{+})+\mathcal{M}(K^{\ast+},\Xi_{c}^{(\prime)0};\rho^{+})\nonumber \\
 & +\mathcal{M}(\Xi_{c}^{(\prime)0},K^{+};\Omega_{c}^{0})+\mathcal{M}(\Xi_{c}^{(\prime)0},K^{\ast+};\Omega_{c}^{0})+\mathcal{M}(\Sigma_{c}^{0},\pi^{+};\Xi_{c}^{(\prime)0})+\mathcal{M}(\Sigma_{c}^{0},\rho^{+};\Xi_{c}^{(\prime)0})\nonumber \\
 & +\mathcal{M}(\Xi_{c}^{(\prime)0},K^{+};\Omega_{c}^{*0})+\mathcal{M}(\Xi_{c}^{(\prime)0},K^{\ast+};\Omega_{c}^{*0})+\mathcal{M}(\Sigma_{c}^{0},\pi^{+};\Xi_{c}^{\prime*0})+\mathcal{M}(\Sigma_{c}^{0},\rho^{+};\Xi_{c}^{\prime*0}),
\\
\mathcal{A}(\Omega_{cc}^{+}\to\Sigma_{c}^{+}\overline{K}^{0}) & =\mathcal{C}_{SD}(\Omega_{cc}^{+}\to\Sigma_{c}^{+}\overline{K}^{0})+\mathcal{M}(\pi^{+},\Xi_{c}^{(\prime)0};K^{\ast+})+\mathcal{M}(\rho^{+},\Xi_{c}^{(\prime)0};K^{+})+\mathcal{M}(\rho^{+},\Xi_{c}^{(\prime)0};K^{\ast+})\nonumber \\
 & +\mathcal{M}(\Xi_{c}^{(\prime)0},\pi^{+};\Sigma_{c}^{0})+\mathcal{M}(\Xi_{c}^{(\prime)0},\rho^{+};\Sigma_{c}^{0})+\mathcal{M}(\Omega_{c}^{0},K^{+};\Xi_{c}^{(\prime)0})+\mathcal{M}(\Omega_{c}^{0},K^{\ast+};\Xi_{c}^{(\prime)0})\nonumber \\
 & +\mathcal{M}(\Xi_{c}^{(\prime)0},\pi^{+};\Sigma_{c}^{*0})+\mathcal{M}(\Xi_{c}^{(\prime)0},\rho^{+};\Sigma_{c}^{*0})+\mathcal{M}(\Omega_{c}^{0},K^{+};\Xi_{c}^{\prime*0})+\mathcal{M}(\Omega_{c}^{0},K^{\ast+};\Xi_{c}^{\prime*0}),
\\
\mathcal{A}(\Omega_{cc}^{+}\to\Lambda_{c}^{+}\overline{K}^{0}) & =\mathcal{C}_{SD}(\Omega_{cc}^{+}\to\Lambda_{c}^{+}\overline{K}^{0})+\mathcal{M}(\pi^{+},\Xi_{c}^{(\prime)0};K^{\ast+})+\mathcal{M}(\rho^{+},\Xi_{c}^{(\prime)0};K^{+})+\mathcal{M}(\rho^{+},\Xi_{c}^{(\prime)0};K^{\ast+})\nonumber \\
 & +\mathcal{M}(\Xi_{c}^{(\prime)0},\pi^{+};\Sigma_{c}^{0})+\mathcal{M}(\Xi_{c}^{(\prime)0},\rho^{+};\Sigma_{c}^{0})+\mathcal{M}(\Omega_{c}^{0},K^{+};\Xi_{c}^{(\prime)0})+\mathcal{M}(\Omega_{c}^{0},K^{\ast+};\Xi_{c}^{(\prime)0})\nonumber \\
 & +\mathcal{M}(\Xi_{c}^{(\prime)0},\pi^{+};\Sigma_{c}^{*0})+\mathcal{M}(\Xi_{c}^{(\prime)0},\rho^{+};\Sigma_{c}^{*0})+\mathcal{M}(\Omega_{c}^{0},K^{+};\Xi_{c}^{\prime*0})+\mathcal{M}(\Omega_{c}^{0},K^{\ast+};\Xi_{c}^{\prime*0}),\\
\mathcal{A}(\Omega_{cc}^{+}\to\Sigma_{c}^{++}K^{-}) & =\mathcal{M}(\Xi_{c}^{(\prime)0},\pi^{+};\Lambda_{c}^{+}/\Sigma_{c}^{+})+\mathcal{M}(\Xi_{c}^{(\prime)0},\rho^{+};\Lambda_{c}^{+}/\Sigma_{c}^{+})+\mathcal{M}(\Omega_{c}^{0},K^{+};\Xi_{c}^{(\prime)+})+\mathcal{M}(\Omega_{c}^{0},K^{\ast+};\Xi_{c}^{(\prime)+})\nonumber \\
 & +\mathcal{M}(\Xi_{c}^{(\prime)0},\pi^{+};\Sigma_{c}^{*+})+\mathcal{M}(\Xi_{c}^{(\prime)0},\rho^{+};\Sigma_{c}^{*+})+\mathcal{M}(\Omega_{c}^{0},K^{+};\Xi_{c}^{\prime*+})+\mathcal{M}(\Omega_{c}^{0},K^{\ast+};\Xi_{c}^{\prime*+}),\\
\mathcal{A}(\Omega_{cc}^{+}\to\Xi_{c}^{+}\pi^{0}) & =\mathcal{C}_{SD}(\Omega_{cc}^{+}\to\Xi_{c}^{+}\pi^{0})+\mathcal{M}(\pi^{+},\Xi_{c}^{(\prime)0};\rho^{+})+\mathcal{M}(\rho^{+},\Xi_{c}^{(\prime)0};\pi^{+})\nonumber \\
 & +\mathcal{M}(K^{+},\Omega_{c}^{0};K^{\ast+})+\mathcal{M}(K^{\ast+},\Omega_{c}^{0};K^{+})+\mathcal{M}(\rho^{+},\Xi_{c}^{(\prime)0};\rho^{+})+\mathcal{M}(K^{\ast+},\Omega_{c}^{0};K^{\ast+}),\nonumber \\
 & +\mathcal{M}(\Xi_{c}^{(\prime)0},\pi^{+};\Xi_{c}^{(\prime)0})+\mathcal{M}(\Xi_{c}^{(\prime)0},\rho^{+};\Xi_{c}^{(\prime)0})+\mathcal{M}(\Xi_{c}^{(\prime)0},\pi^{+};\Xi_{c}^{\prime*0})+\mathcal{M}(\Xi_{c}^{(\prime)0},\rho^{+};\Xi_{c}^{\prime*0}),\end{align}}
 \begin{align}
\mathcal{A}(\Omega_{cc}^{+}\to\Xi_{c}^{+}\eta_{1}) & =\mathcal{C}_{SD}(\Omega_{cc}^{+}\to\Xi_{c}^{+}\eta_{1})+\mathcal{M}(\Omega_{c}^{0},K^{+};\Omega_{c}^{0})+\mathcal{M}(\Omega_{c}^{0},K^{\ast+};\Omega_{c}^{0})+\mathcal{M}(\Xi_{c}^{0},\pi^{+};\Xi_{c}^{0})\nonumber \\
 & +\mathcal{M}(\Xi_{c}^{\prime0},\pi^{+};\Xi_{c}^{\prime0})+\mathcal{M}(\Xi_{c}^{0},\rho^{+};\Xi_{c}^{0})+\mathcal{M}(\Xi_{c}^{\prime0},\rho^{+};\Xi_{c}^{\prime0})\nonumber \\
 & +\mathcal{M}(\Omega_{c}^{0},K^{+};\Omega_{c}^{*0})+\mathcal{M}(\Omega_{c}^{0},K^{\ast+};\Omega_{c}^{*0})+\mathcal{M}(\Xi_{c}^{\prime0},\pi^{+};\Xi_{c}^{\prime*0})+\mathcal{M}(\Xi_{c}^{\prime0},\rho^{+};\Xi_{c}^{\prime*0}),\\
\mathcal{A}(\Omega_{cc}^{+}\to\Xi_{c}^{+}\eta_{8}) & =\mathcal{C}_{SD}(\Omega_{cc}^{+}\to\Xi_{c}^{+}\eta_{8})+\mathcal{M}(K^{+},\Omega_{c}^{0};K^{\ast+})+\mathcal{M}(K^{\ast+},\Omega_{c}^{0};K^{+})+\mathcal{M}(K^{\ast+},\Omega_{c}^{0};K^{\ast+})\nonumber \\
 & +\mathcal{M}(\Omega_{c}^{0},K^{+};\Omega_{c}^{0})+\mathcal{M}(\Omega_{c}^{0},K^{\ast+};\Omega_{c}^{0})+\mathcal{M}(\Xi_{c}^{(\prime)0},\pi^{+};\Xi_{c}^{(\prime)0})+\mathcal{M}(\Xi_{c}^{(\prime)0},\rho^{+};\Xi_{c}^{(\prime)0})\nonumber \\
 & +\mathcal{M}(\Omega_{c}^{0},K^{+};\Omega_{c}^{*0})+\mathcal{M}(\Omega_{c}^{0},K^{\ast+};\Omega_{c}^{*0})+\mathcal{M}(\Xi_{c}^{(\prime)0},\pi^{+};\Xi_{c}^{\prime*0})+\mathcal{M}(\Xi_{c}^{(\prime)0},\rho^{+};\Xi_{c}^{\prime*0}),\\
\mathcal{A}(\Omega_{cc}^{+}\to\Xi_{c}^{\prime+}\pi^{0}) & =\mathcal{C}_{SD}(\Omega_{cc}^{+}\to\Xi_{c}^{\prime+}\pi^{0})+\mathcal{M}(\pi^{+},\Xi_{c}^{(\prime)0};\rho^{+})+\mathcal{M}(\rho^{+},\Xi_{c}^{(\prime)0};\pi^{+})\nonumber \\
 & +\mathcal{M}(K^{+},\Omega_{c}^{0};K^{\ast+})+\mathcal{M}(K^{\ast+},\Omega_{c}^{0};K^{+})+\mathcal{M}(\rho^{+},\Xi_{c}^{(\prime)0};\rho^{+})+\mathcal{M}(K^{\ast+},\Omega_{c}^{0};K^{\ast+})\nonumber \\
 & +\mathcal{M}(\Xi_{c}^{(\prime)0},\pi^{+};\Xi_{c}^{(\prime)0})+\mathcal{M}(\Xi_{c}^{(\prime)0},\rho^{+};\Xi_{c}^{(\prime)0})+\mathcal{M}(\Xi_{c}^{(\prime)0},\pi^{+};\Xi_{c}^{\prime*0})+\mathcal{M}(\Xi_{c}^{(\prime)0},\rho^{+};\Xi_{c}^{\prime*0}),\\
\mathcal{A}(\Omega_{cc}^{+}\to\Xi_{c}^{\prime+}\eta_{1}) & =\mathcal{C}_{SD}(\Omega_{cc}^{+}\to\Xi_{c}^{\prime+}\eta_{1})+\mathcal{M}(\Omega_{c}^{0},K^{+};\Omega_{c}^{0})+\mathcal{M}(\Omega_{c}^{0},K^{\ast+};\Omega_{c}^{0})+\mathcal{M}(\Xi_{c}^{0},\pi^{+};\Xi_{c}^{0})\nonumber \\
 & +\mathcal{M}(\Xi_{c}^{\prime0},\pi^{+};\Xi_{c}^{\prime0})+\mathcal{M}(\Xi_{c}^{0},\rho^{+};\Xi_{c}^{0})+\mathcal{M}(\Xi_{c}^{\prime0},\rho^{+};\Xi_{c}^{\prime0})\nonumber \\
 & +\mathcal{M}(\Omega_{c}^{0},K^{+};\Omega_{c}^{*0})+\mathcal{M}(\Omega_{c}^{0},K^{\ast+};\Omega_{c}^{*0})+\mathcal{M}(\Xi_{c}^{\prime0},\pi^{+};\Xi_{c}^{\prime*0})+\mathcal{M}(\Xi_{c}^{\prime0},\rho^{+};\Xi_{c}^{\prime*0}),
\\
\mathcal{A}(\Omega_{cc}^{+}\to\Xi_{c}^{\prime+}\eta_{8}) & =\mathcal{C}_{SD}(\Omega_{cc}^{+}\to\Xi_{c}^{\prime+}\eta_{8})+\mathcal{M}(K^{+},\Omega_{c}^{0};K^{\ast+})+\mathcal{M}(K^{\ast+},\Omega_{c}^{0};K^{+})+\mathcal{M}(K^{\ast+},\Omega_{c}^{0};K^{\ast+})\nonumber \\
 & +\mathcal{M}(\Omega_{c}^{0},K^{+};\Omega_{c}^{0})+\mathcal{M}(\Omega_{c}^{0},K^{\ast+};\Omega_{c}^{0})+\mathcal{M}(\Xi_{c}^{(\prime)0},\pi^{+};\Xi_{c}^{(\prime)0})+\mathcal{M}(\Xi_{c}^{(\prime)0},\rho^{+};\Xi_{c}^{(\prime)0})\nonumber \\
 & +\mathcal{M}(\Omega_{c}^{0},K^{+};\Omega_{c}^{*0})+\mathcal{M}(\Omega_{c}^{0},K^{\ast+};\Omega_{c}^{*0})+\mathcal{M}(\Xi_{c}^{(\prime)0},\pi^{+};\Xi_{c}^{\prime*0})+\mathcal{M}(\Xi_{c}^{(\prime)0},\rho^{+};\Xi_{c}^{\prime*0}).
\end{align}
 The amplitudes for long-distance dominated doubly Cabibbo suppressed
can be given as follows.  
\begin{align}
\mathcal{A}(\Xi_{cc}^{++}\to\Sigma_{c}^{++}K^{0}) & =\mathcal{C}_{SD}(\Xi_{cc}^{++}\to\Sigma_{c}^{++}K^{0})+\mathcal{M}(K^{\ast+},\Lambda_{c}^{+}/\Sigma_{c}^{+};\pi^{+})+\mathcal{M}(K^{+},\Lambda_{c}^{+}/\Sigma_{c}^{+};\rho^{+})\nonumber \\
 & +\mathcal{M}(K^{\ast+},\Lambda_{c}^{+}/\Sigma_{c}^{+};\rho^{+})+\mathcal{M}(\Lambda_{c}^{+}/\Sigma_{c}^{+},K^{+};\Xi_{c}^{(\prime)+})+\mathcal{M}(\Lambda_{c}^{+}/\Sigma_{c}^{+},K^{\ast+};\Xi_{c}^{(\prime)+})\nonumber\\
 & +\mathcal{M}(\Lambda_{c}^{+}/\Sigma_{c}^{+},K^{+};\Xi_{c}^{\prime*+})+\mathcal{M}(\Lambda_{c}^{+}/\Sigma_{c}^{+},K^{\ast+};\Xi_{c}^{\prime*+}),\\
\mathcal{A}(\Xi_{cc}^{+}\to\Sigma_{c}^{+}K^{0}) & =\mathcal{C}_{SD}(\Xi_{cc}^{+}\to\Sigma_{c}^{+}K^{0})+\mathcal{M}(K^{+},\Sigma_{c}^{0};\rho^{+})+\mathcal{M}(K^{\ast+},\Sigma_{c}^{0};\pi^{+})+\mathcal{M}(K^{\ast+},\Sigma_{c}^{0};\rho^{+})\nonumber \\
 & +\mathcal{M}(\Sigma_{c}^{0},K^{+};\Xi_{c}^{(\prime)0})+\mathcal{M}(\Sigma_{c}^{0},K^{\ast+};\Xi_{c}^{(\prime)0})+\mathcal{M}(\Sigma_{c}^{0},K^{+};\Xi_{c}^{\prime*0})+\mathcal{M}(\Sigma_{c}^{0},K^{\ast+};\Xi_{c}^{\prime*0}),\\
\mathcal{A}(\Xi_{cc}^{+}\to\Lambda_{c}^{+}K^{0}) & =\mathcal{C}_{SD}(\Xi_{cc}^{+}\to\Lambda_{c}^{+}K^{0})+\mathcal{M}(K^{+},\Sigma_{c}^{0};\rho^{+})+\mathcal{M}(K^{\ast+},\Sigma_{c}^{0};\pi^{+})+\mathcal{M}(K^{\ast+},\Sigma_{c}^{0};\rho^{+})\nonumber \\
 & +\mathcal{M}(\Sigma_{c}^{0},K^{+};\Xi_{c}^{(\prime)0})+\mathcal{M}(\Sigma_{c}^{0},K^{\ast+};\Xi_{c}^{(\prime)0})+\mathcal{M}(\Sigma_{c}^{0},K^{+};\Xi_{c}^{\prime*0})+\mathcal{M}(\Sigma_{c}^{0},K^{\ast+};\Xi_{c}^{\prime*0}),\\
\mathcal{A}(\Omega_{cc}^{+}\to\Sigma_{c}^{+}\pi^{0}) & =\mathcal{M}(K^{+},\Xi_{c}^{(\prime)0};K^{\ast+})+\mathcal{M}(K^{\ast+},\Xi_{c}^{(\prime)0};K^{+})+\mathcal{M}(K^{\ast+},\Xi_{c}^{(\prime)0};K^{\ast+})\nonumber \\
 & +\mathcal{M}(\Xi_{c}^{(\prime)0},K^{+};\Xi_{c}^{(\prime)0})+\mathcal{M}(\Xi_{c}^{(\prime)0},K^{\ast+};\Xi_{c}^{(\prime)0})+\mathcal{M}(\Xi_{c}^{(\prime)0},K^{+};\Xi_{c}^{\prime*0})+\mathcal{M}(\Xi_{c}^{(\prime)0},K^{\ast+};\Xi_{c}^{\prime*0}),\\
\mathcal{A}(\Omega_{cc}^{+}\to\Sigma_{c}^{+}\eta_{1}) & =\mathcal{C}_{SD}(\Omega_{cc}^{+}\to\Sigma_{c}^{+}\eta_{1})\nonumber \\
 & +\mathcal{M}(\Xi_{c}^{0},K^{+};\Xi_{c}^{0})+\mathcal{M}(\Xi_{c}^{\prime0},K^{+};\Xi_{c}^{\prime0})+\mathcal{M}(\Xi_{c}^{0},K^{\ast+};\Xi_{c}^{0})+\mathcal{M}(\Xi_{c}^{\prime0},K^{\ast+};\Xi_{c}^{\prime0})\nonumber \\
 & +\mathcal{M}(\Xi_{c}^{\prime0},K^{+};\Xi_{c}^{\prime*0})+\mathcal{M}(\Xi_{c}^{\prime0},K^{\ast+};\Xi_{c}^{\prime*0}),\\
\mathcal{A}(\Omega_{cc}^{+}\to\Sigma_{c}^{+}\eta_{8}) & =\mathcal{C}_{SD}(\Omega_{cc}^{+}\to\Sigma_{c}^{+}\eta_{8})+\mathcal{M}(K^{+},\Xi_{c}^{(\prime)0};K^{\ast+})+\mathcal{M}(K^{\ast+},\Xi_{c}^{(\prime)0};K^{+})+\mathcal{M}(K^{\ast+},\Xi_{c}^{(\prime)0};K^{\ast+})\nonumber \\
 & +\mathcal{M}(\Xi_{c}^{(\prime)0},K^{+};\Xi_{c}^{(\prime)0})+\mathcal{M}(\Xi_{c}^{(\prime)0},K^{\ast+};\Xi_{c}^{(\prime)0})+\mathcal{M}(\Xi_{c}^{(\prime)0},K^{+};\Xi_{c}^{\prime*0})+\mathcal{M}(\Xi_{c}^{(\prime)0},K^{\ast+};\Xi_{c}^{\prime*0}),\\
\mathcal{A}(\Omega_{cc}^{+}\to\Lambda_{c}^{+}\pi^{0}) & =+\mathcal{M}(K^{+},\Xi_{c}^{(\prime)0};K^{\ast+})+\mathcal{M}(K^{\ast+},\Xi_{c}^{(\prime)0};K^{+})+\mathcal{M}(K^{\ast+},\Xi_{c}^{(\prime)0};K^{\ast+})\nonumber \\
 & +\mathcal{M}(\Xi_{c}^{(\prime)0},K^{+};\Xi_{c}^{(\prime)0})+\mathcal{M}(\Xi_{c}^{(\prime)0},K^{\ast+};\Xi_{c}^{(\prime)0})+\mathcal{M}(\Xi_{c}^{(\prime)0},K^{+};\Xi_{c}^{\prime*0})+\mathcal{M}(\Xi_{c}^{(\prime)0},K^{\ast+};\Xi_{c}^{\prime*0}),\\
\mathcal{A}(\Omega_{cc}^{+}\to\Lambda_{c}^{+}\eta_{1}) & =\mathcal{C}_{SD}(\Omega_{cc}^{+}\to\Lambda_{c}^{+}\eta_{1})\nonumber \\
 & +\mathcal{M}(\Xi_{c}^{0},K^{+};\Xi_{c}^{0})+\mathcal{M}(\Xi_{c}^{\prime0},K^{+};\Xi_{c}^{\prime0})+\mathcal{M}(\Xi_{c}^{0},K^{\ast+};\Xi_{c}^{0})+\mathcal{M}(\Xi_{c}^{\prime0},K^{\ast+};\Xi_{c}^{\prime0})\nonumber \\
 & +\mathcal{M}(\Xi_{c}^{\prime0},K^{+};\Xi_{c}^{\prime*0})+\mathcal{M}(\Xi_{c}^{\prime0},K^{\ast+};\Xi_{c}^{\prime*0}),\\
\mathcal{A}(\Omega_{cc}^{+}\to\Lambda_{c}^{+}\eta_{8}) & =\mathcal{C}_{SD}(\Omega_{cc}^{+}\to\Lambda_{c}^{+}\eta_{8})+\mathcal{M}(K^{+},\Xi_{c}^{(\prime)0};K^{\ast+})+\mathcal{M}(K^{\ast+},\Xi_{c}^{(\prime)0};K^{+})+\mathcal{M}(K^{\ast+},\Xi_{c}^{(\prime)0};K^{\ast+})\nonumber \\
 & +\mathcal{M}(\Xi_{c}^{(\prime)0},K^{+};\Xi_{c}^{(\prime)0})+\mathcal{M}(\Xi_{c}^{(\prime)0},K^{\ast+};\Xi_{c}^{(\prime)0})+\mathcal{M}(\Xi_{c}^{(\prime)0},K^{+};\Xi_{c}^{\prime*0})+\mathcal{M}(\Xi_{c}^{(\prime)0},K^{\ast+};\Xi_{c}^{\prime*0}),\\
\mathcal{A}(\Omega_{cc}^{+}\to\Sigma_{c}^{0}\pi^{+}) & =\mathcal{M}(K^{+},\Xi_{c}^{(\prime)0};K^{\ast0})+\mathcal{M}(K^{\ast+},\Xi_{c}^{(\prime)0};K^{0})+\mathcal{M}(K^{\ast+},\Xi_{c}^{(\prime)0};K^{\ast0}),\\
\mathcal{A}(\Omega_{cc}^{+}\to\Sigma_{c}^{++}\pi^{-}) & =\mathcal{M}(\Xi_{c}^{(\prime)0},K^{+};\Xi_{c}^{(\prime)+})+\mathcal{M}(\Xi_{c}^{(\prime)0},K^{\ast+};\Xi_{c}^{(\prime)+})+\mathcal{M}(\Xi_{c}^{(\prime)0},K^{+};\Xi_{c}^{\prime*+})+\mathcal{M}(\Xi_{c}^{(\prime)0},K^{\ast+};\Xi_{c}^{\prime*+}),\\
\mathcal{A}(\Omega_{cc}^{+}\to\Xi_{c}^{+}K^{0}) & =\mathcal{C}_{SD}(\Omega_{cc}^{+}\to\Xi_{c}^{+}K^{0})+\mathcal{M}(K^{+},\Xi_{c}^{(\prime)0};\rho^{+})+\mathcal{M}(K^{\ast+},\Xi_{c}^{(\prime)0};\pi^{+})+\mathcal{M}(K^{\ast+},\Xi_{c}^{(\prime)0};\rho^{+})\nonumber \\
 & +\mathcal{M}(\Xi_{c}^{(\prime)0},K^{+};\Omega_{c}^{0})+\mathcal{M}(\Xi_{c}^{(\prime)0},K^{\ast+};\Omega_{c}^{0})+\mathcal{M}(\Xi_{c}^{(\prime)0},K^{+};\Omega_{c}^{*0})+\mathcal{M}(\Xi_{c}^{(\prime)0},K^{\ast+};\Omega_{c}^{*0}),\\
\mathcal{A}(\Omega_{cc}^{+}\to\Xi_{c}^{\prime+}K^{0}) & =\mathcal{C}_{SD}(\Omega_{cc}^{+}\to\Xi_{c}^{(\prime)+}K^{0})+\mathcal{M}(K^{+},\Xi_{c}^{(\prime)0};\rho^{+})+\mathcal{M}(K^{\ast+},\Xi_{c}^{(\prime)0};\pi^{+})+\mathcal{M}(K^{\ast+},\Xi_{c}^{(\prime)0};\rho^{+})\nonumber \\
 & +\mathcal{M}(\Xi_{c}^{(\prime)0},K^{+};\Omega_{c}^{0})+\mathcal{M}(\Xi_{c}^{(\prime)0},K^{\ast+};\Omega_{c}^{0})+\mathcal{M}(\Xi_{c}^{(\prime)0},K^{+};\Omega_{c}^{*0})+\mathcal{M}(\Xi_{c}^{(\prime)0},K^{\ast+};\Omega_{c}^{*0}).
\end{align}}

\section{Strong couplings}
\label{app:strong}
In general, meson-baryon and meson-meson strong couplings can be obtained from the SU(3)-invariant strong
Hamiltonian. In the present work, we use the SU(3) flavor symmetry to analysis the vector-vector-pseudoscalar (VVP), vector-pseudoscalar-pseudoscalar (VPP), baryon-baryon-pseudoscalar ($BBP/BB^{*}P$) and baryon-baryon-vector ($BBV/BB^{*}V$) couplings. 
Under SU(3) flavor symmetry, the pseudoscalar and vector meson can be written as follows,
\begin{align}
P(J^{P}=0^{-})=\left(\begin{array}{ccc}
\frac{\pi^{0}}{\sqrt{2}}+\frac{\eta_{8}}{\sqrt{6}} & \pi^{+} & K^{+}\\
\pi^{-} & -\frac{\pi^{0}}{\sqrt{2}}+\frac{\eta_{8}}{\sqrt{6}} & K^{0}\\
K^{-} & \overline{K}^{0} & -\sqrt{\frac{2}{3}}\eta_{8}
\end{array}\right)+\frac{1}{\sqrt{3}}\left(\begin{array}{ccc}
\eta_{1} & 0 & 0\\
0 & \eta_{1} & 0\\
0 & 0 & \eta_{1}
\end{array}\right),
\end{align}
\begin{align}
V(J^{P}=1^{-})=\left(\begin{array}{ccc}
\frac{\rho^{0}}{\sqrt{2}}+\frac{\omega}{\sqrt{2}} & \rho^{+} & K^{\ast+}\\
\rho^{-} & -\frac{\rho^{0}}{\sqrt{2}}+\frac{\omega}{\sqrt{2}} & K^{\ast0}\\
K^{\ast-} & \overline{K}^{\ast0} & \phi
\end{array}\right).\label{eq:pv}
\end{align}
And the spin-parity $1/2^{+}$ singly charmed baryons include anti-triplet ${\cal B}_{\mathbf{\bar 3}}$ and sextet ${\cal B}_{\mathbf{6}}$, which can given as follows,
\begin{align}
B_{6}(J^{P}=\frac{1}{2}^{+})=\left(\begin{array}{ccc}
\Sigma_{c}^{++} & \frac{\Sigma_{c}^{+}}{\sqrt{2}} & \frac{\Xi_{c}^{\prime+}}{\sqrt{2}}\\
\frac{\Sigma_{c}^{+}}{\sqrt{2}} & \Sigma_{c}^{0} & \frac{\Xi_{c}^{\prime0}}{\sqrt{2}}\\
\frac{\Xi_{c}^{\prime+}}{\sqrt{2}} & \frac{\Xi_{c}^{\prime0}}{\sqrt{2}} & \Omega_{c}^{0}
\end{array}\right),\quad B_{\bar{3}}(J^{P}=\frac{1}{2}^{+})=\left(\begin{array}{ccc}
0 & \Lambda_{c}^{+} & \Xi_{c}^{+}\\
-\Lambda_{c}^{+} & 0 & \Xi_{c}^{0}\\
-\Xi_{c}^{+} & -\Xi_{c}^{0} & 0
\end{array}\right).\label{eq:b36}
\end{align}
In addition, the spin-parity $3/2^{+}$ singly charmed baryons sextet ${\cal B}_{\mathbf{6}}$ can given as follows,
\begin{align}
B_{6}^{*}(J^{P}=\frac{3}{2}^{+})=\left(\begin{array}{ccc}
\Sigma_{c}^{*++} & \frac{\Sigma_{c}^{*+}}{\sqrt{2}} & \frac{\Xi_{c}^{*\prime+}}{\sqrt{2}}\\
\frac{\Sigma_{c}^{*+}}{\sqrt{2}} & \Sigma_{c}^{*0} & \frac{\Xi_{c}^{*\prime0}}{\sqrt{2}}\\
\frac{\Xi_{c}^{*\prime+}}{\sqrt{2}} & \frac{\Xi_{c}^{*\prime0}}{\sqrt{2}} & \Omega_{c}^{*0}
\end{array}\right).\label{eq:bs6}
\end{align}
Then with the help of Eqs.~(\ref{eq:pv})-(\ref{eq:bs6}), we can obtain the relationships of these strong couplings, as shown in Tabs.~\ref{tab:VPP}-\ref{tab:VBB}.
\begin{table}
  \caption{The strong coupling constants between the vector mesons and two pseudoscalar mesons $g_{VPP}$ can be obtained by using the coupling $g_{\rho\pi\pi}=6.05=-\frac{g_{VPP}}{\sqrt{2}}$~\cite{Cheng:2004ru} and SU(3) relations.}
  \label{tab:VPP}
  \begin{tabular}{cc|cc|cc|cc}
    \hline\hline
  Vertex&g&Vertex&g&Vertex&g&Vertex&g\\\hline
  $\rho^0    \pi^+   \pi^-  $ & $ -\frac{g_{VPP}}{\sqrt{2}}$&
  $\rho^0    \pi^0   \eta_8  $ & $ \frac{g_{VPP}}{\sqrt{6}}$&
  $\rho^0    \pi^0   \eta_{1}  $ & $ \frac{g_{VPP}}{\sqrt{3}}$&
  $\rho^0    \pi^-   \pi^+  $ & $ \frac{g_{VPP}}{\sqrt{2}}$\\\hline
  $\rho^0    \overline K^0   K^0  $ & $ -\frac{g_{VPP}}{\sqrt{2}}$&
  $\rho^0    K^-   K^+  $ & $ \frac{g_{VPP}}{\sqrt{2}}$&
  $\rho^0    \eta_8   \pi^0  $ & $ \frac{g_{VPP}}{\sqrt{6}}$&
  $\rho^0    \eta_{1}   \pi^0  $ & $ \frac{g_{VPP}}{\sqrt{3}}$\\\hline
  $\omega    \pi^+   \pi^-  $ & $ \frac{g_{VPP}}{\sqrt{2}}$&
  $\omega    \pi^0   \pi^0  $ & $ \frac{g_{VPP}}{\sqrt{2}}$&
  $\omega    \pi^-   \pi^+  $ & $ \frac{g_{VPP}}{\sqrt{2}}$&
  $\omega    \overline K^0   K^0  $ & $ \frac{g_{VPP}}{\sqrt{2}}$\\\hline
  $\omega    K^-   K^+  $ & $ \frac{g_{VPP}}{\sqrt{2}}$&
  $\omega    \eta_8   \eta_8  $ & $ \frac{g_{VPP}}{3 \sqrt{2}}$&
  $\omega    \eta_8   \eta_{1}  $ & $ \frac{g_{VPP}}{3}$&
  $\omega    \eta_{1}   \eta_8  $ & $ \frac{g_{VPP}}{3}$\\\hline
  $\omega    \eta_{1}   \eta_{1}  $ & $ \frac{\sqrt{2} g_{VPP}}{3}$&
  $\rho^+    \pi^+   \pi^0  $ & $ \frac{g_{VPP}}{\sqrt{2}}$&
  $\rho^+    \pi^+   \eta_8  $ & $ \frac{g_{VPP}}{\sqrt{6}}$&
  $\rho^+    \pi^+   \eta_{1}  $ & $ \frac{g_{VPP}}{\sqrt{3}}$\\\hline
  $\rho^+    \pi^0   \pi^+  $ & $ -\frac{g_{VPP}}{\sqrt{2}}$&
  $\rho^+    \overline K^0   K^+  $ & $ g_{VPP}$&
  $\rho^+    \eta_8   \pi^+  $ & $ \frac{g_{VPP}}{\sqrt{6}}$&
  $\rho^+    \eta_{1}   \pi^+  $ & $ \frac{g_{VPP}}{\sqrt{3}}$\\\hline
  $K^{*+}    K^+   \pi^0  $ & $ \frac{g_{VPP}}{\sqrt{2}}$&
  $K^{*+}    K^+   \eta_8  $ & $ \frac{g_{VPP}}{\sqrt{6}}$&
  $K^{*+}    K^+   \eta_{1}  $ & $ \frac{g_{VPP}}{\sqrt{3}}$&
  $K^{*+}    K^0   \pi^+  $ & $ g_{VPP}$\\\hline
  $K^{*+}    \eta_8   K^+  $ & $ -\sqrt{\frac{2}{3}} g_{VPP}$&
  $K^{*+}    \eta_{1}   K^+  $ & $ \frac{g_{VPP}}{\sqrt{3}}$&
  $\rho^-    \pi^0   \pi^-  $ & $ \frac{g_{VPP}}{\sqrt{2}}$&
  $\rho^-    \pi^-   \pi^0  $ & $ -\frac{g_{VPP}}{\sqrt{2}}$\\\hline
  $\rho^-    \pi^-   \eta_8  $ & $ \frac{g_{VPP}}{\sqrt{6}}$&
  $\rho^-    \pi^-   \eta_{1}  $ & $ \frac{g_{VPP}}{\sqrt{3}}$&
  $\rho^-    K^-   K^0  $ & $ g_{VPP}$&
  $\rho^-    \eta_8   \pi^-  $ & $ \frac{g_{VPP}}{\sqrt{6}}$\\\hline
  $\rho^-    \eta_{1}   \pi^-  $ & $ \frac{g_{VPP}}{\sqrt{3}}$&
  $K^{*0}    K^+   \pi^-  $ & $ g_{VPP}$&
  $K^{*0}    K^0   \pi^0  $ & $ -\frac{g_{VPP}}{\sqrt{2}}$&
  $K^{*0}    K^0   \eta_8  $ & $ \frac{g_{VPP}}{\sqrt{6}}$\\\hline
  $K^{*0}    K^0   \eta_{1}  $ & $ \frac{g_{VPP}}{\sqrt{3}}$&
  $K^{*0}    \eta_8   K^0  $ & $ -\sqrt{\frac{2}{3}} g_{VPP}$&
  $K^{*0}    \eta_{1}   K^0  $ & $ \frac{g_{VPP}}{\sqrt{3}}$&
  $K^{*-}    \pi^0   K^-  $ & $ \frac{g_{VPP}}{\sqrt{2}}$\\\hline
  $K^{*-}    \pi^-   \overline K^0  $ & $ g_{VPP}$&
  $K^{*-}    K^-   \eta_8  $ & $ -\sqrt{\frac{2}{3}} g_{VPP}$&
  $K^{*-}    K^-   \eta_{1}  $ & $ \frac{g_{VPP}}{\sqrt{3}}$&
  $K^{*-}    \eta_8   K^-  $ & $ \frac{g_{VPP}}{\sqrt{6}}$\\\hline
  $K^{*-}    \eta_{1}   K^-  $ & $ \frac{g_{VPP}}{\sqrt{3}}$&
  $\overline{K}^{*0}   \pi^+   K^-  $ & $ g_{VPP}$&
  $\overline{K}^{*0}   \pi^0   \overline K^0  $ & $ -\frac{g_{VPP}}{\sqrt{2}}$&
  $\overline{K}^{*0}   \overline K^0   \eta_8  $ & $ -\sqrt{\frac{2}{3}} g_{VPP}$\\\hline
  $\overline{K}^{*0}   \overline K^0   \eta_{1}  $ & $ \frac{g_{VPP}}{\sqrt{3}}$&
  $\overline{K}^{*0}   \eta_8   \overline K^0  $ & $ \frac{g_{VPP}}{\sqrt{6}}$&
  $\overline{K}^{*0}   \eta_{1}   \overline K^0  $ & $ \frac{g_{VPP}}{\sqrt{3}}$&
  $\phi   K^+   K^-  $ & $ g_{VPP}$\\\hline
  $\phi   K^0   \overline K^0  $ & $ g_{VPP}$&
  $\phi   \eta_8   \eta_8  $ & $ \frac{2 g_{VPP}}{3}$&
  $\phi   \eta_8   \eta_{1}  $ & $ -\frac{1}{3} \sqrt{2} g_{VPP}$&
  $\phi   \eta_{1}   \eta_8  $ & $ -\frac{1}{3} \sqrt{2} g_{VPP}$\\\hline
  $\phi   \eta_{1}   \eta_{1}  $ & $ \frac{g_{VPP}}{3}$\\
  \hline\hline
  \end{tabular}
\end{table}
\begin{table}
  \caption{The strong couplings between the pseudoscalar meson (P) and two vector mesons (V) can be obtained by using the coupling constant $g_{\omega \rho \pi}=g_{VVP}/\sqrt{2}=-10$~\cite{Nakayama:2006ps} and SU(3) relationships.}
  \label{tab:VVP}
  \begin{tabular}{cc|cc|cc|cc}
    \hline\hline
    Vertex&g&Vertex&g&Vertex&g&Vertex&g\\\hline
$\pi^+    \rho^0   \rho^+  $ & $ -\frac{g_{VVP}}{\sqrt{2}}$&
$\pi^+    \omega   \rho^+  $ & $ \frac{g_{VVP}}{\sqrt{2}}$&
$\pi^+    \rho^+   \rho^0  $ & $ \frac{g_{VVP}}{\sqrt{2}}$&
$\pi^+    \rho^+   \omega  $ & $ \frac{g_{VVP}}{\sqrt{2}}$\\\hline
$\pi^+    \overline{K}^{*0}  K^{*+}  $ & $ g_{VVP}$&
$\pi^0    \rho^0   \omega  $ & $ \frac{g_{VVP}}{\sqrt{2}}$&
$\pi^0    \omega   \rho^0  $ & $ \frac{g_{VVP}}{\sqrt{2}}$&
$\pi^0    \rho^+   \rho^-  $ & $ -\frac{g_{VVP}}{\sqrt{2}}$\\\hline
$\pi^0    \rho^-   \rho^+  $ & $ \frac{g_{VVP}}{\sqrt{2}}$&
$\pi^0    K^{*-}   K^{*+}  $ & $ \frac{g_{VVP}}{\sqrt{2}}$&
$\pi^0    \overline{K}^{*0}  K^{*0}  $ & $ -\frac{g_{VVP}}{\sqrt{2}}$&
$\pi^-    \rho^0   \rho^-  $ & $ \frac{g_{VVP}}{\sqrt{2}}$\\\hline
$\pi^-    \omega   \rho^-  $ & $ \frac{g_{VVP}}{\sqrt{2}}$&
$\pi^-    \rho^-   \rho^0  $ & $ -\frac{g_{VVP}}{\sqrt{2}}$&
$\pi^-    \rho^-   \omega  $ & $ \frac{g_{VVP}}{\sqrt{2}}$&
$\pi^-    K^{*-}   K^{*0}  $ & $ g_{VVP}$\\\hline
$K^+    K^{*+}   \rho^0  $ & $ \frac{g_{VVP}}{\sqrt{2}}$&
$K^+    K^{*+}   \omega  $ & $ \frac{g_{VVP}}{\sqrt{2}}$&
$K^+    K^{*0}   \rho^+  $ & $ g_{VVP}$&
$K^+    \phi  K^{*+}  $ & $ g_{VVP}$\\\hline
$K^0    K^{*+}   \rho^-  $ & $ g_{VVP}$&
$K^0    K^{*0}   \rho^0  $ & $ -\frac{g_{VVP}}{\sqrt{2}}$&
$K^0    K^{*0}   \omega  $ & $ \frac{g_{VVP}}{\sqrt{2}}$&
$K^0    \phi  K^{*0}  $ & $ g_{VVP}$\\\hline
$\overline K^0    \rho^0   \overline{K}^{*0} $ & $ -\frac{g_{VVP}}{\sqrt{2}}$&
$\overline K^0    \omega   \overline{K}^{*0} $ & $ \frac{g_{VVP}}{\sqrt{2}}$&
$\overline K^0    \rho^+   K^{*-}  $ & $ g_{VVP}$&
$\overline K^0    \overline{K}^{*0}  \phi $ & $ g_{VVP}$\\\hline
$K^-    \rho^0   K^{*-}  $ & $ \frac{g_{VVP}}{\sqrt{2}}$&
$K^-    \omega   K^{*-}  $ & $ \frac{g_{VVP}}{\sqrt{2}}$&
$K^-    \rho^-   \overline{K}^{*0} $ & $ g_{VVP}$&
$K^-    K^{*-}   \phi $ & $ g_{VVP}$\\\hline
$\eta_8    \rho^0   \rho^0  $ & $ \frac{g_{VVP}}{\sqrt{6}}$&
$\eta_8    \omega   \omega  $ & $ \frac{g_{VVP}}{\sqrt{6}}$&
$\eta_8    \rho^+   \rho^-  $ & $ \frac{g_{VVP}}{\sqrt{6}}$&
$\eta_8    K^{*+}   K^{*-}  $ & $ -\sqrt{\frac{2}{3}} g_{VVP}$\\\hline
$\eta_8    \rho^-   \rho^+  $ & $ \frac{g_{VVP}}{\sqrt{6}}$&
$\eta_8    K^{*0}   \overline{K}^{*0} $ & $ -\sqrt{\frac{2}{3}} g_{VVP}$&
$\eta_8    K^{*-}   K^{*+}  $ & $ \frac{g_{VVP}}{\sqrt{6}}$&
$\eta_8    \overline{K}^{*0}  K^{*0}  $ & $ \frac{g_{VVP}}{\sqrt{6}}$\\\hline
$\eta_8    \phi  \phi $ & $ -\sqrt{\frac{2}{3}} g_{VVP}$&
$\eta_{1}    \rho^0   \rho^0  $ & $ \frac{g_{VVP}}{\sqrt{3}}$&
$\eta_{1}    \omega   \omega  $ & $ \frac{g_{VVP}}{\sqrt{3}}$&
$\eta_{1}    \rho^+   \rho^-  $ & $ \frac{g_{VVP}}{\sqrt{3}}$\\\hline
$\eta_{1}    K^{*+}   K^{*-}  $ & $ \frac{g_{VVP}}{\sqrt{3}}$&
$\eta_{1}    \rho^-   \rho^+  $ & $ \frac{g_{VVP}}{\sqrt{3}}$&
$\eta_{1}    K^{*0}   \overline{K}^{*0} $ & $ \frac{g_{VVP}}{\sqrt{3}}$&
$\eta_{1}    K^{*-}   K^{*+}  $ & $ \frac{g_{VVP}}{\sqrt{3}}$\\\hline
$\eta_{1}    \overline{K}^{*0}  K^{*0}  $ & $ \frac{g_{VVP}}{\sqrt{3}}$&
$\eta_{1}    \phi  \phi $ & $ \frac{g_{VVP}}{\sqrt{3}}$\\\hline\hline
    \end{tabular}
\end{table}
\begin{table}
    \caption{The spin $1/2$ singly charmed baryons ground states include anti-triplet ${\cal B}_{\mathbf{\bar 3}}$ and sextet ${\cal B}_{\mathbf{6}}$. The strong couplings of ${\cal B}_{\mathbf{\bar 3}}{\cal B}_{\mathbf{\bar 3}}P$, ${\cal B}_{\mathbf{6}}{\cal B}_{\mathbf{\bar 3}}P$ and ${\cal B}_{\mathbf{6}}{\cal B}_{\mathbf{6}}P$ can be obtained by using the coupling constant $g_{\Xi_{c}^{+}\Xi_{c}^{+}\pi^{0}} = 0.7=\frac{g_{PBB1}}{\sqrt{2}}$, $g_{\Xi_{c}^{\prime+}\Xi_{c}^{+}\pi^{0}} = 3.1=\frac{g_{PBB2}}{2}$ and ${\Sigma_{c}^{+}\Sigma_{c}^{0}\pi^{+}} = 8.0=\frac{g_{PBB3}}{\sqrt{2}}$~\cite{Aliev:2010yx}, respectively. Since the matrix forms of the spin-3/2 singly charmed baryon ground state ${\cal B}_{\mathbf{6}}^{*}$ and the 1/2 sextet ${\cal B}_{\mathbf{6}}$ under symmetry are similar, the derived relations are identical. Then the couplings ${\cal B}_{\mathbf{6}}^{*}{\cal B}_{\mathbf{\bar 3}}P$ and ${\cal B}_{\mathbf{6}}^{*}{\cal B}_{\mathbf{6}}P$ can be get from $g_{\Sigma_{c}^{*0}\Lambda_{c}^{+}\pi^{-}} = 3.9={g_{PB^{*}B2}}$ and $g_{\Sigma_{c}^{*+}\Sigma_{c}^{0}\pi^{+}} = 4.3=\frac{g_{PB^{*}B3}}{\sqrt{2}}$~\cite{Aliev:2010ev}, respectively.}
    \label{tab:PBB}
    \begin{tabular}{cc|cc|cc|cc}\hline\hline
    Vertex&g&Vertex&g&Vertex&g&Vertex&g\\\hline
    $\Lambda_c^+   \Lambda_c^+  \eta_8  $ & $ \sqrt{\frac{2}{3}} g_{PBB1}$&
    $\Lambda_c^+   \Lambda_c^+  \eta_{1}  $ & $ \frac{2 g_{PBB1}}{\sqrt{3}}$&
    $\Lambda_c^+   \Xi_c^+  K^0  $ & $ g_{PBB1}$&
    $\Lambda_c^+   \Xi_c^0  K^+  $ & $ -g_{PBB1}$\\\hline
    $\Xi_c^+   \Lambda_c^+  \overline K^0  $ & $ g_{PBB1}$&
    $\Xi_c^+   \Xi_c^+  \pi^0  $ & $ \frac{g_{PBB1}}{\sqrt{2}}$&
    $\Xi_c^+   \Xi_c^+  \eta_8  $ & $ -\frac{g_{PBB1}}{\sqrt{6}}$&
    $\Xi_c^+   \Xi_c^+  \eta_{1}  $ & $ \frac{2 g_{PBB1}}{\sqrt{3}}$\\\hline
    $\Xi_c^+   \Xi_c^0  \pi^+  $ & $ g_{PBB1}$&
    $\Xi_c^0   \Lambda_c^+  K^-  $ & $ -g_{PBB1}$&
    $\Xi_c^0   \Xi_c^+  \pi^-  $ & $ g_{PBB1}$&
    $\Xi_c^0   \Xi_c^0  \pi^0  $ & $ -\frac{g_{PBB1}}{\sqrt{2}}$\\\hline
    $\Xi_c^0   \Xi_c^0  \eta_8  $ & $ -\frac{g_{PBB1}}{\sqrt{6}}$&
    $\Xi_c^0   \Xi_c^0  \eta_{1}  $ & $ \frac{2 g_{PBB1}}{\sqrt{3}}$&\\
    \hline
    $\Sigma_{c}^{++}   \Lambda_c^+  \pi^+  $ & $ -g_{PBB2}$&
    $\Sigma_{c}^{++}   \Xi_c^+  K^+  $ & $ -g_{PBB2}$&
    $\Sigma_{c}^{+}   \Lambda_c^+  \pi^0  $ & $ g_{PBB2}$&
    $\Sigma_{c}^{+}   \Xi_c^+  K^0  $ & $ -\frac{g_{PBB2}}{\sqrt{2}}$\\\hline
    $\Sigma_{c}^{+}   \Xi_c^0  K^+  $ & $ -\frac{g_{PBB2}}{\sqrt{2}}$&
    $\Sigma_{c}^{0}   \Lambda_c^+  \pi^-  $ & $ g_{PBB2}$&
    $\Sigma_{c}^{0}   \Xi_c^0  K^0  $ & $ -g_{PBB2}$&
    $\Xi_{c}^{\prime+}   \Lambda_c^+  \overline K^0  $ & $ -\frac{g_{PBB2}}{\sqrt{2}}$\\\hline
    $\Xi_{c}^{\prime+}   \Xi_c^+  \pi^0  $ & $ \frac{g_{PBB2}}{2}$&
    $\Xi_{c}^{\prime+}   \Xi_c^+  \eta_8  $ & $ \frac{\sqrt{3} g_{PBB2}}{2}$&
    $\Xi_{c}^{\prime+}   \Xi_c^0  \pi^+  $ & $ \frac{g_{PBB2}}{\sqrt{2}}$&
    $\Xi_{c}^{\prime0}   \Lambda_c^+  K^-  $ & $ \frac{g_{PBB2}}{\sqrt{2}}$\\\hline
    $\Xi_{c}^{\prime0}   \Xi_c^+  \pi^-  $ & $ \frac{g_{PBB2}}{\sqrt{2}}$&
    $\Xi_{c}^{\prime0}   \Xi_c^0  \pi^0  $ & $ -\frac{g_{PBB2}}{2}$&
    $\Xi_{c}^{\prime0}   \Xi_c^0  \eta_8  $ & $ \frac{\sqrt{3} g_{PBB2}}{2}$&
    $\Omega_{c}^{0}   \Xi_c^+  K^-  $ & $ g_{PBB2}$\\\hline
    $\Omega_{c}^{0}   \Xi_c^0  \overline K^0  $ & $ g_{PBB2}$\\
    \hline
    $\Sigma_{c}^{++}   \Sigma_{c}^{++}  \pi^0  $ & $ \frac{g_{PBB3}}{\sqrt{2}}$&
    $\Sigma_{c}^{++}   \Sigma_{c}^{++}  \eta_8  $ & $ \frac{g_{PBB3}}{\sqrt{6}}$&
    $\Sigma_{c}^{++}   \Sigma_{c}^{++}  \eta_{1}  $ & $ \frac{g_{PBB3}}{\sqrt{3}}$&
    $\Sigma_{c}^{++}   \Sigma_{c}^{+}  \pi^+  $ & $ \frac{g_{PBB3}}{\sqrt{2}}$\\\hline
    $\Sigma_{c}^{++}   \Xi_{c}^{\prime+}  K^+  $ & $ \frac{g_{PBB3}}{\sqrt{2}}$&
    $\Sigma_{c}^{+}   \Sigma_{c}^{++}  \pi^-  $ & $ \frac{g_{PBB3}}{\sqrt{2}}$&
    $\Sigma_{c}^{+}   \Sigma_{c}^{+}  \eta_8  $ & $ \frac{g_{PBB3}}{\sqrt{6}}$&
    $\Sigma_{c}^{+}   \Sigma_{c}^{+}  \eta_{1}  $ & $ \frac{g_{PBB3}}{\sqrt{3}}$\\\hline
    $\Sigma_{c}^{+}   \Sigma_{c}^{0}  \pi^+  $ & $ \frac{g_{PBB3}}{\sqrt{2}}$&
    $\Sigma_{c}^{+}   \Xi_{c}^{\prime+}  K^0  $ & $ \frac{g_{PBB3}}{2}$&
    $\Sigma_{c}^{+}   \Xi_{c}^{\prime0}  K^+  $ & $ \frac{g_{PBB3}}{2}$&
    $\Sigma_{c}^{0}   \Sigma_{c}^{+}  \pi^-  $ & $ \frac{g_{PBB3}}{\sqrt{2}}$\\\hline
    $\Sigma_{c}^{0}   \Sigma_{c}^{0}  \pi^0  $ & $ -\frac{g_{PBB3}}{\sqrt{2}}$&
    $\Sigma_{c}^{0}   \Sigma_{c}^{0}  \eta_8  $ & $ \frac{g_{PBB3}}{\sqrt{6}}$&
    $\Sigma_{c}^{0}   \Sigma_{c}^{0}  \eta_{1}  $ & $ \frac{g_{PBB3}}{\sqrt{3}}$&
    $\Sigma_{c}^{0}   \Xi_{c}^{\prime0}  K^0  $ & $ \frac{g_{PBB3}}{\sqrt{2}}$\\\hline
    $\Xi_{c}^{\prime+}   \Sigma_{c}^{++}  K^-  $ & $ \frac{g_{PBB3}}{\sqrt{2}}$&
    $\Xi_{c}^{\prime+}   \Sigma_{c}^{+}  \overline K^0  $ & $ \frac{g_{PBB3}}{2}$&
    $\Xi_{c}^{\prime+}   \Xi_{c}^{\prime+}  \pi^0  $ & $ \frac{g_{PBB3}}{2 \sqrt{2}}$&
    $\Xi_{c}^{\prime+}   \Xi_{c}^{\prime+}  \eta_8  $ & $ -\frac{g_{PBB3}}{2 \sqrt{6}}$\\\hline
    $\Xi_{c}^{\prime+}   \Xi_{c}^{\prime+}  \eta_{1}  $ & $ \frac{g_{PBB3}}{\sqrt{3}}$&
    $\Xi_{c}^{\prime+}   \Xi_{c}^{\prime0}  \pi^+  $ & $ \frac{g_{PBB3}}{2}$&
    $\Xi_{c}^{\prime+}   \Omega_{c}^{0}  K^+  $ & $ \frac{g_{PBB3}}{\sqrt{2}}$&
    $\Xi_{c}^{\prime0}   \Sigma_{c}^{+}  K^-  $ & $ \frac{g_{PBB3}}{2}$\\\hline
    $\Xi_{c}^{\prime0}   \Sigma_{c}^{0}  \overline K^0  $ & $ \frac{g_{PBB3}}{\sqrt{2}}$&
    $\Xi_{c}^{\prime0}   \Xi_{c}^{\prime+}  \pi^-  $ & $ \frac{g_{PBB3}}{2}$&
    $\Xi_{c}^{\prime0}   \Xi_{c}^{\prime0}  \pi^0  $ & $ -\frac{g_{PBB3}}{2 \sqrt{2}}$&
    $\Xi_{c}^{\prime0}   \Xi_{c}^{\prime0}  \eta_8  $ & $ -\frac{g_{PBB3}}{2 \sqrt{6}}$\\\hline
    $\Xi_{c}^{\prime0}   \Xi_{c}^{\prime0}  \eta_{1}  $ & $ \frac{g_{PBB3}}{\sqrt{3}}$&
    $\Xi_{c}^{\prime0}   \Omega_{c}^{0}  K^0  $ & $ \frac{g_{PBB3}}{\sqrt{2}}$&
    $\Omega_{c}^{0}   \Xi_{c}^{\prime+}  K^-  $ & $ \frac{g_{PBB3}}{\sqrt{2}}$&
    $\Omega_{c}^{0}   \Xi_{c}^{\prime0}  \overline K^0  $ & $ \frac{g_{PBB3}}{\sqrt{2}}$\\\hline
    $\Omega_{c}^{0}   \Omega_{c}^{0}  \eta_8  $ & $ -\sqrt{\frac{2}{3}} g_{PBB3}$&
    $\Omega_{c}^{0}   \Omega_{c}^{0}  \eta_{1}  $ & $ \frac{g_{PBB3}}{\sqrt{3}}$\\\hline
    \hline
    \end{tabular}
\end{table}
\begin{table}
\caption{similar with Tab.~\ref{tab:PBB}, the strong couplings of ${\cal B}_{\mathbf{\bar 3}}{\cal B}_{\mathbf{\bar 3}}V$, ${\cal B}_{\mathbf{6}}{\cal B}_{\mathbf{\bar 3}}V$ and ${\cal B}_{\mathbf{6}}{\cal B}_{\mathbf{6}}V$ can be obtained by using the coupling constant $g_{\Xi_{c}^{0}\Lambda_{c}^{+}K^{*-}} = \{4.6,6\}= g_{VBB1}$, $g_{\Sigma_{c}^{0}\Lambda_{c}^{+}\rho^{-}} = \{2.6,16\}= g_{VBB2}$ and $g_{\Sigma_{c}^{+}\Sigma_{c}^{0}\rho^{+}} = \{4.0,27\}=\frac{g_{VBB3}}{\sqrt{2}}$~\cite{Aliev:2010nh}, respectively. Then the couplings ${\cal B}_{\mathbf{6}}^{*}{\cal B}_{\mathbf{\bar 3}}V$ and ${\cal B}_{\mathbf{6}}^{*}{\cal B}_{\mathbf{6}}V$ can be get from $g_{\Sigma_{c}^{*+}\Lambda_{c}^{+}\rho^{0}} = 10=g_{VB^{*}B2}$ and $g_{\Sigma_{c}^{*0}\Sigma_{c}^{+}\rho^{-}} = 5.77=\frac{g_{VB^{*}B3}}{\sqrt{2}}$~\cite{Lin:2017mtz}, respectively.}
\label{tab:VBB}
\begin{tabular}{cc|cc|cc|cc}
\hline\hline
Vertex&g&Vertex&g&Vertex&g&Vertex&g\\\hline
    $\Lambda_c^+   \Lambda_c^+  \omega  $ & $ -\sqrt{2} g_{VBB1}$&
    $\Lambda_c^+   \Xi_c^+  K^{*0}  $ & $ -g_{VBB1}$&
    $\Lambda_c^+   \Xi_c^0  K^{*+}  $ & $ g_{VBB1}$&
    $\Xi_c^+   \Lambda_c^+  \overline{K}^{*0} $ & $ -g_{VBB1}$\\\hline
    $\Xi_c^+   \Xi_c^+  \rho^0  $ & $ -\frac{g_{VBB1}}{\sqrt{2}}$&
    $\Xi_c^+   \Xi_c^+  \omega  $ & $ -\frac{g_{VBB1}}{\sqrt{2}}$&
    $\Xi_c^+   \Xi_c^+  \phi $ & $ -g_{VBB1}$&
    $\Xi_c^+   \Xi_c^0  \rho^+  $ & $ -g_{VBB1}$\\\hline
    $\Xi_c^0   \Lambda_c^+  K^{*-}  $ & $ g_{VBB1}$&
    $\Xi_c^0   \Xi_c^+  \rho^-  $ & $ -g_{VBB1}$&
    $\Xi_c^0   \Xi_c^0  \rho^0  $ & $ \frac{g_{VBB1}}{\sqrt{2}}$&
    $\Xi_c^0   \Xi_c^0  \omega  $ & $ -\frac{g_{VBB1}}{\sqrt{2}}$\\\hline
    $\Xi_c^0   \Xi_c^0  \phi $ & $ -g_{VBB1}$
    \\\hline
    $\Sigma_{c}^{++}   \Lambda_c^+  \rho^+  $ & $ -g_{VBB2}$&
    $\Sigma_{c}^{++}   \Xi_c^+  K^{*+}  $ & $ -g_{VBB2}$&
    $\Sigma_{c}^{+}   \Lambda_c^+  \rho^0  $ & $ g_{VBB2}$&
    $\Sigma_{c}^{+}   \Xi_c^+  K^{*0}  $ & $ -\frac{g_{VBB2}}{\sqrt{2}}$\\\hline
    $\Sigma_{c}^{+}   \Xi_c^0  K^{*+}  $ & $ -\frac{g_{VBB2}}{\sqrt{2}}$&
    $\Sigma_{c}^{0}   \Lambda_c^+  \rho^-  $ & $ g_{VBB2}$&
    $\Sigma_{c}^{0}   \Xi_c^0  K^{*0}  $ & $ -g_{VBB2}$&
    $\Xi_{c}^{\prime+}   \Lambda_c^+  \overline{K}^{*0} $ & $ -\frac{g_{VBB2}}{\sqrt{2}}$\\\hline
    $\Xi_{c}^{\prime+}   \Xi_c^+  \rho^0  $ & $ \frac{g_{VBB2}}{2}$&
    $\Xi_{c}^{\prime+}   \Xi_c^+  \omega  $ & $ \frac{g_{VBB2}}{2}$&
    $\Xi_{c}^{\prime+}   \Xi_c^+  \phi $ & $ -\frac{g_{VBB2}}{\sqrt{2}}$&
    $\Xi_{c}^{\prime+}   \Xi_c^0  \rho^+  $ & $ \frac{g_{VBB2}}{\sqrt{2}}$\\\hline
    $\Xi_{c}^{\prime0}   \Lambda_c^+  K^{*-}  $ & $ \frac{g_{VBB2}}{\sqrt{2}}$&
    $\Xi_{c}^{\prime0}   \Xi_c^+  \rho^-  $ & $ \frac{g_{VBB2}}{\sqrt{2}}$&
    $\Xi_{c}^{\prime0}   \Xi_c^0  \rho^0  $ & $ -\frac{g_{VBB2}}{2}$&
    $\Xi_{c}^{\prime0}   \Xi_c^0  \omega  $ & $ \frac{g_{VBB2}}{2}$\\\hline
    $\Xi_{c}^{\prime0}   \Xi_c^0  \phi $ & $ -\frac{g_{VBB2}}{\sqrt{2}}$&
    $\Omega_{c}^{0}   \Xi_c^+  K^{*-}  $ & $ g_{VBB2}$&
    $\Omega_{c}^{0}   \Xi_c^0  \overline{K}^{*0} $ & $ g_{VBB2}$
    \\\hline
    $\Sigma_{c}^{++}   \Sigma_{c}^{++}  \rho^0  $ & $ \frac{g_{VBB3}}{\sqrt{2}}$&
    $\Sigma_{c}^{++}   \Sigma_{c}^{++}  \omega  $ & $ \frac{g_{VBB3}}{\sqrt{2}}$&
    $\Sigma_{c}^{++}   \Sigma_{c}^{+}  \rho^+  $ & $ \frac{g_{VBB3}}{\sqrt{2}}$&
    $\Sigma_{c}^{++}   \Xi_{c}^{\prime+}  K^{*+}  $ & $ \frac{g_{VBB3}}{\sqrt{2}}$\\\hline
    $\Sigma_{c}^{+}   \Sigma_{c}^{++}  \rho^-  $ & $ \frac{g_{VBB3}}{\sqrt{2}}$&
    $\Sigma_{c}^{+}   \Sigma_{c}^{+}  \omega  $ & $ \frac{g_{VBB3}}{\sqrt{2}}$&
    $\Sigma_{c}^{+}   \Sigma_{c}^{0}  \rho^+  $ & $ \frac{g_{VBB3}}{\sqrt{2}}$&
    $\Sigma_{c}^{+}   \Xi_{c}^{\prime+}  K^{*0}  $ & $ \frac{g_{VBB3}}{2}$\\\hline
    $\Sigma_{c}^{+}   \Xi_{c}^{\prime0}  K^{*+}  $ & $ \frac{g_{VBB3}}{2}$&
    $\Sigma_{c}^{0}   \Sigma_{c}^{+}  \rho^-  $ & $ \frac{g_{VBB3}}{\sqrt{2}}$&
    $\Sigma_{c}^{0}   \Sigma_{c}^{0}  \rho^0  $ & $ -\frac{g_{VBB3}}{\sqrt{2}}$&
    $\Sigma_{c}^{0}   \Sigma_{c}^{0}  \omega  $ & $ \frac{g_{VBB3}}{\sqrt{2}}$\\\hline
    $\Sigma_{c}^{0}   \Xi_{c}^{\prime0}  K^{*0}  $ & $ \frac{g_{VBB3}}{\sqrt{2}}$&
    $\Xi_{c}^{\prime+}   \Sigma_{c}^{++}  K^{*-}  $ & $ \frac{g_{VBB3}}{\sqrt{2}}$&
    $\Xi_{c}^{\prime+}   \Sigma_{c}^{+}  \overline{K}^{*0} $ & $ \frac{g_{VBB3}}{2}$&
    $\Xi_{c}^{\prime+}   \Xi_{c}^{\prime+}  \rho^0  $ & $ \frac{g_{VBB3}}{2 \sqrt{2}}$\\\hline
    $\Xi_{c}^{\prime+}   \Xi_{c}^{\prime+}  \omega  $ & $ \frac{g_{VBB3}}{2 \sqrt{2}}$&
    $\Xi_{c}^{\prime+}   \Xi_{c}^{\prime+}  \phi $ & $ \frac{g_{VBB3}}{2}$&
    $\Xi_{c}^{\prime+}   \Xi_{c}^{\prime0}  \rho^+  $ & $ \frac{g_{VBB3}}{2}$&
    $\Xi_{c}^{\prime+}   \Omega_{c}^{0}  K^{*+}  $ & $ \frac{g_{VBB3}}{\sqrt{2}}$\\\hline
    $\Xi_{c}^{\prime0}   \Sigma_{c}^{+}  K^{*-}  $ & $ \frac{g_{VBB3}}{2}$&
    $\Xi_{c}^{\prime0}   \Sigma_{c}^{0}  \overline{K}^{*0} $ & $ \frac{g_{VBB3}}{\sqrt{2}}$&
    $\Xi_{c}^{\prime0}   \Xi_{c}^{\prime+}  \rho^-  $ & $ \frac{g_{VBB3}}{2}$&
    $\Xi_{c}^{\prime0}   \Xi_{c}^{\prime0}  \rho^0  $ & $ -\frac{g_{VBB3}}{2 \sqrt{2}}$\\\hline
    $\Xi_{c}^{\prime0}   \Xi_{c}^{\prime0}  \omega  $ & $ \frac{g_{VBB3}}{2 \sqrt{2}}$&
    $\Xi_{c}^{\prime0}   \Xi_{c}^{\prime0}  \phi $ & $ \frac{g_{VBB3}}{2}$&
    $\Xi_{c}^{\prime0}   \Omega_{c}^{0}  K^{*0}  $ & $ \frac{g_{VBB3}}{\sqrt{2}}$&
    $\Omega_{c}^{0}   \Xi_{c}^{\prime+}  K^{*-}  $ & $ \frac{g_{VBB3}}{\sqrt{2}}$\\\hline
    $\Omega_{c}^{0}   \Xi_{c}^{\prime0}  \overline{K}^{*0} $ & $ \frac{g_{VBB3}}{\sqrt{2}}$&
    $\Omega_{c}^{0}   \Omega_{c}^{0}  \phi $ & $ g_{VBB3}$\\\hline
    \hline
    \end{tabular}
\end{table}

\end{document}